\documentclass[twoside,11pt] {article}

\newcommand{\mb}[1]{{\mbox{\boldmath{$#1$}}}}

\begin{document}

\thispagestyle{plain}
\setcounter{page}{1}

\input amssym.tex

\newcommand{\Lor}{L^{\uparrow}_{+}}
\undefine\circledS
\newsymbol\circledS 1073

\newtheorem{lem}{Lemma}
\newtheorem{defin}{Definition}
\newtheorem{theor}{Theorem}
\newtheorem{rem}{Remark}
\newtheorem{prop}{Proposition}
\newtheorem{cor}{Corollary}
\newenvironment{demo}
{\bgroup\par\smallskip\noindent{\it Proof: }}{\rule{0.5em}{0.5em}
\egroup}

\title{Symmetries and supersymmetries of the Dirac operators in curved
spacetimes\thanks
{Final version of Ref. \cite{NOVA} containing new results,
extended discussions and some corrections.}}

\author{Ion I. Cot\u aescu \thanks{E-mail:~~~cota@physics.uvt.ro}\\
{\small \it West University of Timi\c soara,}\\
       {\small \it V. P\^ arvan Ave. 4, RO-1900 Timi\c soara, Romania}
\and
Mihai Visinescu \thanks{E-mail:~~~mvisin@theory.nipne.ro}\\
{\small \it Department of Theoretical Physics,}\\
{\small \it National Institute for Physics and Nuclear Engineering,}\\
{\small \it P.O.Box M.G.-6, RO-077125 Magurele, Bucharest, Romania}}
\date{}

\maketitle

\begin{abstract}

It is shown that the main geometrical objects involved in all the symmetries or
supersymmetries of the Dirac operators in curved manifolds of arbitrary
dimensions are the Killing vectors and the Killing-Yano tensors of any ranks.
The general theory of external symmetry transformations associated to the usual
isometries is presented, pointing out that these leave the standard Dirac
equation invariant providing the correct spin parts of the group generators.
Furthermore, one analyzes the new type of symmetries generated by the
covariantly constant Killing-Yano tensors  that realize  certain square roots
of the metric tensor. Such a Killing-Yano tensor produces simultaneously a
Dirac-type operator and the generator of a one-parameter Lie group connecting
this operator with the standard Dirac one. In this way the Dirac operators are
related among themselves through continuous transformations associated to
specific discrete ones. It is shown that the groups of this continuous symmetry
can be only $U(1)$ or $SU(2)$, as those of the (hyper-)K\"ahler spaces, but
arising even in cases when the requirements for these special geometries are
not fulfilled. Arguments are given that for the non-K\" ahlerian manifolds it
is convenient to enlarge this $SU(2)$ symmetry up to a $SL(2, {\Bbb C})$ one
through complexification. In other respects, it is pointed out that the
Dirac-type operators can form ${\cal N}=4$ superalgebras whose automorphisms
combine external symmetry transformations with those of the mentioned $SU(2)$
or $SL(2, {\Bbb C})$ groups. The discrete symmetries are also studied obtaining
the discrete groups  ${\Bbb Z}_4$ and ${\Bbb Q}$. To exemplify, the Euclidean
Taub-NUT space with its Dirac-type operators is presented in much details,
pointing out that there is an infinite-loop superalgebra playing the role of a
closed dynamical algebraic structure. As a final topic, we go to
consider the properties of the Dirac-type
operators of the Minkowski spacetime.

Pacs 04.62.+v

\end{abstract}

\tableofcontents

\section{Introduction}

The quantum physics in curved backgrounds needs to use algebras of operators
acting on spaces of vector, tensor or spinor fields whose properties depend on
the geometry of the manifolds where these objects are defined. A crucial
problem is to find the symmetries having geometrical sources able to produce
conserved quantities related to operators with specific algebraic properties.
The problem is not trivial since, beside the evident geometrical symmetry
given by isometries, there are different types of hidden symmetries frequently
associated with supersymmetries that deserve to be carefully studied.

The isometries are related to the existence of the Killing vectors that
give rise to the orbital operators of the scalar quantum theory commuting
with that of the free field equation. In the theories with spin these operators
get specific spin terms whose form is strongly dependent on the local
non-holonomic frames we choose by fixing the gauge. For the Dirac
field, these spin parts are known  in any chart and arbitrary tetrad gauge
fixing of the four-dimensional manifolds \cite{CML,DGB}. However, with these
results one cannot say that this problem is generally solved, for fields with
any spin obeying different free or coupled field equations. For this reason the
theory of isometries was extended to that of the external symmetry which allows
one to pick up well-defined conserved quantities in theories with matter fields
of {\em any spin} \cite{ES,CDS}. Here we extend the theory of external
symmetries to the Dirac theory in manifolds of any dimensions.

Another type of geometrical objects related to the so called hidden symmetries
or several specific supersymmetries are the Killing-Yano (K-Y) tensors \cite{Y}
and the {St\" ackel-Killing} (S-K) tensors of any rank. The K-Y tensors play
an important role in theories with spin and especially in the Dirac theory on
curved spacetimes where they produce first order differential operators, called
Dirac-type operators, which anticommute with the standard Dirac one, $D$
\cite{CML,MaCa}. Another virtue of the K-Y tensors is that they enter as square
roots in the structure of several second rank {S-K} tensors that
generate conserved quantities in classical mechanics or conserved operators
which commute with $D$. The construction of Ref. \cite{CML} depends upon
the remarkable fact that the {S-K} tensors  must have  square
root in terms of K-Y tensors in order to eliminate the quantum anomaly
and produce operators commuting with $D$ \cite{PF}.
These  attributes of the K-Y tensors lead to an efficient mechanism of
supersymmetry especially when the {S-K} tensor is proportional
with the metric tensor and the corresponding roots are covariantly constant
K-Y tensors.  Then each tensor of this type, $f^i$, gives rise to a Dirac-type
operator,  $D^i$,  representing a supercharge of a non-trivial superalgebra
$\{ D^i , D^j \} \propto D^2 \delta_{ij}$ \cite{CV2}.
It was shown that $D^i$  can be produced by covariantly constant K-Y tensors
having not only real-valued components but also complex ones \cite{K1,K2}.
This represents an extension of the K\" ahlerian manifolds that seems to be
productive for the Dirac theory since it permits to construct superalgebras
of Dirac-type operators even in the Minkowski spacetime which is not
K\" ahlerian, having only complex-valued covariantly constant K-Y tensors
\cite{K2,CV7}. For this reason, in what follows we shall consider such more
general tensors, called  {\em unit roots} (instead of complex structures)
since all of them are roots of the metric tensor. We note that
the complex structures defining K\" ahlerian geometries are special
automorphisms of the tangent fiber bundle while the unit roots we consider
here are automorphisms of the {\em complexified} tangent fiber bundle. A part
of this paper is devoted to the theory of the Dirac-type operators generated by
unit roots \cite{CV7,CV6}.

It is known that in four-dimensional manifolds the standard Dirac operator
and the Dirac-type ones can be related among themselves through continuous
or discrete transformations \cite{CV6,K2}. It is interesting that there are
only two possibilities, namely either transformations of the $U(1)$ group
associated with the discrete group ${\Bbb Z}_4$ or $SU(2)$ transformations
and discrete ones of the quaternionic group ${\Bbb Q}$ \cite{CV6,K2,CV7}.
Particularly, in the case when the roots are real-valued (complex structures)
the first type of symmetry is proper to  K\" ahler manifolds while the second
largest one is characteristic for hyper-K\" ahler geometries \cite{CV6}. The
problem is what happens in the case of manifolds with larger number of
dimensions allowing complex-valued roots. We have shown that, in general, there
are no larger symmetries of this type \cite{CV7} but here we point out how
these could be embedded with the isometries. Other new result is that in
non-K\" ahlerian manifolds with complex-valued unit roots this operation
requires to enlarge the $SU(2)$ symmetry up to a $SL(2,{\Bbb C})$ one using
complexification.

The typical example is the Euclidean Taub-NUT space which is a hyper-K\" ahler
manifold possessing three covariantly constant K-Y tensors with real-valued
components which constitute a hypercomplex structure generating a ${\cal N}=4$
superalgebra of Dirac-type operators \cite{CV2}, in a similar way as in
semi-classical spinning models \cite{VV1,VV2}. Moreover, each involved K-Y
tensor is a unit root of the metric tensor as it results from the definition of
the K\" ahlerian geometries (given in Appendix A). It is worth pointing out
that the Euclidean Taub-NUT space has, in addition, a non-covariantly constant
K-Y tensor related to its specific hidden symmetry showed off by the existence
of a conserved Runge-Lenz operator that can be constructed with the help of the
Dirac-type operators produced by the four K-Y tensors of this space
\cite{CV3,CV4}. In Euclidean Taub-NUT space there are no quantum anomalies
\cite{PW} and one obtains a rich algebra of conserved observables \cite{CV5}
that offers many possibilities to choose sets of commuting operators defining
quantum modes \cite{CV1,CV2}. On the other hand, one can select or build
superalgebras, dynamical algebras typical for the Keplerian problems
\cite{CV4}, or even interesting infinite-dimensional algebras or superalgebras.
Our last objective here is to present the complete Dirac theory in the Taub-NUT
background including our new results concerning the twisted infinite-loop
superalgebra of the Dirac theory on Taub-NUT background.

The paper is organized as follows.

We start in the second section with the
construction of a simple version of the Dirac theory in manifolds of any
dimensions, introducing the group of the external symmetry in non-holonomic
local frames as the universal covering group of the isometry one \cite{ES}. In
this way we can define the spinor representation  recovering the specific form
of its generators  Ref. \cite{CML} in a suitable context that allows us to use
the Noether theorem for deriving conserved quantities \cite{CDS}.

In the next
section we present the theory of the Dirac-type operators produced by the unit
roots. The continuous and discrete symmetries of these operators are studied
showing that there exists either an $U(1)$ symmetry associated to a single unit
root or a $SU(2)$ one of of the Dirac-type operators produced by triplets of
unit roots. Moreover, we point out that the these triplets give rise to
triplets of Dirac-type operators, $D^i$, $i=1,2,3$ anticommuting with $D$ and
among themselves too, forming thus a basis of a ${\cal N}=4$ superalgebra.
Furthermore, we show that in the case of the hyper-K\" ahler manifolds, the
automorphisms of these superalgebras combine the mentioned $SU(2)$ specific
transformations with those of a representation of the group of external
symmetry induced by $SO(3)$ rotations among the triplet elements. For the
superalgebras generated in non-K\" ahlerian manifolds the problem is more
complicated since there the isometry transformations are induced by the
complexified group $O(3)_c$  which forces us to consider complexified groups
and Lie algebras.  The discrete  symmetries  associated with the continuous
ones, ${\Bbb Z}_4$ and  ${\Bbb Q}$, as well as the parity and the charge
conjugation are also presented.

The Section 4 is devoted to the theory of
the Dirac operators in the Euclidean Taub-NUT space. Adopting a group
theoretical point of view, we start with the integral form of the isometry
transformations we have recently obtained and the orbital angular momentum
operator that deals with them \cite{CV8}. Moreover, we show why in the usual
gauge the whole theory presents a $SO(3)$ global symmetry and we review the
principal operators of the scalar Klein-Gordon, Pauli and Dirac theory. Some
important algebraic features are pointed out giving a special attention to an
association among Pauli and Dirac conserved operators \cite{CV5} that
simplifies the algebraic calculus leading to possible infinite-dimensional
superalgebras. Different discrete quantum Dirac modes are constructed with the
help of  new types of spherical harmonics and spinors \cite{CV1,CV2,CV4}.
The last two subsections are devoted to the construction of a twisted infinite-loop
superalgebra which takes over here the role of the closed dynamical algebraic
structure associated to the Dirac theory on Taub-NUT manifolds.

Finally, in a short section we present some conclusions and in
three appendices we briefly discuss the K\" ahlerian manifolds, the
properties of the Dirac-type operators of the Minkowski spacetime and a
representation of our infinite-loop superalgebra.

\section{The external symmetry of the Dirac field}

The relativistic covariance  in the sense of general relativity is
too general to play the same role as the Lorentz or Poincar\' e covariance in
special relativity  \cite{W}. In other respects, the gauge covariance of
the theories with spin represents another kind of general symmetry that is not
able to produce itself conserved observable  \cite{SW}. Therefore, if
we look for sources of symmetries able to generate conserved quantities, we
have to concentrate first on isometries that point out the spacetime
symmetry  giving us the specific Killing vectors  \cite{SW,WALD,ON}. The
physical fields minimally coupled with the gravitational one take over
this symmetry, transforming according to different representations of the
isometry group. In the case of the scalar vector or tensor fields these
representations are completely defined by the well-known  rules of the
general coordinate transformations  since the isometries are in fact
particular coordinate transformations. However, the behavior under isometries
of the fields with half integer spin is more complicated since their
transformation rules explicitly depend on the gauge fixing. The specific
theory of this type of transformations is the recent theory of external
symmetry we present in this section  \cite{ES}.

\subsection{Clifford algebra and the gauge group}

The theory of the Dirac spinors in arbitrary dimensions depends on the choice
of the manifold and Clifford algebra. Bearing in mind that the irreducible
representations of the Clifford algebra can have only an odd number of
dimensions, we consider a $2l+1$-dimensional pseudo-Riemannian manifold
$M_{2l+1}$ whose flat metric $\tilde\eta$ (of its pseudo-Euclidean model) has
the signature $( m_{+}, m_{-})$ where $ m_{+} + m_{-}=m=2l+1$. This is the
{\em maximal} manifold that can be associated to the $2l+1$-dimensional
Clifford algebra  \cite{Clif} acting on the $2^l$-dimensional space
$\Psi$ of the complex spinors $\psi=\tilde\varphi_1 \otimes \tilde\varphi_2 ...
\otimes\tilde\varphi_l$ built using  complex two-dimensional Pauli
spinors $\tilde\varphi$. In this algebra we start with the standard Euclidean
basis formed by the hermitian
matrices $\tilde\gamma^A= (\tilde\gamma^{A})^{+}$ ($A,B,...=1,2,...,m$) that
obey  $\{\tilde\gamma^{A},\, \tilde\gamma^{B} \} =2\delta^{A B}{\mb 1}$ where
${\mb 1}$ is the identity matrix. Furthermore, we define
the suitable basis corresponding to the metric $\tilde\eta$ as
\begin{equation}
\gamma^A=\left\{
\begin{array}{lll}
\tilde\gamma^A&{\rm for}& A=1,2,...,  m_{+}\\
i\tilde\gamma^A&{\rm for}& A= m_{+}+1,  m_{+}+2,..., m
\end{array}\right.\,,
\end{equation}
such that
\begin{equation}\label{ACOM}
\{ \gamma^{A},\, \gamma^{B} \}
=2\tilde\eta^{A B}{\mb 1} \,.
\end{equation}
Since the first $ m_{+}$ matrices $\gamma^A$ remain hermitian while the
$ m_{-}$ last ones become anti-hermitian, it seems that the unitaryness
of the theory is broken. However, this can be restored replacing the usual
Hermitian adjoint with the generalized Dirac adjoint  \cite{Pro}.
\begin{defin}
We say that $\overline{\psi}=\psi^{+}\gamma$ is the generalized Dirac adjoint
of the field $\psi$ if the hermitian matrix  $\gamma=\gamma^{+}$  satisfies
the condition $(\gamma)^2={\mb 1}$ and all the matrices $\gamma^A$ are either
self-adjoint or anti self-adjoint with respect to this operation,
i.e. $\overline{\gamma}^A=\gamma (\gamma^A)^{+}\gamma=\pm \gamma^A$.
\end{defin}
It is clear that the matrix $\gamma$ play here the role of {\em metric
operator} giving the generalized Dirac adjoint of any square matrix $X$ as
$\overline{X}=\gamma X^{+} \gamma$.
\begin{theor}
The metric operator can be represented as the product
$\gamma= \epsilon \,\gamma^{1}\gamma^{2}...\gamma^{m_{+}}$
with the phase factor
\begin{equation}
\epsilon =\left\{
\begin{array}{lll}
(i)^{\frac{{m}_{+}-1}{2}}&{\rm for~ odd}&{m}_{+}<m
\\
(i)^{\frac{{m}_{+}}{2}}&{\rm for~ even}&{m}_{+}<m
\end{array}\right.\,.
\end{equation}
\end{theor}
\begin{demo}
In the special case of the Euclidean metric (when $ m_{-}=0$) we have the
trivial solution $\gamma={\mb 1}$. Otherwise, the algebraic properties of the
matrix $\gamma$  depends on  $m_{+}$ such that for $ m_{+}$ taking odd values
we have  the following superalgebra
\begin{eqnarray}
[\gamma,\,\gamma^{A}]=0& {\rm for}& A=1,2,...,m_{+}\,,
\nonumber\\
\{\gamma,\,\gamma^{A}\}=0& {\rm for}& A=m_{+}+1,...,m\,,
\end{eqnarray}
while for even $ m_{+}$ the situation is reversed. Consequently, one
can verify that
\begin{equation}
\overline{\gamma^{A}}=\left\{
\begin{array}{lll}
\gamma^{A}& {\rm for~ odd}& m_{+}\\
-\gamma^{A}& {\rm for~ even}& m_{+}
\end{array}\right.\,,\quad  A=1,2,...,m\,,
\end{equation}
which means that from the point of view of the Dirac adjoint all the
matrices $\gamma^A$ have the same behavior, being either self-adjoint
or anti self-adjoint. Thus the unitaryness of the theory is guaranteed.
\end{demo}\\
In what follows we consider that $m_{+}$ is odd representing the number of
time-like coordinates and match all the phase factors according to self-adjoint
gamma-matrices.
\begin{rem}\label{rem1}
When these matrices are anti self-adjoint (because of an even $m_{+}$) it
suffices to change $\gamma^A\to \pm i\gamma^A$ in the formulas of all the
operators one defines. The same procedure is indicated when one works
with self-adjoint gamma-matrices but metrics of reversed signature, with
$m_{+}$ space-like coordinates.
\end{rem}

The isometry group $G(\tilde\eta)=O(m_{+}, m_{-})$ of the metric $\tilde\eta$,
with the mentioned signature, is the {\em gauge} group of the theory defining
the principal fiber bundle. This is a pseudo-orthogonal group that
admits an universal covering group ${\mb G}(\tilde \eta)$ which is simply
connected and has the same Lie algebra we denote by ${\mb g}(\tilde \eta)$.
The group ${\mb G}(\tilde \eta)$ is the model of the spinor fiber
bundle that completes the spin structure we need. In order to avoid
complications due to the presence of these two groups we consider here that
the basic piece is the group ${\mb G}(\tilde\eta)$, denoting by $[\omega]$
their elements in the standard {\em covariant} parametrization given by the
skew-symmetric real parameters $\omega_{AB}=-\omega_{BA}$. Then the
identity element of ${\mb G}(\tilde\eta)$ is $1=[0]$ and the inverse of
$[\omega]$ with respect to the group multiplication reads
$[\omega]^{-1}=[-\omega]$.
\begin{defin}
We say that the gauge group is the {\em vector} representation of
${\mb G}(\tilde \eta)$ and denote $G(\tilde \eta)=
vect[{\mb G}(\tilde \eta)]$.
The representation $spin[{\mb G}(\tilde\eta)]$ carried by the space
$\Psi$ and generated by the spin operators
\begin{equation}\label{SAB}
S^{AB}=\frac{i}{4}\left[\gamma^{A},\,\gamma^{B}
\right]
\end{equation}
is called the {\em spinor} representation of ${\mb G}(\tilde\eta)$. The
spin operators are the basis generators of the spinor representation
$spin[{\mb g}(\tilde\eta)]$ of the Lie algebra ${\mb g}(\tilde\eta)$.
\end{defin}
In general, the spinor representation is reducible. Its generators are
self-adjoint, $\overline{S}^{AB} =S^{AB}$,  and satisfy
\begin{eqnarray}
[S^{AB},\,\gamma^{C}]&=&
i(\tilde\eta^{BC}\gamma^{A}-
\tilde\eta^{AC}\gamma^{B})\,,\label{Sgg}\\
{[} S_{AB},\,S_{CD} {]}&=&i(
\tilde\eta_{AD}\,S_{BC}-
\tilde\eta_{AC}\,S_{BD}+
\tilde\eta_{BC}\,S_{AD}-
\tilde\eta_{BD}\,S_{AC})\,,\label{SSS}
\end{eqnarray}
as it results from Eqs. (\ref{ACOM}) and (\ref{SAB}). It is obvious that
Eq. (\ref{SSS}) gives just the canonical commutation rules of a Lie algebra
isomorphic with that of the groups $G(\tilde \eta)$ or ${\mb G}(\tilde\eta)$.
The spinor and vector representations are related between themselves through
the following
\begin{theor}
For any real or complex valued skew-symmetric tensor $\omega_{AB}=-\omega_{BA}$
the matrix
\begin{equation}\label{TeS}
T(\omega)=e^{-iS(\omega)}\,,\quad S(\omega)=\frac{1}{2}
\omega_{AB} S^{AB}\,,
\end{equation}
transforms the gamma-matrices according to the rule
\begin{equation}\label{TgT}
[T(\omega)]^{-1}\gamma^{A}T(\omega)=\Lambda^{A\,\cdot}
_{\cdot\,B}(\omega)\gamma^{B}\,,
\end{equation}
where
\begin{equation}\label{Lam}
\Lambda^{A\,\cdot}_{\cdot\,B}(\omega)=
\delta^{A}_{B}
+\omega^{A\,\cdot}_{\cdot\,B}
+\frac{1}{2}\,\omega^{A\,\cdot}_{\cdot\,C}
\omega^{C\,\cdot}_{\cdot\,B}+...
+\frac{1}{n!}\,\underbrace{\omega^{A\,\cdot}_{\cdot\,C}
\omega^{C\,\cdot}_{\cdot\, C'}
...\omega^{D\,\cdot}_{\cdot\,B}}_{n}+...\,.
\end{equation}
\end{theor}
\begin{demo}
All these results can be obtained using Eqs. (\ref{ACOM}) and (\ref{SAB}).
\end{demo}\\
The real components $\omega_{AB}$ are the parameters of the covariant basis of
the Lie algebra ${\mb g}(\tilde \eta)$ giving all the  transformation
matrices $T(\omega) \in spin[{\mb G}(\tilde\eta)]$ and
$\Lambda(\omega)\in vect[{\mb G}(\tilde\eta)]$. Hereby we see that the
spinor representation $spin[{\mb G}(\tilde\eta)]$ is unitary since for
$\omega \in {\Bbb R}$ the generators $S(\omega)\in spin[{\mb g}(\tilde\eta)]$
are self-adjoint, $\overline{S}(\omega)=S(\omega)$, and the matrices
$T(\omega)$ are unitary with respect to the Dirac adjoint satisfying
$\overline{T}(\omega)=[T(\omega)]^{-1}$.

The covariant parameters $\omega$ can also take complex values.  Then this
parametri\-za\-ti\-on spans the {\em complexified} group of
${\mb G}(\tilde \eta)$, denoted by ${\mb G}_c(\tilde\eta)$, and the
corres\-pon\-ding vector and
(non-unitary) spinor representations. Obviously, in this case the Lie
algebra is the complexified algebra ${\mb g}_c(\tilde\eta)$.
We note that from the mathematical point of view
$G(\tilde\eta)=vect[{\mb G}(\tilde \eta)]$ is the group of automorphisms of
the tangent
fiber bundle ${\cal T}(M_m)$ of $M_m$ while the transformations of
$G_c(\tilde\eta)=vect[{\mb G}_c(\tilde \eta)]$ are automorphisms of the
complexified tangent fiber bundle ${\cal T}(M_m)\otimes {\Bbb C}$.

\subsection{The Dirac theory}

With these preparations, the gauge-covariant theory of the Dirac field can be
formulated on any submanifold of $M_{m=2l+1}$, like, for example,
in the usual (1+3)-dimensional spacetimes immersed in the five-dimensional
manifold of the Kaluza-Klein theory (with $l=2$). We consider the
general case of the Dirac theory on any submanifold $M_{n}\subset  M_{m}$  of
dimension $n\le m$ whose flat metric $\eta$ is a part (or restriction) of the
metric $\tilde\eta$, having the signature $(n_{+},n_{-})$, with
$n_{+}\le  m_{+}$,  $n_{-}\le  m_{-}$ and  $n_{+}+n_{-}=n$, such that
the gauge group is $G(\eta)=vect[{\mb G}(\eta)]=O(n_{+}, n_{-})$.
In $M_{n}$ we choose  a local chart (i.e. natural frame) with coordinates
$x^{\mu}$, $\alpha,...,\mu,\nu,...=1,2,...,n$, and introduce local orthogonal
non-holonomic frames using the gauge fields (or "vilbeins") $e(x)$ and $\hat
e(x)$, whose components
are labeled by local (hated) indices, $\hat\alpha,...\hat\mu,\hat\nu,...
=1,2,...,n$, that represent a subset of the  Latin capital ones, eventually
renumbered. The local indices have to be raised or lowered by the metric
$\eta$. The fields $e$ and $\hat e$ accomplish the conditions
\begin{equation}\label{eeee}
e_{\hat\alpha}^{\mu}\hat e_{\nu}^{\hat\alpha}=\delta_{\mu}^{\nu}\,,\quad
e_{\hat\alpha}^{\mu}\hat e_{\mu}^{\hat\beta}=\delta_{\hat\alpha}^{\hat\beta}
\end{equation}
and  orthogonality relations as
$g_{\mu\nu} e_{\hat\alpha}^{\mu} e^{\nu}_{\hat\beta}=
\eta_{\hat\alpha \hat\beta}$. With their help  the metric tensor of $M_n$ can
be put in the form   $g_{\mu\nu}(x)=\eta_{\hat\alpha \hat\beta}
\hat e ^{\hat\alpha}_{\mu}(x) \hat e^{\hat\beta}_{\nu}(x)$.
\begin{defin}\label{phys}
We call physical spacetimes the manifolds $M_n=M_{d+1}$ having only one
time-like coordinate $x^0=t$ and  $d=n-1$ space-like ones,
${\mb x}=(x^1,x^2, ... x^d)$, with  metrics of the signature $(1,d)$. In
addition, we assume that these manifolds are orientable and time-orientable
\cite{WALD}.
\end{defin}

The next step is to choose a suitable representation of the $n$ matrices
$\gamma^{\hat\alpha}$ obeying Eq. (\ref{ACOM}) and to calculate the spin
matrices $S^{\hat\alpha\hat\beta}$ defined by Eq. (\ref{SAB}). Now these are
the basis generators of the spinor representation $spin[{\mb g}(\eta)]$ of
the Lie algebra ${\mb g}(\eta)$, corresponding to the metric $\eta$.
If $n<m$ there are many matrices, $\gamma^{n+1},..., \gamma^{m}$,
which anticommutes with all the $n$ matrices $\gamma^{\hat\alpha}$ one
uses for the Dirac theory in $M_n$. We can select one of these extra
gamma-matrices denoting it by  $\gamma^{ch}$ and matching its phase factor
such that $(\gamma^{ch})^2={\mb 1}$ and $({\gamma}^{ch})^{+}= \gamma^{ch}$.
This matrix obeying
\begin{equation}
\left\{\gamma^{ch},\, \gamma^{\hat\mu}\right\}=0\,, \quad \hat\mu=1,2,...,n\,,
\end{equation}
is called the {\em chiral} matrix since it plays the same role as the
matrix $\gamma^5$ in the usual Dirac theory, helping us to distinguish between
even and odd matrices or matrix operators.
\begin{defin}
One says that a matrix operator acting on $\Psi$ is {\em even}
whenever it commutes with $\gamma^{ch}$ and is {\em odd} if it anticommutes
with this matrix.
\end{defin}
The matrix $\gamma$ can be either even (for even $m_{+}$) or odd (when
$m_{+}$ is odd). These two different situations lead to self-adjoint or
anti self-adjoint chiral matrices, such that it is convenient to use
the chiral phase factor $\epsilon_{ch}=\pm 1$ giving
$\overline{\gamma}^{ch}=\epsilon_{ch}
\gamma^{ch}$. In general, when $m>n$ we can define one or even many chiral
matrices different from $\gamma$ but for $n=m=2l+1$ we must take
$\gamma^{ch}=\gamma$. In any case, the space of the Dirac spinors can be split
in its left and right-handed parts, $\Psi=\Psi_L\oplus \Psi_R$  ($\Psi_L=
P_L\Psi$, $\Psi_R=P_R\Psi$), using the traditional projection operators
\begin{equation}\label{PLR}
P_L=\frac{1}{2}\left({\mb 1}-\gamma^{ch}\right)\,,\quad
P_R=\frac{1}{2}\left({\mb 1}+\gamma^{ch}\right)\,.
\end{equation}
In general, any matrix operator $X :\Psi\to \Psi$ can be written as the sum
$X=X_{even}+X_{odd}$, between its even part $X_{even}=P_LXP_L + P_RXP_R$ and
the odd one $X_{odd}=P_LXP_R + P_RXP_L$.
When one intends to exploit this mechanism it is convenient to use the
{\em chiral} representation of the gamma-matrices where
\begin{equation}
\gamma^{ch}=\left(
\begin{array}{cc}
-{\mb 1}_L&0\\
0&{\mb 1}_R
\end{array}\right)
\end{equation}
is diagonal and all the matrices $\gamma^{\hat\mu}$ are off-diagonal.
The notations  ${\mb 1}_L$ and ${\mb 1}_R$ stand for the identity matrices on
the spaces of spinors $\Psi_L$ and $\Psi_R$ respectively. The
gamma-matrices and the metric operator $\gamma$ in the chiral representation
must be calculated in each concrete case separately since they depend on the
metric signature. In what follows we assume that in the physical spacetimes
defined by Definition \ref{phys} the matrix $\gamma =\gamma^0$ in the
chiral representation is hermitian. Otherwise, we proceed as indicated in
Remark
\ref{rem1}.

The gauge-covariant theory of the free spinor field $\psi\in \Psi$ of the mass
$m_0$, defined on $M_n$, is based on the gauge invariant action
\begin{equation}\label{action}
{\cal S}[e,\psi]=\int\, d^{n}x\sqrt{g}\left\{
\frac{i}{2}[\overline{\psi}\gamma^{\hat\alpha}\nabla_{\hat\alpha}\psi-
(\overline{\nabla_{\hat\alpha}\psi})\gamma^{\hat\alpha}\psi] -
m_0\overline{\psi}\psi\right\}\,,
\end{equation}
where  $g=|\det(g_{\mu\nu})|$ and
$\nabla_{\mu}=\hat e_{\mu}^{\hat\alpha}\nabla_{\hat\alpha}=
\tilde\nabla_{\mu}+\Gamma_{\mu}^{spin}$
are the covariant derivatives formed by the usual ones, $\tilde\nabla_{\mu}$
(acting in natural indices), and the spin connection
\begin{equation}
\Gamma_{\mu}^{spin}=\frac{i}{2}
e^{\beta}_{\hat\nu}
(\hat e^{\hat\sigma}_{\alpha}\Gamma^{\alpha}_{\beta\mu}-
\hat e^{\hat\sigma}_{\beta,\mu} )
S^{\hat\nu\,\cdot}_{\cdot\,\hat\sigma}\,,
\end{equation}
giving $\nabla_{\mu}\psi=(\partial_{\mu}+\Gamma_{\mu}^{spin})\psi$. The
action (\ref{action}) produces the Dirac equation $D\psi =m_0\psi$ involving
the {\em standard} Dirac operator that can be expressed in terms of
point-dependent Dirac matrices as
\begin{equation}\label{Dirac}
D=i\gamma^{\mu}\nabla_{\mu}\,,\quad
\gamma^{\mu}(x)=e^{\mu}_{\hat\alpha}(x)\gamma^{\hat\alpha}\,.
\end{equation}
Now we can convince ourselves that our definition of the generalized Dirac
adjoint is correct since $\overline{\gamma^{\mu}}=\gamma^{\mu}$ and
$\overline{\Gamma}_{\mu}^{spin}=-\Gamma_{\mu}^{spin}$
such that the Dirac operator results to be self-adjoint, $\overline{D}=D$.
Moreover,  the quantity $\overline{\psi}\psi$ has to be derived as a
scalar, i.e. $\nabla_{\mu}(\overline{\psi}\psi)=
\overline{\nabla_{\mu}\psi}\,\psi+\overline{\psi}\,\nabla_{\mu}\psi=
\partial_{\mu}(\overline{\psi}\psi)$,
while the quantities $\overline{\psi}\gamma^{\alpha}\gamma^{\beta}...\psi$
behave as tensors of different ranks.

Thus we  reproduced  the main features of the familiar tetrad gauge covariant
theories with spin in (1+3)-dimensions from which we can take over all the
results arising from similar formulas. In this way we find that
the point-dependent matrices $\gamma^{\mu}(x)$ and
$S^{\mu\nu}(x)=e^{\mu}_{\hat\alpha}(x)e^{\nu}_{\hat\beta}(x)
S^{\hat\alpha\hat\beta}$ have similar properties as (\ref{ACOM}), (\ref{SAB}),
(\ref{Sgg}) and (\ref{SSS}), but written in natural indices and with $g(x)$
instead of the flat metric. Using this algebra and the standard notations for
the Riemann-Christoffel curvature tensor,  $R_{\alpha\beta \mu\nu}$,
Ricci tensor, $R_{\alpha\beta}=R_{\alpha \mu \beta\nu}g^{\mu\nu}$, and scalar
curvature, $R=R_{\mu\nu}g^{\mu\nu}$, we  recover the useful formulas
\begin{eqnarray}
&&\nabla_{\mu}(\gamma^{\nu}\psi)=\gamma^{\nu}\nabla_{\mu}\psi\,,
\label{Nabg}\\
&&[\nabla_{\mu},\,\nabla_{\nu}]\psi=
\textstyle\frac{1}{4}
R_{\alpha\beta\mu\nu}\gamma^\alpha\gamma^{\beta}\psi \,,
\end{eqnarray}
and the identity $R_{\alpha\beta\mu\nu}\gamma^{\beta}
\gamma^{\mu}\gamma^{\nu}=-2R_{\alpha\nu}\gamma^{\nu}$
that allow one to calculate
\begin{equation}
D^2=-\nabla^2+\textstyle{\frac{1}{4}}R\,{\mb 1}\,, \quad
\nabla^2=g^{\mu\nu}\nabla_{\mu}\nabla_{\nu}\,.
\end{equation}
It remains to complete the operator algebra with new observables from which
we have to select complete sets of commuting observables for defining quantum
modes.

Another important problem is to find the conserved quantities associated
with the symmetries of the theory arising from the action (\ref{action}).
The internal symmetries of the Lagrangian density for $m_0\not =0$ reduce to
the abelian unitary transformations
\begin{equation}
\psi\to \psi'=U(\xi_{em})\psi=e^{-i\xi_{em}}\psi\,,\quad
\xi_{em}\in{\Bbb R}\,,
\end{equation}
of the group $U(1)_{em}$. Whenever $m_0=0$ a supplementary symmetry is
given by the transformations of the {\em chiral} group $U(1)_{ch}$ that read
\begin{equation}
\psi\to \psi'=U(\xi_{ch})\psi=\left\{
\begin{array}{lll}
e^{ -i\xi_{ch} \gamma^{ch}}\psi& {\rm if}& \epsilon_{ch}=-1\\
e^{ \xi_{ch} \gamma^{ch}}\psi& {\rm if}& \epsilon_{ch}=1
\end{array}\right.\,,
\end{equation}
depending on the real parameters $\xi_{ch}$. Here it is crucial the
operators  $U(\xi_{ch})$ be self-adjoint since then
$\overline{U}(\xi_{ch})\gamma^{\mu}U(\xi_{ch})=\gamma^{\mu}$ and the Lagrangian
density remains invariant.
\begin{theor}[Noether]
The $U(1)_{em}$ internal symmetry produces the vector current
\begin{equation}
{\frak J}^{\mu}_{vect}=\overline{\psi}\gamma^{\mu}\psi\,,
\end{equation}
that is conserved obeying $\nabla_{\mu}{\frak J}^{\mu}_{vect} =0$. For $m_0=0$
the chiral symmetry leads to the conservation of the {\em axial} current,
\begin{equation}
{\frak J}^{\mu}_{ax}=\overline{\psi}\gamma^{\mu}\gamma^{ch}\psi\,,
\quad \nabla_{\mu}{\frak J}^{\mu}_{ax}=0\,.
\end{equation}
\end{theor}
\begin{demo}
In both cases one starts with infinitesimal transformations and uses the
standard method.
\end{demo}\\
Thus the notion of conservation of the vector currents gets the same meaning
as in special relativity. This give us the possibility to use the Stokes's
theorem for defining specific conserved {\em charges}  \cite{W}. Indeed,
supposing that $M_n=M_{d+1}$ is a physical spacetime (with $x^0=t$)
and $\sigma(t)$ is a Cauchy surface (i.e. space volumes of dimension $d$), we
can define the {\em electric} and respectively {\em chiral} time-independent
charges as
\begin{equation}
{\frak Q}_{em}=\int_{\sigma}d^{d}x\,\sqrt{g}\,
\overline{\psi}\gamma^{0}\psi\,, \quad
{\frak Q}_{ch}=\int_{\sigma}d^{d}x\,\sqrt{g}\,
\overline{\psi}\gamma^{0}\gamma^{ch}\psi\,.
\end{equation}
This result justifies the definition of the time-independent relativistic
scalar product of the space $\Psi$ of the spinors defined on $M_{d+1}$
\cite{BD,W}.
\begin{defin}\label{psc}
In physical spacetimes $M_{d+1}$ the relativistic scalar product
$\left<~,~\right>:\,\Psi\times \Psi \to
{\Bbb C}$ is
\begin{equation}\label{relsp}
\left<\psi,\psi'\right>=\int_{\sigma}d^{d}x\,\sqrt{g}\,
\overline{\psi}\gamma^{0}\psi'\,.
\end{equation}
\end{defin}
According to this definition we can write ${\frak Q}_{em}
=\left<\psi,\psi\right>$ and ${\frak Q}_{ch}=
\left<\psi, \gamma^{ch}\psi\right>$, opening thus the way to the physical
interpretation of the relativistic quantum mechanics that is the starting point
to the canonical quantization of the spinor field.

Other conserved current and conserved charges arise from the external
symmetries corresponding to the isometries of $M_n$ that will be studied in
the next sections.

\subsection{The gauge and relativistic covariance}

The use of the covariant derivatives assures the covariance of the whole theory
under the  gauge transformations,
\begin{eqnarray}
\hat e^{\hat\alpha}_{\mu}(x)&\to& \hat e'^{\hat\alpha}_{\mu}(x)=
\Lambda^{\hat\alpha\,\cdot}_{\cdot\,\hat\beta}[A(x)]
\,\hat e^{\hat\beta}_{\mu}(x)\,,\\
e_{\hat\alpha}^{\mu}(x)&\to&  {e'}_{\hat\alpha}^{\mu}(x)=
\Lambda_{\hat\alpha\,\cdot}^{\cdot\,\hat\beta}[A(x)]
\,e_{\hat\beta}^{\mu}(x)\,,\label{gauge}\\
\psi(x)&\to&\psi'(x)=T [A(x)]\,\psi(x)\,,
\end{eqnarray}
produced by the mappings $A: M_n\to {\mb G}(\eta)$ the values of which
are {\em local} transformations $A(x)=[\omega(x)]\in {\mb G}(\eta)$
determined by the set of {\em real} functions $\omega_{\hat\mu\hat\nu}=
-\omega_{\hat\nu\hat\mu}$ defined on $M_n$. In other words $A$ denotes
sections of the spinor fiber bundle that can be organized as
a group, ${\cal G}(M_n)$, with respect to the multiplication $\times$ defined
as $(A'\times A)(x)=A'(x)A(x)$. We use the notations $Id$ for the mapping
identity, $Id(x)=1\in {\mb G}(\eta)$, and $A^{-1}$ for the inverse of
$A$ which satisfies $(A^{-1})(x)=[A(x)]^{-1}$.

The general gauge-covariant theory of Dirac spinors outlined here must
be also covariant under  general coordinate transformation of $M_n$ which, in
the {\em passive} mode,
\footnote{We prefer the term of coordinate transformation instead of
diffeomorphism since we adopt this viewpoint.}
can be seen as changes of the local charts corresponding to the same domain of
$M_n$  \cite{WALD,ON}. If $x$ and $x'$ are the coordinates of a given point in
two different charts then there is a mapping $\phi$ between these charts giving
the coordinate transformation $x\to x'=\phi(x)$. These transformations form
the group ${\cal G}_{\phi}(M_n)$ with respect to the composition of mappings,
$\,\circ\,$, defined as usual, i.e. $(\phi'\circ\phi)(x)=\phi'[\phi(x)]$. We
denote the identity map of this group by $id$ and the inverse mapping of
$\phi$ by $\phi^{-1}$.

The coordinate transformations change all the components carrying  natural
indices including those of the gauge fields \cite{SW} changing thus the
positions of the local frames with respect to the natural ones. If we assume
that the physical experiment makes reference to the axes of the local frame
then it could appear situations when several correction of the positions of
these frames should be needed before (or after) a general coordinate
transformation. Obviously, these have to be made with the help of suitable
gauge transformations associated to the coordinate ones.
\begin{defin}
The {\em combined} transformation $(A,~\phi)$ is the gauge transformation given
by the section $A\in {\cal G}(M_n)$ followed by the coordinate transformation
$\phi\in {\cal G}_{\phi}(M_n)$.
\end{defin}
In this new notation the pure gauge transformations appear as $(A,id)\in
{\cal G}(M_n)$ while the coordinate transformations will be denoted from now
by $(Id,\phi)\in {\cal G}_{\phi}(M_n)$.
The  effect of a combined transformation $(A,\phi)$ upon our basic elements
is, $x\to x'=\phi(x)$,
\begin{equation}\label{genTP}
\psi(x)\to \psi'(x')=T[A(x)]\psi(x)\,,
\end{equation}
and $e(x)\to \hat e'(x')$ where $e'$ are the transformed gauge fields  of the
components
\begin{equation}
e'^{\mu}_{\hat\alpha}[\phi(x)]=\Lambda^{\cdot\,\hat\beta}
_{\hat\alpha\,\cdot}[A(x)]\,e^{\nu}_{\hat\beta}(x)
\frac{\partial\phi^{\mu}(x)}{\partial x^{\nu}}
\end{equation}
while the components of $\hat e'$ have to be calculated according to Eqs.
(\ref{eeee}). Thus we have written down the most general transformation laws
that leave the action invariant in the sense that
${\cal S}[\psi',e']={\cal S}[\psi,e]$.

The association among the transformations of the gauge group  and coordinate
transformation  leads to a new group with a specific multiplication. In order
to find how looks this new operation  it is convenient to use the composition
among the mappings $A$ and $\phi$ (taken only in this order) giving the new
mappings $A\circ\phi$ defined as $(A\circ \phi)(x)=A[\phi(x)]$. The calculation
rules $Id\circ \phi=Id$, $A\circ id=A$ and
$(A'\times A)\circ \phi=(A'\circ \phi)\times (A\circ \phi)$ are obvious.
In this context one can demonstrate
\begin{theor}
The set of combined transformations of $M_n$, $\tilde{\cal G}(M_n)$,
form a group with respect to the multiplication $*$ defined as
\begin{equation}\label{comp}
(A',\phi')*(A,\phi)=
\left((A'\circ\phi)\times A,\phi'\circ\phi\right)\,.
\end{equation}
\end{theor}
\begin{demo}
First of all we observe that the operation $*$ is well-defined and
represents the composition among the combined transformations since these can
be  expressed, according to their definition, as  $(A,\phi)=(Id,\phi)*(A,id)$.
Furthermore,  one can verify the result calculating the effect of
this product upon the field $\psi$.
\end{demo}\\
Now the identity is $(Id,id)$  while the inverse of a pair $(A,\phi)$ reads
\begin{equation}\label{compAphi}
(A,\phi)^{-1}=(A^{-1}\circ\phi^{-1},\phi^{-1})\,.
\end{equation}
In addition, one can demonstrate that the group of combined transformations is
the semi-direct product $\tilde{\cal G}(M_n)={\cal G}(M_n) {\rm ~\circledS~}
{\cal G}_{\phi}(M_n)$ between the group of sections which is the invariant
subgroup and that of coordinate transformations \cite{ES}. The same construction
starting with the group ${\mb G}_c(\eta)$ instead of ${\mb G}(\eta)$ yields the
complexified group  of combined transformations, $\tilde{\cal G}_c(M_n)$.

The use of the combined transformations is justified only in theories
where there are physical reasons to use some local frames  since in natural
frames the effect of the combined transformations  on the vector and tensor
fields reduces to that of their coordinate transformations. However, the
physical systems involving spinors can be described exclusively in local frames
where our theory is essential. Therein, the vector representation
$vect[\tilde{\cal G}(M_n)]$ is the usual one  \cite{SW,W}.
\begin{defin}
The spinor representation of $\tilde{\cal G}(M_n)$ has values in the space of
the linear operators ${U}: \Psi\to \Psi$ such that for each $(A,\phi)$ there
exists an operator ${U}(A,\phi)\in spin[\tilde{\cal G}(M_n)]$ having the action
\begin{equation}
{U}(A,\phi)\psi=[T(A)\psi]\circ \phi^{-1}=[T(A\circ \phi^{-1})]
(\psi\circ \phi^{-1})\,.
\end{equation}
\end{defin}
This rule gives the transformations (\ref{genTP}) in each point of $M_n$ if we
put $\psi'={U}(A,\phi)\psi$ and then calculate the value of $\psi'$ in the
point $x'$.  The Dirac operator derived from ${\cal S}$
{\em covariantly} transforms as
\begin{equation}\label{ero}
(A,\phi)~ :~~ D(x) \to D'(x')=
T[A(x)]D(x)\overline{T}[A(x)] \,,
\end{equation}
where $D'={U}(A,\phi)D[{U}(A,\phi)]^{-1}$. In general, the combined
transformations change the form of the Dirac operator which depends on the
gauge one uses ($D'\not=D$). We note that for the gauge transformations with
$\phi=id$ (when $x'=x$) the action of $U(A,id)$ reduces to the linear
transformation given by the matrix $T(A)$.

\subsection{Isometries and the external symmetry}

In general, the symmetry of the manifold $M_n$ is given by its isometry group,
$I(M_n)\subset {\cal G}_{\phi}(M_n)$, whose transformations, $x\to x'(x)$, are
coordinate transformation which leave the metric tensor invariant in any chart
 \cite{SW,WALD,ON},
\begin{equation}\label{giso}
g_{\alpha\beta}(x')\frac{\partial x'^\alpha}{\partial x^\mu}
\frac{\partial x'^\beta}{\partial x^\nu}=g_{\mu\nu}(x)\,.
\end{equation}
The isometry group is formed by sets of coordinate transformations,
$x\to x'=\phi_{\xi}(x)$, depending on $N$ independent real parameters, $\xi^a$
($a,b,c...=1,2,...,N$), such that $\xi=0$ corresponds to the identity map,
$\phi_{\xi=0}=id$. These transformations form a Lie group equipped with  the
composition rule
\begin{equation}\label{compphi}
\phi_{\xi'}\circ \phi_{\xi}=\phi_{p(\xi',\xi)}\,,
\end{equation}
where the functions $p$  define the group multiplication. These  satisfy
$p^{a}(0,\xi)=p^{a}(\xi,0)=\xi^{a}$ and
$p^{a}(\xi^{-1},\xi)=p^{a}(\xi,\xi^{-1})=0$ where $\xi^{-1}$ are the
parameters of the inverse mapping of $\phi_{\xi}$,
$\phi_{\xi^{-1}}=\phi^{-1}_{\xi}$. Moreover, the structure constants,
$c_{abc}$, of this group can be calculated in the usual way  \cite{HAM},
\begin{equation}\label{c}
c_{abc}=\left.\left(\frac{\partial p^{c}(\xi,\xi')}{\partial \xi^{a}\partial
\xi'^{b}}-
\frac{\partial p^{c}(\xi,\xi')}{\partial \xi^{b}\partial \xi'^{a}}
\right)\right|_{\xi=\xi'=0}\,.
\end{equation}
These define the commutation relations of the basis generators of the Lie
algebra of $I(M_n)$, denoted from now by $i(M_n)$. For small values of the
group parameters the infinitesimal transformations,
$x^{\mu}\to x'^{\mu}=x^{\mu}+\xi^{a}k_{a}^{\mu}(x)+\cdots$,
are given by the Killing vectors $k_{a}$ whose components,
\begin{equation}\label{ka}
k_{a}^{\mu}=\left.{\frac{\partial \phi^{\mu}_{\xi}}
{\partial\xi^{a}}}\right|_{\xi=0}\,,
\end{equation}
satisfy the Killing equations  $k_{a\, (\mu;\nu)}\equiv k_{a\, \mu;\nu}+
k_{a\, \nu;\mu}=0$ and the identities
\begin{equation}\label{kkc}
k^{\mu}_{a}k^{\nu}_{b,\mu}
-k^{\mu}_{b}k^{\nu}_{a,\mu}+c_{abc}k^{\nu}_{c}=0\,.
\end{equation}

The simplest representation of $I(M_n)$ is the {\em natural} one carried by the
space of the {\em scalar} fields $\vartheta$ which transform as
$\vartheta \to \vartheta' =\vartheta\circ\phi_{\xi}^{-1}$. This rule defines
the operator-valued representation of the group $I(M_n)$ generated by the
operators,
\begin{equation}\label{genL}
L_{a}=-ik_{a}^{\mu}\partial_{\mu}\,, \quad a=1,2,...,N\,,
\end{equation}
which are completely determined by the Killing vectors. From Eq. (\ref{kkc})
we see that they obey the commutation rules
\begin{equation}\label{comL}
[L_{a}, L_{b}]=ic_{abc}L_{c}\,,
\end{equation}
given by the structure constants of the Lie algebra $i(M_n)$.

In the theories involving fields with spin, an isometry can change the
relative positions of the local frames with respect to the natural ones.
This fact may be an impediment when one intends to study the symmetries of
these theories in local frames. For this reason it is natural to
suppose that the good symmetry transformations we need are  isometries
preceded by appropriate gauge transformations which should assure that not
only the form of the metric tensor would be conserved but the form of the
gauge field components too. However, these transformations are nothing other
than {\em particular} combined transformations whose coordinate
transformations are isometries.
\begin{defin}
The {\em external symmetry} transformations, $(A_{\xi},\phi_{\xi})$,
are par\-ti\-cu\-lar co\-m\-bi\-ned transformations involving isometries,
$(Id,\phi_{\xi})\in I(M_n)$, and cor\-res\-pon\-ding gauge transformations,
$(A_{\xi}, id)\in {\cal G}(M_n)$, necessary to {\em preserve the gauge}.
\end{defin}
This requirement is accomplished only if we assume that, for  given
gauge fields $e$ and $\hat e$, $A_{\xi}$  is defined by
\begin{equation}\label{Axx}
\Lambda^{\hat\alpha\,\cdot}_{\cdot\,\hat\beta}[A_{\xi}(x)]=
\hat e_{\mu}^{\hat\alpha}[\phi_{\xi}(x)]\frac{\partial \phi^{\mu}_{\xi}(x)}
{\partial x^{\nu}}\,e^{\nu}_{\hat\beta}(x)\,,
\end{equation}
with the supplementary condition $A_{\xi=0}(x)= 1\in {\mb G}(\eta)$.
Since $\phi_{\xi}$ is an isometry Eq. (\ref{giso}) guarantees that
$\Lambda[A_{\xi}(x)]\in vect[{\mb G}(\eta)]$ and, implicitly,
$A_{\xi}(x)\in {\mb G}(\eta)$. Then the transformation laws of our fields are
\begin{equation}\label{es}
(A_{\xi},\phi_{\xi}):\qquad
\begin{array}{rlrcl}
x&\to&x'&=&\phi_{\xi}(x)\\
e(x)&\to&e'(x')&=&e[\phi_{\xi}(x)]\\
\hat e(x)&\to&\hat e'(x')&=&\hat e[\phi_{\xi}(x)]\\
\psi(x)&\to&\psi'(x')&=&T[A_{\xi}(x)]\psi(x)\,.
\end{array}
\qquad
\end{equation}
The mean virtue of these transformations is that they leave {\em invariant}
the form of the Dirac operator, $D'=D$.
\begin{theor}
The set of the external symmetry transformations $(A_{\xi},\phi_{\xi})$ form
the Lie group $S(M_n)\subset \tilde{\cal G}(M_n)$ with respect to the
operation $*$. This group, will be called the group of the external symmetry
of $M_n$.
\end{theor}
\begin{demo}
Starting with Eq. (\ref{Axx}) after a little calculation we find that
\begin{equation}\label{compA}
(A_{\xi'}\circ\phi_{\xi})\times A_{\xi}=A_{p(\xi',\xi)}\,,
\end{equation}
and, according to Eqs. (\ref{comp}) and (\ref{compphi}), we obtain
\begin{equation}\label{mult}
(A_{\xi'},\phi_{\xi'})*(A_{\xi},\phi_{\xi})=
(A_{p(\xi',\xi)},\phi_{p(\xi',\xi)})\,,
\end{equation}
and $(A_{\xi=0},\phi_{\xi=0})=(Id,id)$.
\end{demo}\\
From Eq. (\ref{mult}) we understand that $S(M_n)$ is {\em locally
isomorphic} with $I(M_n)$ and, therefore, the Lie algebra $s(M_n)$ of
of the group $S(M_n)$ is isomorphic with  $i(M_n)$ having the same structure
constants.  There are arguments that the group $S(M_n)$ must be isomorphic with the
universal covering group of $I(M_n)$ since it has  anyway the topology induced
by ${\mb G}(\eta)$ which is simply connected. In general, the number of group
parameters of $I(M_n)$ or $S(M_n)$ (which is equal to the number of the
independent Killing vectors of $M_n$) can be $0\le N\le \frac{1}{2}n(n+1)$
\cite{SW}.

\subsection{The spinor representation of $S(M_n)$}

The last of Eqs. (\ref{es}) giving the transformation law of the field
$\psi$ defines the operator-valued representation
$(A_{\xi},\phi_{\xi})\to {U}_{\xi}$ of the group $S(M_n)$,
\begin{equation}\label{rep}
({U}_{\xi}\psi)[\phi_{\xi}(x)]=T[A_{\xi}(x)]\psi(x)\,,
\end{equation}
which is the spinor representation
$spin[S(M_n)]\subset spin[\tilde{\cal G}(M_n)]$ of the group $S(M_n)$.
This representation has unitary transformation matrices in the sense of the
Dirac adjoint ($\overline{T}=T^{-1}$) and its transformations leaves the
operator $D$ invariant,
\begin{equation}\label{invE}
{U}_{\xi}D{U}_{\xi}^{-1}=D\,.
\end{equation}
Since $A_{\xi}(x)\in {\mb G}(\eta)$ we say that  $spin[S(M_n)]$ is
{\em induced} by the representation $spin[{\mb G}(\eta)]$  \cite{BR,MAK}.
\begin{theor}
The basis generators of the spinor representation $spin[s(M_n)]$ of the Lie
algebra $s(M_n)$ are
\begin{equation}\label{X}
X_{a} =\left. i{\frac{\partial {U}_{\xi}}{\partial \xi^{a}}}\right |_{\xi=0}=
L_{a}+ S_a= L_a +
\frac{1}{2}\,\Omega^{\hat\alpha\hat\beta}_{a}
S_{\hat\alpha\hat\beta}\,,
\end{equation}
where
\begin{equation}\label{Om}
\Omega^{\hat\alpha\hat\beta}_{a}
=\left( \hat e^{\hat\alpha}_{\mu}\,k_{a,\nu}^{\mu}
+\hat e^{\hat\alpha}_{\nu,\mu}
k_{a}^{\mu}\right)e^{\nu}_{\hat\lambda}\eta^{\hat\lambda\hat\beta}\,.
\end{equation}
\end{theor}
\begin{demo}
For small values of $\xi^{a}$, the covariant parameters $\omega$ of the element
$A_{\xi}(x)\equiv [\omega_{\xi}(x)]\in {\mb G}(\eta)$ can be written as
$\omega^{\hat\alpha\hat\beta}_{\xi}(x)=
\xi^{a}\Omega^{\hat\alpha\hat\beta}_{a}(x)+\cdots$. Then,
from Eq. (\ref{Axx}) we can calculate the quantities
\begin{equation}\label{Om1}
S_a(x)=\left.i{\frac{\partial A_{\xi}(x)}{\partial \xi^{a}}}\right |_{\xi=0}=
\frac{1}{2}\,\Omega^{\hat\alpha\hat\beta}_{a}(x)
S_{\hat\alpha\hat\beta}\,, \quad
\Omega^{\hat\alpha\hat\beta}_{a}\equiv \left.{\frac{\partial
\omega^{\hat\alpha\hat\beta}_{\xi}}
{\partial\xi^a}}\right |_{\xi=0}\,,
\end{equation}
which yields the desired result.
\end{demo}\\
We must specify that the functions $\Omega^{\hat\alpha\hat\beta}_{a}$
are antisymmetric if and only if $k_{a}$ are Killing vectors. This indicates
that the association among isometries and the gauge transformations defined
by Eq. (\ref{Axx}) is correct.
\begin{rem}
The generators (\ref{X}) can be written in {\em covariant} form as
\begin{equation}
X_{a}=-ik^{\mu}_{a}\nabla_{\mu}+\frac{1}{2}\,
k_{a\, \mu;\nu}\,e^{\mu}_{\hat\alpha}\,e^{\nu}_{\hat\beta}\,
S^{\hat\alpha\hat\beta} \,.
\end{equation}
\end{rem}
In Ref. \cite{ES} we have shown that similar formula can be written for any
spin, generalizing thus the important result derived in  Ref. \cite{CML}
for the Dirac field in $M_4$.
\begin{theor}
The operators (\ref{X}) are self-adjoint with respect to the Dirac adjoint and
satisfy the commutation rules
\begin{equation}\label{comX}
[X_{a}, X_{b}]=ic_{abc}X_{c}\,, \quad
[D,X_{a}]=0\,, \quad a,b...=1,2,...,N\,,
\end{equation}
where $c_{abc}$ are the structure constants of the isomorphic Lie algebras
$s(M_n) \sim i(M_n)$.
\end{theor}
\begin{demo}
First we observe that $X_a=\overline{X}_a$ are self-adjoint since all the
gamma-matrices have this property. Furthermore, in order to demonstrate Eqs.
(\ref{comX}), we derive  Eq. (\ref{compA}) with respect to $\xi$ and $\xi'$
and from Eqs. (\ref{c}) and (\ref{Om}), after a few manipulations, we obtain
the identities
\begin{equation}\label{idOM}
\eta_{\hat\alpha\hat\beta}\left(
\Omega_{a}^{\hat\alpha\hat\mu}\Omega_{b}^{\hat\beta\hat\nu}
-\Omega_{b}^{\hat\alpha\hat\mu}\Omega_{a}^{\hat\beta\hat\nu}\right)+
k^{\mu}_{a}\Omega_{b,\mu}^{\hat\mu\hat\nu}
-k^{\mu}_{b}\Omega_{a,\mu}^{\hat\mu\hat\nu}+c_{abc}\Omega_{c}^{\hat\mu\hat\nu}
=0\,,
\end{equation}
leading to
\begin{equation}
[S_{a},S_{b}]+[L_{a},S_{b}]-[L_{b},S_{a}]
=ic_{abc}S_{c}\,,
\end{equation}
and, according to Eq. (\ref{comL}),
we find the expected commutation rules. The commutators with the operator
$D$ result from Eq. (\ref{invE}).
\end{demo}\\
The natural consequence is
\begin{cor}\label{adjX}
The operators $U_{\xi}\in spin[S(M_n)]$ transform the basis generators $X_a$
according to the adjoint representation of $S(M_n)$,
\begin{equation}
U_{\xi}X_aU^{-1}_{\xi}=Adj(\xi)_{a b}X_b\,,
\end{equation}
defined as
\begin{equation}
Adj(\xi)=e^{i\xi^a adj(X_a)}\,, \quad adj(X_a)_{bc}=-ic_{abc}\,,
\end{equation}
where $adj(X_a)$ are the basis generators of the adjoint representation of
$s(M_n)$.
\end{cor}
\begin{demo} This is a general result of the group representation theory
 \cite{BR}. We note that here the phase factors are chosen such that
the commutators
\begin{equation}
[adj(X_a),\, adj(X_b)]=ic_{abc} \, adj(X_c)
\end{equation}
keep  the form (\ref{comX}).
\end{demo}\\
Whenever the field $\psi$ obeys convenient conditions at the boundary of
$\sigma$ then these operators are Hermitian with respect to the relativistic
scalar product (\ref{relsp}) and the representation $spin[S(M_n)]$ is
{\em unitary} (with $X_a^{\dagger}=X_a$ and $U^{\dagger}=U^{-1}$). In this case
one can define quantum modes correctly, using the set of commuting operators
formed by the Casimir operators of $spin[s(M_n)]$, the generators of its Cartan
subalgebra and the Dirac operator, $D$.

The non-covariant form (\ref{X}) of the generators $X_a$ helps us to understand
the meaning of the notion of {\em manifest} covariance in curved manifolds
where the representations of $S(M_n)$ are induced by those of ${\mb G}(\eta)$,
depending thus on the gauge fixing. For this reason, in general, their
generators have point-dependent spin terms that do not commute with the orbital
parts. However, one may find several special gauge fixings where some spin
terms become point-independent.
\begin{defin}
When the generators $S_{a}(x)$, $a=1,2,...,N'$ ($N'\le N$), of a subgroup
$G_1\subset S(M_n)$ are independent on $x$ obeying $[S_{a},\, L_{b}]=0$, for
all $a=1,2,...,N'$  and $b=1,2,...,N$, we say that $\psi$ behaves manifestly
covariant with respect to this subgroup.
\end{defin}
The point-independent operators $S_{a},\, a=1,2,...,N'$,  are then just
the generators of an usual  linear representation of $G_1$.
One knows many examples of curved spacetimes for which one can choose
suitable local frames where the spinor fields transform manifestly
covariant with respect to different subgroups of $S(M_n)$ or even to this whole
group. Particularly, the frames of the flat spacetimes where the fields with
spin transform manifestly covariant under the transformations of the $S(M_n)$
group are nothing other than the usual {\em inertial} frames of the special
relativity.

From the physical point of view, our approach is useful since this allows
one to derive the conserved quantities predicted by the
Noether theorem.
\begin{theor}[Noether]
The basis generators $X_a \in spin[s(M_n)]$
give rise to conserved currents, ${\frak J}^{\mu}[X_a]$, which satisfy
\begin{equation}
{\frak J}^{\mu}[X_a]_{;\mu}=0 \,.
\end{equation}
\end{theor}
\begin{demo}
The conserved currents have to be calculated from the action (\ref{action})
in the usual way, starting with the infinitesimal transformations generated by
$X_a$. One finds
\begin{equation}\label{TetX}
{\frak J}^{\mu}[X_a]=
-\frac{i}{2}\left[\overline{\psi}\gamma^{\hat\alpha}e^{\mu}_{\hat\alpha}
\partial_{\nu}\psi-
(\overline{\partial_{\nu}\psi})\gamma^{\hat\alpha}e^{\mu}_{\hat\alpha}\psi
\right] k_a^{\nu}+\frac{1}{4}
\,\overline{\psi}\{\gamma^{\hat\alpha}, S^{\hat\beta \hat\gamma} \}\psi\,
e^{\mu}_{\hat\alpha} \,\Omega_{a\,\hat\beta \hat\gamma}\,,
\end{equation}
where $\psi$ satisfies the Dirac equation. Notice that the first term here
involves a part of the stress-energy tensor of the Dirac field  \cite{SW,BD}.
\end{demo}
\begin{cor}
If $M_n=M_{d+1}$ is a physical spacetime then every
basis generator $X_a$
defines its specific time-independent quantity
\begin{equation}
{\frak Q}_a=\int_{\sigma}d^{d}x\,\sqrt{g}\,{\frak J}^{0}[X_a]=
\frac{1}{2}\left(\left<\psi,X_a\psi\right>
+\left<X_a\psi,\psi\right>\right)\,.
\end{equation}
\end{cor}
\begin{demo}
Bearing in mind that the time coordinate of Definitions \ref{phys} and
\ref{psc}
was $t=x^{0}$ and using Eqs. (\ref{TetX}) and (\ref{X}), one can arrange the
terms in order to obtain this formula.
\end{demo}\\
Whenever the operators $X_{a}$ are Hermitian with respect to the relativistic
scalar product (\ref{relsp}) one can write
${\frak Q}_a=\left<\psi, X_a\psi\right>$. In the relativistic quantum
mechanics this quantity has to be interpreted as the expectation value of the
observable $X_a$ in the state described by the spinor $\psi$. Of course, at the
level of the quantum field theory ${\frak Q}_a$ becomes the one-particle
operator
which takes over the role of the generator $X_a$  \cite{CDS}.

Hence we have built a complete theory of the external symmetries related to
the genuine isometries defined as coordinate transformations of $M_n$ which
preserve the metric tensor. This is very close to the theory of the
Poincar\' e group of the Minkowski spacetime, producing conserved quantities
through the Noether theorem in a similar manner as in special relativity.

\section{Dirac-type operators related to K-Y tensors}

Our theory of the external symmetry is not suitable for the study of other
types of symmetries having more subtle geometrical origins as the so called
hidden symmetries encapsulated in the existence of the S-K and K-Y tensors.
In the classical theory, the hidden symmetries are arising from more general
isometries defined in the whole phase space which cannot be reduced to pure
coordinate transformations. For this reason the previous group theoretical
methods seem to be not appropriate for obtaining new conserved quantities or
operators commuting with $D$, produced by the {S-K} or K-Y tensors fields at
the quantum level. Here new specific mechanisms have to be exploited for
analyzing the hidden symmetries or several new types of supersymmetries.

\subsection{Operators produced by S-K and K-Y tensors}

It is obvious that in the classical theory only the S-K tensors,
$ k^{(r)}$, can give rise directly to new conserved quantities since
these
are completely symmetric tensors of a given rank, $r\ge 2$, whose components,
$ k^{(r)}_{\mu_1\mu_2...\mu_r}$,
satisfy the generalized Killing equation,
\begin{equation}\label{S-K}
k^{(r)}_{(\mu_1\mu_2...\mu_r;\mu)=0}\,.
\end{equation}
They allow one to construct the quantities
$ k^{(r)}_{\mu_1\mu_2...\mu_r}\dot{x}^{\mu_1}\dot{x}^{\mu_2}...
\dot{x}^{\mu_r}$ that are conserved along the geodesics. Unfortunately, this
property does not hold in the quantum theory because of the {\em gravitational
anomaly} presented in manifolds with non-vanishing Ricci tensor since the
operators $K^{(r)} =  k^{(r)\,\mu_1\mu_2...\mu_r}
\nabla_{\mu_1}\nabla_{\mu_2}...\nabla_{\mu_r}$ do not commute with the usual
Laplace operator of $M_n$, $\nabla^2$, as it might be expected. Particularly,
in the case of the second order operators one finds  \cite{KKK}
\begin{equation}\label{gravan}
\left[K^{(2)}, \nabla^2\right]= {\frac{4}{3}}
\left({ k^{(2)}\,}^{\mu \,[\nu}
R^{\sigma ]\, \cdot}_{\cdot\,\,\nu}\right)_{;\sigma}\nabla_\mu\,.
\end{equation}

The next interesting geometrical object connected with higher order
symmetries of a manifold after the S-K tensors is the K-Y tensors.
A differential $r-$form $f$ is called a K-Y tensor if its covariant
derivative $f^{(r)}_{\mu_1\mu_2...(\mu_r;\lambda)}$ is totally
antisymmetric. Equivalently, a tensor is called a K-Y tensor of rank $r$
if it is totally antisymmetric and satisfies the equation
\begin{equation}\label{i3}
f^{(r)}_{\mu_1\mu_2...(\mu_r;\lambda)} \equiv
f^{(r)}_{\mu_1\mu_2...\mu_r;\lambda} +
f^{(r)}_{\mu_1\mu_2...\mu;\lambda}=0\,.
\end{equation}

The K-Y tensors were first introduced from purely mathematical reasons
\cite{Y}, but subsequently it was realized their profound connection with the
supersymmetric classical and quantum mechanics on curved manifolds where such
tensors do exist \cite{GRH}. Thus it seems that, at least from the point of
view of the quantum theory of (super)symmetry, the natural generalization of
the Killing vectors are the K-Y tensors rather than the S-K ones.

These two generalizations (\ref{S-K}) and (\ref{i3}) of the usual Killing
vector equation could be related. Let $f^{(r)}_{\mu_1\mu_2...\mu_r}$
be a K-Y tensor, then the symmetric tensor
\begin{equation}\label{KYY}
k^{(2)}_{\mu\nu} = f^{(r)}_{\mu\mu_2...\mu_r}
f^{(r)¬¬\mu_2...\mu_r}_\nu
\end{equation}
is a second rank S-K tensor and it is sometimes refers to this S-K tensor as
the associated tensor to $f$. However, the converse statement is not true in
general since not all S-K tensors of rank $2$ are associated to a K-Y tensor.
It is worth notice that the gravitational anomaly (\ref{gravan}) is absent
in the case of the second rank
S-K tensors that can be expressed as symmetrized contractions of K-Y tensors
as in (\ref{KYY}) \cite{CML}.

It was surprising to see that the K-Y tensors are naturally related to the
Dirac theory in curved manifolds since all of them are able to produce
first-order differential operators which commutes or anticommutes with $D$.
The Killing vectors considered K-Y tensors of rank $r=1$ give rise to the
operators $X_a$ defined by Eq. (\ref{X}). This result was reported in
\cite{CML} simultaneously with the operators built using second rank
K-Y tensors  \cite{MLS, KML}. A recent generalization  \cite{MaCa} yields
\begin{theor}
Given a K-Y tensor $f^{(r)}$ of an arbitrary rank $r=1,2,...$, the
operator
\begin{eqnarray}\label{Yf}
Y[f^{(r)}]&=&(-1)^r i
\gamma^{\mu_1}\gamma^{\mu_2}\cdots \gamma^{\mu_{r-1}}\left(
f^{(r)~~~~~~~~\mu_r}_{\mu_1\mu_2...\mu_{r-1}\,\cdot}\nabla_{\mu_r}\right.
\nonumber\\
&&\left.-\frac{1}{2(r+1)}
f^{(r)}_{\mu_1\mu_2...\mu_r;\mu} \gamma^{\mu_r}\gamma^{\mu}\right)
\end{eqnarray}
commute with $D$ if $r$ is odd and anticommute with $D$ if $r$ is even.
\end{theor}
\begin{demo}
We delegate the proof to the Ref. \cite{MaCa}.
\end{demo}\\
In general, one can construct new operators commuting
with $D$ using the operators (\ref{Yf}) built with the help of arbitrary K-Y
tensors. Indeed, given two K-Y tensors of any rank, $\tilde f^{(r_1)}$ and
$f^{(r_2)}$, the new second order operator
$K^{(2)}=\{Y[f^{(r_1)}],\, Y[f^{(r_2)}]\}$ commutes with $D$
whenever $r_1+r_2$ is an even number. Moreover, in this way we
obtain the corresponding factorized {S-K} tensor of the second rank
that gives rise to the operator $K^{(2)}$ freely of quantum anomaly.
In this manner one can generate new types of operators that help one
to investigate the hidden symmetries and to obtain large sets of conserved
operators that may constitute new (super)algebras. In other respects, the
implication of the K-Y tensors in the quantum theory suggests us that such
tensors with complex-valued components would be also useful
even if from the classical viewpoint these are pointless.

Of a particular interest are the operators built with the help of
the second rank K-Y tensors, $f$, with real or complex-valued components
$ f_{\mu\nu} = -  f_{\nu\mu}$ which satisfies the equation
(\ref{i3}) for $r=2$.
\begin{defin}
The operators
\begin{equation}\label{df}
D_{f} = i\gamma^\mu \left(f_{\mu\,\cdot}^{\cdot\,\nu}\nabla_\nu -
\textstyle{1\over 6}f_{\mu\nu;\rho}\gamma^\nu \gamma^\rho \right)\,,
\end{equation}
given by the second rank K-Y tensors, $f$, are
called  Dirac-type operators.
\end{defin}
These are non-standard Dirac operators which obey $\{D_{f},\,D\}=0$ and
can be involved in new types of genuine or hidden (super)symmetries. Remarkable
superalgebras of Dirac-type operators can be produced by special second-order
K-Y tensors that represent square roots of the metric tensor.

\subsection{Roots and their Dirac-type operators}

Let us start with some technical details and the basic definitions. Given
$\rho$ an arbitrary tensor field of rank 2  defined on a domain of $M_n$,
we denote with the same symbol $\left<\rho\right>$ the equivalent
matrices with the elements $\rho^{\mu\,\cdot}_{\cdot\,\nu}$ in natural
frames and
$\rho^{\hat\alpha\,\cdot}_{\cdot\,\hat\beta}=\hat e_{\mu}^{\hat\alpha}
\rho^{\mu\,\cdot}_{\cdot\, \nu}e^{\nu}_{\hat\beta}$
in local frames. We say that $\rho$ is non-singular on
$M_n$ if det$\,\left<\rho\right>\not=0$  on a domain of $M_n$ where the metric
is non-singular. This tensor is said irreducible on $M_n$ if its matrix is
irreducible.
\begin{defin}\label{Def1}
The  non-singular real or complex-valued K-Y tensor $f$ of rank 2 defined on
$M_n$ which satisfies
\begin{equation}\label{fi}
f^{\mu\,\cdot}_{\cdot\,\alpha} f_{\mu\beta}=g_{\alpha\beta}\,,
\end{equation}
is called an unit root of the metric tensor of $M_n$, or simply an unit root
of $M_n$.
\end{defin}
It was shown that any K-Y tensor that satisfy Eq.  (\ref{fi}) is covariantly
constant  \cite{K1},
\begin{equation}\label{cc}
f_{\mu\nu;\sigma}=0\,.
\end{equation}
Since Eq.  (\ref{fi}) can be written as
$f^{\mu\,\cdot}_{\cdot\,\alpha}f^{\alpha\,\cdot}_{\cdot\,\nu}=
-\delta^{\mu}_{\nu}$
this takes the matrix form
\begin{equation}\label{f2I}
\left<f\right>^2=-1_n\,,
\end{equation}
where the notation $1_n$ stands for the $n\times n$ identity matrix.
Hereby we see that the unit roots are matrix representations of several
{\em complex units} (similar to $i\in {\Bbb C}$) with usual properties as,
for example, $\left<f\right>^{-1}=-\left<f\right>$. The unit roots having
only {\em real}-valued components are called {\em complex structures}
and represent automorphisms of the tangent fiber bundle ${\cal T}(M_n)$ of
$M_n$. In local frames these appear as particular point-dependent
transformations of the gauge group $G(\eta)=vect[{\mb G}(\eta)]$.
The manifold possessing such structures are said to have a K\" ahlerian
geometry (see the Appendix A). However, the unit roots considered here
are beyond this case since these are defined as automorphisms of the
complexified fiber bundle ${\cal T}(M_n)\otimes {\Bbb C}$, being thus
transformations of the complexified group
$G_c(\eta)=vect[{\mb G}_c(\eta)]$.

As in the case of the complex structures of the K\" ahlerian geometries, the
matrices of the unit roots have specific algebraic properties resulted from
Eq.  (\ref{f2I}). These can be pointed out in local frames (where the matrix
elements are $f^{\hat\alpha\,\cdot}_{\cdot\,\hat\beta}=
\hat e_{\mu}^{\hat\alpha} f^{\mu\,\cdot}_{\cdot\, \nu} e^{\nu}_{\hat\beta}$)
using gauge transformations of $G(\eta)$.
\begin{lem}
The matrix  of any root of $M_n$ is equivalent with a matrix completely
reducible in $2\times 2$ diagonal blocks.
\end{lem}
\begin{demo}
The matrix $\left<f\right>$ which satisfies Eq.  (\ref{f2I}) has only
two-dimensional invariant
subspaces spanned by pairs of vectors $z$ and ${\left<f\right>}z$.
On these subspaces, Eq.  (\ref{f2I}) is solved in local frames by two types
of $2\times 2$ unimodular blocks without diagonal elements: either
skew-symmetric blocks with factors $\pm 1$, when the involved dimensions are
of the same signature, or symmetric ones with pure imaginary phase factors,
$\pm i$, if the signatures are opposite. However, the diagonalization
procedure cannot be continued using transformations of $G(\eta)$ since
these preserve the form of the $2\times 2$ blocks which are proportional with
the generators of the subgroups $SO(2)$ or $SO(1,1)$ acting on the
corresponding invariant subspaces. Notice that other transformations of
$G_c(\eta)$ are not useful since these do not leave the metric invariant.
\end{demo}\\
This selects the geometries allowing unit roots.
\begin{cor}
The unit roots are allowed only by manifolds $M_n$ with an even number of
dimensions, $n=2k, \, k\le l$.
\end{cor}
\begin{demo}
If $n$ is odd then the $2\times 2$ blocks do not cover all dimensions, so that
${\rm det}\hat{\left<f\right>}=0$ and $f$ is no more an unit root of $M_n$.
\end{demo}
\begin{cor}\label{coco}
The unit roots of $M_n$ have real matrices only when the metric $\eta$
has a signature with even $n_{+}$ and $n_{-}$. Otherwise the unit roots have
only complex-valued matrices. In both cases the matrices of the unit roots are
unimodular, i.e. {\rm det}$\left<f\right>=1$.
\end{cor}
It is clear that for the real-valued unit roots (i.e., complex structures)
one can construct the {\em symplectic} 2-forms
$\tilde\omega =\frac{1}{2}f_{\mu\nu}dx^{\mu}\land dx^{\nu}$ which are
closed and  non-degenerate.

The above properties indicate that the unit roots are defined up to
sign. Therefore, if two unit roots $f_1$ and $f_2$ do not obey the
condition $f_1=\pm f_2$ then these will be considered {\em different}
between themselves. We denote by ${\mb R}_1(M_n)$ the set of all different
unit roots of the manifold $M_n$. On the other hand, when an unit root $f$
is multiplied by an arbitrary {\em real} number $\alpha \not=0$, we say that
$\rho(x)=\alpha f(x)$ is a {\em root} of norm $\|\rho\|=|\alpha|$. Thus we can
associate to any unit root $f$ the one-dimensional linear real space
$L_f =\{\rho\,|\,\rho=\alpha f,\, \alpha\in {\Bbb R}\}$ in which each  non
vanishing element is a root. According to Corollary \ref{coco}, when the metric
$\eta$ is pseudo-Euclidean,
the unit roots can have complex matrix elements and in that case the unit
root $f$ and its {\em adjoint}, $f^*$, are different. This last one generates
its own linear real space $L_{f^*}$  of adjoint roots  which satisfy
$\left[ \,\left<\rho \right>^* ,\,\left<\rho'\right>\,\right] =0,\, \forall
\rho,\, \rho' \in L_f$ since the matrices of $f$ and $f^*$ commutes with each
other, having same diagonal blocks up to signs.

The whole set of roots of $M_n$ defined as
\begin{equation}\label{setR}
{\mb R}(M_n)=\bigcup_{f\in {\mb R}_1(M_n)}(L_f-\{0\})
\end{equation}
seems to have special algebraic structure since it does not have the
element zero and, in general, it is not certain that a linear
combination of roots is a root too. To convince this, it is enough to observe
that the sum of the roots $\rho$ and  $\rho^*$ is no more a root since
det$(\left<\rho\right>+\left<\rho\right>^*)=0$ when $\rho^*\not=\rho$ because
of the reduction of the pure imaginary diagonal $2\times 2$ blocks. Moreover,
the product of the matrices of two different roots gives a nonsingular matrix
but that may be not of a root. Thus we understand that ${\mb R}(M_n)+\{0\}$
cannot be organized as a global linear space or algebra even though,
according to the definition (\ref{setR}), it naturally includes linear parts
as  $L_f$ or $L_{f^*}$. In other respects, we know examples
indicating that ${\mb R}(M_n)$  may contain subsets which are parts of some
linear spaces with one or three dimensions, isomorphic with Lie algebras
 \cite{CV6,K2}. In any event, the algebraic properties and the topology of
${\mb R}(M_n)$ seem to be complicated depending on the
topological structure of the set of unit roots ${\mb R}_1(M_n)\subset
{\mb R}(M_n)$.

The K-Y tensor gives rise to  Dirac-type operators of the form
(\ref{df}) which have  an important property formulated in
 \cite{K1}.
\begin{theor}\label{DtyD}
The Dirac-type operator $D_f$ produced by the K-Y tensor $f$
satisfies the  condition
\begin{equation}\label{D2D2}
(D_{f})^2=D^2
\end{equation}
if and only if $f$ is an unit root.
\end{theor}
\begin{demo}
The arguments of Ref.  \cite{K1} show that the condition Eq.  (\ref{D2D2}) is
equivalent  with Eqs.  (\ref{fi}) and (\ref{cc}).
Moreover, we note that for $f\in {\mb R}_1(M_n)$  the square of
the Dirac-type operator
\begin{equation}\label{Dirf}
D_f=if_{\mu\,\cdot}^{\cdot\,\nu}\gamma^{\mu}\nabla_{\nu}\,,
\end{equation}
has to be calculated exploiting  the identity
$0=f_{\mu\nu;\alpha;\beta} -f_{\mu\nu;\beta;\alpha}=
f_{\mu\sigma} R^{\sigma}_{\cdot\,\nu\alpha\beta}+
f_{\sigma\nu} R^{\sigma}_{\cdot\,\mu\alpha\beta}$,
which gives
\begin{equation}\label{D2D}
R_{\mu\nu\alpha\beta}f^{\mu\,\cdot}_{\cdot\,\sigma} f^{\nu\,\cdot}_{\cdot\,
\tau}
=R_{\sigma\tau\alpha\beta}
\end{equation}
and leads to Eq.  (\ref{D2D2}).
\end{demo}\\
Thus we  conclude that the equivalence of the condition (\ref{D2D2})
with Eqs.  (\ref{fi}) and (\ref{cc}) holds in any geometry of dimension $n=2k$
allowing roots. When $f^*\not=f$ then $D_{f^*}$ is different from $D_f$ even
if $(D_f)^2=(D_{f^*})^2=D^2$. These operators are no longer self-adjoint,
obeying $\overline{D_f}=D_{f^*}$ and
\begin{equation}\label{DDO}
\left\{ D_{f},\,D\right\}=0\,,\quad
\left\{ D_{f^*},\,D\right\}=0\,.
\end{equation}

\subsection{Continuous symmetries generated by unit roots}

Now we shall reach an interesting points of our study, showing that there are
continuous transformations able to relate the operators $D_f$ and $D$ to each
other. We know that in many particular cases  \cite{CV6, K2, CV7} this is
possible and now we intend to point out that this is a general property of
theories involving roots. To this end we introduce a new useful point-dependent
matrix.
\begin{defin}
Given the unit root $f$, the matrix
\begin{equation}\label{SfS}
\Sigma_{f}=\frac{1}{2}f_{\mu\nu}S^{\mu\nu}
\end{equation}
is the spin-like operator associated to $f$.
\end{defin}
This is a matrix that acts on the space of spinors $\Psi$ and, therefore,
can be interpreted as a generator of the spinor representation
$spin[{\mb G}_c(\eta)]$ since the components of $f$ are, in general,
complex-valued functions.  It has the obvious property $\overline{\Sigma_{f}}=
\Sigma_{f^*}$ while from (\ref{Nabg}) and (\ref{fi}) one
obtains that it is covariantly constant in the sense that
$\nabla_{\nu}(\Sigma_f\psi)=\Sigma_f \nabla_{\nu} \psi$.
Hereby we find that the Dirac-type operator (\ref{Dirf}) can be written as
\begin{equation}\label{DDS}
D_f=i\left[D,\, \Sigma_f\right] \,,
\end{equation}
where $D$ is the standard Dirac operator defined by Eq.  (\ref{Dirac}).
Moreover, from Eqs.  (\ref{DDO}) we deduce
$[\Sigma_f,\, D^2]=[\Sigma_f,\,(D_{f})^2]=0$
and similarly for $\Sigma_{f^*}$.
\begin{defin}
We say that ${G}_{f}=\{(A_{\rho},id)\,|\,\rho=\alpha f,\, \alpha
\in {\Bbb R}\} \subset {\mb G}_c(\eta)$ is the one-parameter Lie
group associated to the unit root $f\in {\mb R}_1(M_n)$.
\end{defin}
The spinor representation of this group, $spin(G_f)$, is the restriction
to $G_f$ of the representation $spin[\tilde{\cal G}_c(M_n)]$. Therefore, the
operators ${U}_{\rho} \in spin(G_f)$,  have the action
\begin{equation}\label{Taf1}
({U}_{\rho}\psi)(x) = [T(\rho)\psi](x)
= T[\alpha f(x)]\,\psi(x) \,,
\end{equation}
where the transformation matrices
\begin{equation}\label{Taf}
T(\alpha f) =e^{-i\alpha \Sigma_f} \in spin[{\mb G}_c(\eta)]
\end{equation}
depend on the group parameter $\alpha \in{\Bbb R}$.  Hence we defined the new
mappings $A_{\rho} : M_n \to {\mb G}_c(\eta)$ representing sections of the
complexified spinor fiber bundle such that
$A_\rho(x)=[\rho(x)]\in {\mb G}_c(\eta)$. Since the matrices (\ref{Taf}) are
just those defined by Eq.
(\ref{TeS}) where we replace $\omega$ by the roots $\rho=\alpha f\in L_f$,
their action on the point-dependent Dirac matrices results from Eq.
(\ref{TgT}) to be,
\begin{equation}\label{TgT1}
[T(\alpha f)]^{-1}\gamma^{\mu}T(\alpha f)=
\Lambda^{\mu\,\cdot}_{\cdot\,\nu}(\alpha f)\gamma^{\nu}\,,
\end{equation}
where
$\Lambda^{\mu\,\cdot}_{\cdot\,\nu}=
e^{\mu}_{\hat\alpha}\Lambda^{\hat\alpha\,\cdot}_{\cdot\,\hat\beta}
\hat e^{\hat\beta}_{\nu}$ are matrix elements with natural indices of the
matrix
\begin{equation}\label{Lamaf}
\Lambda(\alpha f)=e^{\alpha \left<f\right>}=1_n\cos\alpha+
\left<f\right>\sin\alpha \,,
\end{equation}
calculated according to Eqs. (\ref{Lam})  and (\ref{f2I}). We note that this
is a matrix representation of the usual {\em Euler formula} of the complex
numbers. Now it is obvious that in local frames
$\left<f\right>=\Lambda(\frac{\pi}{2}f)\in vect({\mb G}_c)$, as mentioned above.
\begin{theor}\label{TDT}
The operators ${U}_{\rho}\in spin(G_f)$, with $\rho=\alpha f$,
have the following action in the linear space spanned by the operators
$D$ and $D_f$:
\begin{eqnarray}
&&~{U}_{\rho}D({U}_{\rho})^{-1}
=T(\alpha f) D [T(\alpha f)]^{-1}=D\cos\alpha+D_f\sin\alpha\,,\label{T1}\\
&&{U}_{\rho}D_f({U}_{\rho})^{-1}
=T(\alpha f) D_f [T(\alpha f)]^{-1}=-D\sin\alpha+D_f\cos\alpha\,.
\label{T2}
\end{eqnarray}
\end{theor}
\begin{demo}
From Eq.  (\ref{Lamaf}) we obtain the matrix elements
$\Lambda^{\mu\,\cdot}_{\cdot\,\nu}(\alpha f)=
\cos\alpha\, \delta^{\mu}_{\nu}+\sin\alpha\,
f^{\mu\,\cdot}_{\cdot\,\nu}$
which lead to the above result since $\Sigma_f$ as well as $T(\alpha f)$
are covariantly constant.
\end{demo}\\
From this theorem it results that $\alpha\in [0,2\pi]$ and, consequently,
the group $G_f\sim U(1)$ is {\em compact}. Therefore, it must be
a subgroup of the maximal compact subgroup of ${\mb G}_c$. In addition,
from Eq.  (\ref{Lamaf}) we see that $L_{f}\sim so(2)$ is the Lie algebra of the
vector representation of $G_f$ that is the compact group
$vect(G_f)=\{\Lambda(\alpha f)|\alpha\in [0,2\pi]\} \sim U(1)$.
Note that the transformations (\ref{T1}) and (\ref{T2})
leave invariant the operator $D^2=(D_f)^2$  because this commutes
with the spin-like operator $\Sigma_f$ which generates these transformations.

Particularly, if $M_n$ allows real-valued unit roots (i.e. complex structures)
this is an usual K\" ahler manifold. In general, when $f$ has complex
components (and $f^*\not=f$) then $L_{f^*}\sim so(2)$ is a different linear
space representing the Lie algebra of $vect({G}_{f^*})$. These two Lie
algebras are complex conjugated to each other but remain isomorphic since
they are real algebras.  The relation among the transformation matrices of
$spin({G}_{f})$ and $spin({G}_{f^*})$ is
$\overline{T}(\alpha f)=T(-\alpha f^*)=[T(\alpha f^*)]^{-1}$
which means that when $f^*\not=f$ the representation $spin(G_f)$ is
no more unitary in the sense of the generalized Dirac adjoint.

The conclusion is that an unit root gives rise  simultaneously to a Dirac-type
operator $D_f$ which satisfies Eq.  (\ref{D2D2}) and the one-parameter
Lie group $G_f$ one needs to relate $D$ and $D_f$ to each
other.

\subsection{Symmetries due to families of unit roots}

The next step is to investigate if there could appear higher symmetries
given by non-abelian Lie groups with many parameters, embedding different
abelian groups $G_f$ produced by some sets of unit roots which have to
form bases of linear spaces isomorphic with the Lie algebras of these
non-abelian groups. Such Lie algebras must include many one-dimensional Lie
algebras $L_f$ being thus subsets of ${\mb R}(M_n)+\{0\}$ where we know that
the linear properties are rather exceptions. Therefore, we must look for
special {\em families} of unit roots, ${\mb f}=\{f^i\,|\,i=1,2,...,N_f\}\subset
{\mb R}_1(M_n)$, having supplementary  properties which should guarantee
simultaneously that: (I) the linear space $L_{\mb f}=\{\rho\,|\,
\rho=\rho_i f^i,\, \rho_i\in {\Bbb R}\}$ is isomorphic with a real Lie algebra,
and (II) each element $\rho\in L_{\mb f}-\{0\}$ is a root (of an arbitrary
norm).

The first condition is accomplished only if the set $\{ T(\rho)\,|\, \rho \in
L_{\mb f}\}$ includes a Lie group with $N_f$ parameters. This means that the
operators $\Sigma^i=\Sigma_{f^i},\, i=1,2,...N_f $ must be (up to constant
factors) the basis-generators of a Lie algebra with some real structure
constants $c_{ijk}$. Then according to Eqs.  (\ref{SSS}) and (\ref{SfS}), we
can write
\begin{equation}\label{SSffS}
\left[\Sigma^i, \Sigma^j\right]=
\textstyle{\frac{i}{2} \left[\,\left<f^i\right>,\,\left<f^j\right>\,
\right]_{\mu\nu}} S^{\mu\nu}= ic_{ijk}\Sigma^k   \,,
\end{equation}
obtaining a necessary condition for ${\mb f}$ be a family of unit roots,
\begin{equation}\label{ffcf}
{\textstyle [\,\left<f^i\right>,\,\left<f^j\right>\, ]}=c_{ijk}
{\textstyle \left< \right.} f^k {\textstyle \left. \right>}\,.
\end{equation}
The condition (II) is accomplished only when $\left<\rho\right>^2$ is equal
up to a
positive factor (i.e. the squared norm) with $-1_n$. This requires to have
\begin{equation}\label{ffkf}
 \left\{\,\textstyle{\left<f^i\right>,\,\left<f^j\right>}\,\right\}
=-2\kappa_{ij} 1_n\,.
\end{equation}
where $\kappa$ is a {\em positive definite} metric that can be brought in
canonical form $\kappa_{ij}=\delta_{ij}$ through a suitable choice of the
unit roots.  If ${\mb f}$ satisfy simultaneously Eqs. (\ref{ffcf}) and
(\ref{ffkf}) then $L_{\mb f}$ is just the Lie algebra of the group
$\{\Lambda(\rho)\,|\, \rho\in L_{\mb f}\}$ the matrices of which read
\begin{equation}\label{LX}
\Lambda(\rho)=\textstyle{e^{\rho_i \left<f^i\right>}=1_n\cos\|\rho\|+
\nu_i \left<f^i\right>}\sin\|\rho\| \,,
\end{equation}
where $\|\rho\|=\sqrt{\rho_i\rho_i}$ (when we take $\kappa_{ij}=\delta_{ij}$)
and $\nu_i =\rho_i/\|\rho\|$. All these results lead to the following
\begin{theor}
If the set ${\mb f}=\{f^i\,|\,i=1,2,...,N_f\} \in {\mb R}_1(M_n)$ is a family
of unit roots then the matrices $1_n$ and $\left<f^i\right>,\,
i=1,2,...,N_f,$ form the basis of a matrix representation of a
finite-dimensional associative algebra over ${\Bbb R}$.
\end{theor}
\begin{demo}
Since ${\mb f}$ is a family of unit roots in the sense of above definition,
$f^i$ must satisfy Eqs. (\ref{ffcf}) and (\ref{ffkf}) with the canonical metric.
Hereby it results that the set of the real linear combinations
$\rho_{0}1_n+\rho_{i}\left<f^i\right>$ forms an associative algebra
with respect to the matrix multiplication
that can be calculated by adding the commutator and anticommutator.
Moreover, this algebra is a division one since there exists the zero element
(with $\rho_0=0,\,\rho_i=0$), the unit element is $1_n$ and each element
different from zero has the inverse
$(\rho_{0}1_n+\rho_{i}\left<f^i\right>)^{-1}=
(\rho_{0}1_n-\rho_{i}\left<f^i\right>)
/({\rho_0}^2+\rho_i\rho_i)$.
Obviously, this real algebra is finite possessing a basis of dimension
$N_f+1$ where $\left<f^i\right>$ play the role of complex
units. Eq. (\ref{LX}) can be interpreted as a matrix representation of the
Euler formula.
\end{demo}\\
This theorem severely restricts the  existence of the families of unit roots.
Indeed, according to the Frobenius theorem there are only two finite real
algebras able to give suitable representations in spaces of roots, namely the
algebra ${\Bbb C}$ of complex numbers and the {\em quaternion} algebra,
${\Bbb H}$. In the first case we have {\em isolated} unit roots $f$  and
representations of the ${\Bbb C}$ algebra generated by the matrices $1_n$
and $\left<f\right>$ (which play the role of $i\in {\Bbb C}$) related to the
continuous symmetry  group $G_f\sim U(1)$ we studied in the previous
section.

Here we focus on the second possibility leading to families of unit roots with
$N_f=3$ that constitute matrix representations of the quaternion units.
\begin{theor}\label{bumbum}
The unique type of family of unit roots with $N_f>1$ having the properties (I)
and (II) are the triplets ${\mb f}=\{ f^1,f^2,f^3\}\subset {\mb R}_1(M_n)$ which
satisfy
\begin{equation}\label{algf}
{{\textstyle  \left<f^i\right>\,\left<f^j\right>}=-\delta_{ij} 1_n} +
\varepsilon_{ijk} {\textstyle
\left< \right.} f^k {\left. \textstyle \right>}\,,
\quad i,j,k...=1,2,3\,.
\end{equation}
\end{theor}
\begin{demo}
Taking into account that $\varepsilon_{ijk}$ is the anti-symmetric tensor
with $\varepsilon_{123}=1$ we recognize that Eqs. (\ref{algf}) are the
well-known multiplication rules of the quaternion units or similar algebraic
structures (e.g. the Pauli matrices). Consequently, the matrices
$\left<f^i\right>$ and $1_n $ generate a matrix representation of
${\Bbb H}$. Other choices are forbidden by the Frobenius theorem.
\end{demo}\\
If the unit roots $f^i$ have only real-valued components
we recover the {\em hypercomplex structures} defining hyper-K\" ahler
geometries (presented in the Appendix A).

Eqs.  (\ref{algf}) combined with the previous results (\ref{SSffS})-(\ref{LX})
provide all the features of the specific continuous symmetry associated to
${\mb f}$.
\begin{defin}
We say that $G_{\mb f}=\{(A_{\rho},id)\,|\,\rho \in L_{\mb f}\}\sim SU(2)
\subset {\mb G}_c(\eta)$
is the Lie group associated to the triplet ${\mb f}\subset {\mb R}_1(M_n)$.
\end{defin}
The spinor and the vector representations of this group are determined by the
representations of its Lie algebra, $g_{\mb f}$, resulted from
Theorem \ref{bumbum}.
\begin{cor}
The basis-generators of $vect(g_{\mb f})=L_{\mb f}$ are
$\frac{i}{2} f^i$ while the
basis-ge\-ne\-ra\-tors of the algebra $spin(g_{\mb f})\sim su(2)$ read
$\hat s_i=\frac{1}{2}\Sigma^i$ $(i=1,2,3)$.
\end{cor}
\begin{demo}
From Eqs.  (\ref{algf}) and (\ref{ffcf}) we deduce that
$c_{ijk}=2\varepsilon_{ijk}$.
Furthermore, from  Eqs.  (\ref{SSffS}) and (\ref{ffcf})
we obtain the standard commutation rules of $SU(2)$ generators,
\begin{equation}\label{sss}
\left[ \hat s_i,\,\hat s_j \right]=i\varepsilon_{ijk}\hat s_k\,,
\end{equation}
and similarly for  $\frac{i}{2}\left<f^i\right>$.
\end{demo}\\
Now $vect(G_{\mb f})=\{\Lambda(A_{\rho})\,|\,\rho\in L_{\mb f},\,
\|\rho\|\le 2\pi\}$ is the compact group formed by the
matrices $\Lambda(A_\rho)=\Lambda(\rho)$ of the form (\ref{LX}) constructed
using the elements of the Lie algebra $L_{\mb f}\sim su(2)\sim so(3)$. The
transformation matrices giving the action of the operators
$U_{\rho}\in spin(G_{\mb f})$,
\begin{equation}\label{TxiS}
T(\rho)=e^{-i\rho_i \Sigma^i} =e^{-2i\rho_i \hat s_i}\,,
\quad \rho=\rho_i f^i\in L_{\mb f},
\end{equation}
have to be calculated directly from Eq.  (\ref{Taf}) replacing
$\alpha=\pm\|\rho\|=\pm\sqrt{\rho_i\rho_i}$ and $f=\pm \rho/ \|\rho\|$. Then
the transformations (\ref{TgT1}) can be expressed in terms
of the parameters $\rho_i$ using the matrices $\Lambda({\rho})$.
Hereby we observe that $\rho_i$ are nothing other than
the analogous of the well-known Cayley-Klein parameters but ranging in a larger
spherical domain (where $\|\rho\|\le 2\pi$) such that they cover two times the
group $G_{\mb f}\sim SU(2)$, as we can convince ourselves calculating
\begin{equation}\label{LfLf}
\textstyle{\Lambda(\rho)\left<f^i\right>\Lambda^T(\rho)=
{\frak R}_{ij}(2\vec{\rho})
\left<f^j\right>}\,, \quad \forall\, \Lambda_{\rho}\in vect(G_{\mb f})\,,
\end{equation}
where ${\frak R}(2\vec{\rho})\in O(3)$ is the rotation of the Cayley-Klein
parameters $2\rho_i$. These arguments lead to the conclusion  that
$vect(G_{\mb f})\sim SU(2)$  \cite{CV6}.
On the other hand, since the rotations (\ref{LfLf}) change the basis of
$L_{\mb f}$ leaving Eqs.  (\ref{algf}) invariant, we understand that these
form the group $Aut(L_{\mb f})$, of the automorphisms of the Lie algebra
$L_{\mb f}$ considered as a real algebra.

In the case of triplets involving only real-valued unit roots when the
geometry is  hyper-K\" ahler, every family of real unit roots (i.e., a
hypercomplex structure) ${\mb f}$ has its own Lie algebra $L_{\mb f}\sim su(2)$.
These algebras cannot be embedded in a larger one because of the restrictions
imposed by the Frobenius theorem.  An example of hyper-K\" ahler manifold
is the Euclidean Taub-NUT space which is equipped with only one family of
real unit roots  \cite{CV6, CV7}. The manifolds with
pseudo-Euclidean metric with odd $n_{+}$ and $n_{-}$ have only pairs of
{\em adjoint} triplets, ${\mb f}$ and ${\mb f}^*$, the last one being formed
by the adjoints of the unit roots of ${\mb f}$. The spaces $L_{\mb f}$ and
$L_{{\mb f}^*}$ are isomorphic between themselves (as real vector spaces)
and all the results concerning the symmetries generated by ${\mb f}^*$ can be
taken over from those of ${\mb f}$ using complex conjugation. Moreover, we
must specify that the set $L_{\mb f}\bigcup L_{{\mb f}^*}$ is no more a linear
space since the linear operations among the elements of $L_{\mb f}$ and
$L_{{\mb f}^*}$ are not allowed. An  example is the Minkowski spacetime
which has a pair of conjugated triplets of complex-valued unit roots  \cite{K2}.
Both these examples of manifolds possessing triplets with the properties
(\ref{algf}) are of dimension four. The results we know  indicate that similar
properties may have other manifolds of dimension $n=4k,\, k=1,2,3,...$
where we expect to find many such triplets  \cite{GaMo}. The main geometric
feature of all these manifolds is given by
\begin{theor}\label{ricci}
If a manifold $M_n$ allows a triplet of unit roots then this must be
Ricci flat (having $R_{\mu\nu}=0$).
\end{theor}
\begin{demo}
As in the case of any hyper-K\" ahler manifold, using Eqs.  (\ref{D2D}) and
(\ref{algf}) we calculate the expression
$R_{\mu\nu\alpha\beta}f^{1\,\alpha\beta}=
R_{\mu\nu\sigma\beta}f^{3\, \sigma\,\cdot}_{~\,\,\cdot\,\alpha}
(\left<f^3\right>\left<f^{1}\right>)^{\alpha\beta}=
R_{\mu\nu\sigma\beta}f^{3\, \sigma\,\cdot}_{~\,\,\cdot\,\alpha}
f^{2\,\alpha\beta}=
-R_{\mu\nu\alpha\beta}f^{1\,\alpha\beta}$ which vanishes.  Furthermore,
permutating the first  three indices of $R$ we find the identity
\begin{equation}\label{2Rf0}
2R_{\mu\alpha\nu\beta}f^{1\,\alpha\beta}=
R_{\mu\nu\alpha\beta}f^{1\,\alpha\beta}=0\,.
\end{equation}
Finally, writing
$R_{\mu\nu}=
R_{\mu\alpha\nu\beta}f^{1\,\alpha\,\cdot}_{~\,\cdot\,\tau} f^{1\,\beta\tau}=
-R_{\mu\alpha\sigma\beta}f^{1\,\sigma\,\cdot}_{~\,\cdot\,\nu}
f^{1\,\alpha\beta}
=0$, we draw the conclusion that the manifold is Ricci flat. The
same procedure holds for $f^2$ or $f^3$ leading  to identities similar
to (\ref{2Rf0}). Note that the manifolds possessing only single unit roots
(as the K\" ahler ones) are not forced to be Ricci flat.
\end{demo}

Starting with a triplet ${\mb f}=\{f^1,f^2,f^3\}\subset {\mb R}_1(M_n)$
satisfying (\ref{algf}) one can construct a rich set of Dirac-type
operators of the form $D(\vec{\nu})=\nu_i D^i$ where $\vec{\nu}$ is an unit
vector (with $\vec{\nu}^2=1$) and $D^i= D_{f^i}=i[D,\,\Sigma^i]$, $i=1,2,3$,
play the role of a {\em basis}. This set is {\em compact} and isomorphic with
the sphere of unit roots $S^2_{\mb f}=\{f_{\vec{\nu}}\,|\,f_{\vec{\nu}}=
\nu_i f^i,\, \vec{\nu}^2=1\} \subset L_{\mb f}$, since $D(\vec{\nu})=
D_{f_{\vec{\nu}}}$ for any $f_{\vec{\nu}}\in {S}_{\mb f}^2$. Moreover,
each operator $D(\vec{\nu})$ can be related to $D$ through the transformations
(\ref{T1}) and (\ref{T2}) of the one-parameter
subgroup ${G}_{f_{\vec\nu}}\subset G_{\mb f} \sim SU(2)$  defined by
$f_{\vec{\nu}}$.
\begin{theor}
The operators $U_{\rho}\in spin(G_{\mb f})$ transform the
Dirac operators  $D$, $D^i$ ($i=1,2,3$) as
\begin{eqnarray}
{U}_{\rho}D[{U}_{\rho}]^{-1}&=&
T(\rho)D[T(\rho)]^{-1}=D\cos \|\rho\|+\nu_i D^i\sin\|\rho\|\,,\label{TDT10}\\
{U}_{\rho}D^i[{U}_{\rho}]^{-1}&=&
T(\rho)D^i[T(\rho)]^{-1}\nonumber\\
&=&D^i\cos \|\rho\|-
(\nu_i D+\varepsilon_{ijk} \nu_j D^k)\sin\|\rho\|\,, \label{TDT20}
\end{eqnarray}
where $\rho=\rho_i f^i=\|\rho\| \nu_i f^i$.
\end{theor}
\begin{demo}
We consider the result of Theorem \ref{TDT} for each one-parameter subgroup of
$G_{\mb f}$ generated by the unit roots $f_{\vec{\nu}}=\nu_if^i$. The
straightforward calculation starting with Eq. (\ref{TxiS}) is also efficient.
\end{demo}\\
The previous results indicate that the set ${\mb R}_1(M_n)$, of unit roots
producing Dirac-type operators, has an interesting topological structure
involving either single $f$  producing isolated  Dirac-type operators  or unit
spheres  ${S}^2_{\mb f}$ leading to compact sets of Dirac-type operators. In
order to show off this structure one needs to exploit the mechanisms of our
theory based on the fact that the linear spaces $L_f$ or $L_{\mb f}$ are
isomorphic with the Lie algebras of the symmetry  groups of the Dirac-type
operators generated by spin-like operators.

In the non-K\" ahlerian manifolds equipped with pairs of adjoint triplets
${\mb f}$ and ${\mb f}^*$,  the Dirac-type operators $D^i$ and
$D_{(f^i)^*}=\overline{D}^{i}$ are related to each other through the
Dirac adjoint. However, when an extended symmetry dealing with the physical
needs would be necessary, we may consider the complexified group
$(G_{\mb f})_c\sim SL(2,{\Bbb C})$ of $G_{\mb f}$. Then we have to work with
more complicated groups and Lie algebras since the complexification doubles
the number of generators of the spinor or vector representations. For example,
the generators of the complexified spinor representation, $spin(g_{\mb f})_c$,
are $\hat s_i$ and $(\pm)i\hat s_i$ and similarly for $vect(g_{\mb f})_c$.

\subsection{Supersymmetry and isometries in K\" ahlerian manifolds}

Beside the types of continuous  symmetries we have studied, the presence of
the unit roots gives rise to  supersymmetries related to the external
symmetries in an interesting manner. In order to avoid the complications
due to the presence of the pair of adjoint triplets we restrict ourselves
to discuss in this section only K\" ahlerian manifolds.

In a K\" ahler manifold, a complex structure $f=f^*$ generates its
own ${\cal N}=2$ {\em real} superalgebra.
\begin{defin}
Given an isolated unit root $f$, we say that the set ${\mb
  d}_f=\{D(\lambda)|D(\lambda)\\=\lambda_0 D+\lambda_1
  D_f\,;\lambda_0,\lambda_1\in {\Bbb R}\}$ represent the ${\cal N}=2$
 real D-superalgebra generated by the unit root $f$.
\end{defin}
When $f^*\not = f$ the D-superalgebra ${\bf d}_{f^*}$ differs from
${\mb
  d}_f$ and, in general, these can not be embedded in a larger superalgebra.
The basis of this D-superalgebra, $D$ and $D_f$ (obeying
$\{D,D_f\}=0$, $(D_f)^2=D^2$)  can be changed through the
transformations (\ref{T1}) and (\ref{T2}) that preserve the
anticommutation relations. These form the group of automorphisms
of ${\mb d}_f$, $Aut({\mb d}_f)\sim SO(2)$. If the manifold  has a
non-trivial isometry group $I(M_n)$ then an arbitrary isometry
$x\to x'=\phi_{\xi}(x)$ transforms $f$ as a second rank tensor,
\begin{equation}\label{ffprim}
f_{\mu\nu}(x)\to f'_{\mu\nu}(x')
\frac{\partial x^{\prime\,\mu}}{\partial x_\alpha}
\frac{\partial x^{\prime\,\nu}}{\partial x_\beta}=
f_{\alpha\beta}(x)\,.
\end{equation}
When there is only one $f$ we are forced to put $f'=f$ which means this remains
{\em invariant} under isometries.
\begin{theor}\label{invf}
If a K\" ahler manifold $M_n$ with the external symmetry group $S(M_n)$ has a
single complex structure, $f$, then every generator $X\in spin[s(M_n)]$
commutes  with $D_f$.
\end{theor}
\begin{demo}
We calculate first the derivatives with respect to $\xi^a$ of Eq. (\ref{ffprim})
for $f'=f$ and $\xi=0$. Then, taking into account that $f$ is covariantly
constant we can write $f_{\alpha\lambda}k_{~;\beta}^{\lambda}=
f_{\beta\lambda}k_{~;\alpha}^{\lambda}$ for each Killing vector field $k$
defined by Eq.  (\ref{ka}). This identity yields
\begin{equation}
[X,\, \Sigma_f]=0\,,\quad [X,\,D_{f}]=0\,, \quad \forall\, X\in spin[s(M_n)]\,,
\end{equation}
which means that the operators $\Sigma_f$ and $D_f$ are invariant under
isometries.
\end{demo}

The case of the hyper-K\" ahler manifolds is  more complicated since
a triplet ${\mb f}$ gives rise to self-adjoint Dirac-type operators
$D^i=\overline{D}^i$ which anticommute with $D$ and present the continuous
symmetry discussed in the previous section. In these conditions a new algebraic
structure is provided by
\begin{theor}
If a triplet ${\mb f}\subset {\mb R}_1(M_n)$ accomplishes  Eqs.  (\ref{algf})
then the corresponding Dirac-type operators satisfy
\begin{equation}\label{4sup}
\left\{D^i,\,D^j\right\}=2\delta_{ij} D^2\,,\quad
\left\{D^i,\,D\right\}=0\,.
\end{equation}
\end{theor}
\begin{demo}
If $i=j$ we take over the result of Theorem \ref{DtyD}. For $i\not=j$
we take into account that $M_n$ is Ricci flat  finding  that $D^i$ and $D^j$
anticommute. The second relation was demonstrated earlier for any
unit root.
\end{demo}\\
Thus it is clear that the operators $D$ and $D^i$ ($i=1,2,3$) form
a basis of a four-dimensional real superalgebra of Dirac
operators.
\begin{defin}
The set ${\mb d}_{\mb f}=\{D(\lambda)|
D(\lambda)=\lambda_0D+\lambda_iD^i; \lambda_0, \lambda_i\in{\Bbb
R}\}$ is the ${\cal N}=4$ D-superalgebra generated by the triplet
${\mb f}$.
\end{defin}
This D-superalgebra contains the subset ${\mb d}_{\mb
f}^1=\{D(\nu)\,|\, {\nu_0}^2+\vec{\nu}^2=1\}$ of the Dirac
operators which have the property  $D(\nu)^2=D^2$. The set ${\mb
d}_{\mb f}^1$ has the topology of the sphere $S^3$ including the
sphere $S^2$ of the Dirac-type operators
$D(\vec{\nu})=D(0,\vec{\nu})$.

Furthermore, it is natural to study the group of automorphisms of
this D-superalgebra, $Aut({\mb d}_{\mb f})$, and its Lie algebra,
$aut({\mb d}_{\mb f})$. Obviously, these automorphisms have to be
linear transformations among $D$ and $D^i$ preserving their
anticommutation rules. The transformation matrices $T(\rho)$
commute with $D^2$, leaving Eqs. (\ref{4sup}) invariant under
transformations (\ref{TDT10}) and (\ref{TDT20}) which appear thus
as automorphisms of ${\mb d}_{\mb f}$ forming a $SU(2)$ subgroup
of $Aut({\mb d}_{\mb f})$. However, we need more automorphisms in
order to complete the group $Aut({\mb d}_{\mb f})$ with more
ordinary or invariant subgroups, isomorphic with $SU(2)$ or
$O(3)$. These supplemental automorphisms must transform the
operators $D^i$ among themselves preserving their anticommutators
as well as the form of $D$. Therefore, these may be produced by
the transformations of $S(M_n)$ since these leave the operator $D$
invariant.

In what concerns the transformation of the triplets ${\mb f}$ under isometries
we have two possibilities, either to consider that all the complex structures
$f^i\in ${\mb f}$ $ are invariant under isometries or to
assume that the isometries transform the components of the triplet among
themselves, $f^{\prime\,i}=\hat{\frak R}_{ij}f^j$, such that Eqs.  (\ref{algf})
remain invariant. The first hypothesis is not suitable since we need more
transformations in order to fill in the group $Aut({\mb d}_{\mb f})$ when
we do not have other sources of symmetry. Therefore, we must adopt the second
viewpoint assuming that the components of ${\mb f}$ are transformed as
\begin{equation}\label{ffprim1}
f^{j}_{\mu\nu}(x')\frac{\partial x^{\prime\,\mu}}{\partial x_\alpha}
\frac{\partial x^{\prime\,\nu}}{\partial x_\beta}=
\hat{\frak R}_{kj}(\xi,x)f^k_{\alpha\beta}(x)\,,
\end{equation}
by $3\times 3$ real {\em orthogonal} matrices  $\hat{\frak R}\in O(3)$
that leave Eqs. (\ref{algf}) invariant. Their matrix elements can be put in
the equivalent forms
\begin{eqnarray}
\hat{\frak R}_{ij}(\xi,x)&= &\frac{1}{n}
f^{i\,\alpha\beta}(x)
\frac{\partial \phi_{\xi}^{\mu}(x)}{\partial x_\alpha}
\frac{\partial \phi_{\xi}^{\nu}(x)}{\partial x_\beta}
f^{j}_{\mu\nu}[\phi_{\xi}(x)] \nonumber\\
&=& \frac{1}{n}
f^{i\,\hat\alpha\hat\beta}(x)
\Lambda^{\hat\mu\,\cdot}_{\cdot\,\hat\alpha}[A_{\xi}(x)]
\Lambda^{\hat\nu\,\cdot}_{\cdot\,\hat\beta}[A_{\xi}(x)]
f^{j}_{\hat\mu\hat\nu}[\phi_{\xi}(x)]\,.\label{rotroot}
\end{eqnarray}
The last formula is suitable for calculations in local frames where we must
consider the external symmetry using the gauge transformations
$\Lambda[A_{\xi}(x)]\in vect[{\mb G}(\eta)]$ defined by Eq. (\ref{Axx}) and
associated to isometries for preserving the gauge.

The matrices $\hat{\frak R}$ might be point-dependent and depend on the
parameters $\xi^a$ of $I(M_n)$. This means that the canonical covariant
parameters $\hat\omega_{ij}=-\hat\omega_{ji}$ giving the expansion
$\hat{\frak R}_{ij}(\hat\omega)= \delta_{ij}+\hat\omega_{ij}+\cdots$
are also depending on these variables. Then, for small values of the parameters
$\xi^a$ the covariant parameters can be developed as $\hat\omega_{ij}=
\xi^a\hat c_{aij}+...$ emphasizing thus the quantities $\hat c_{aij}$
we shall see that do not depend on coordinates.
\begin{theor}
Let $M_n$ be a hyper -K\" ahler manifold having the hypercomplex structure
${\mb f}=\{f^1,f^2,f^3\}$ and a non-trivial isometry group
$I(M_n)$ with parameters $\xi^a$ and the corresponding Killing vectors $k_a$
as defined by Eq. (\ref{ka}). Then the basis-generators $X_a\in spin[s(M_n)]$
and $\hat s_i=\frac{1}{2}\Sigma^i \in spin(g_{\mb f})\sim su(2)$ satisfy
\begin{equation}\label{Xscs}
[X_a,\, \hat s_i]=i\hat c_{aij} \hat s_j\,, \quad a=1,2,...,N\,,
\end{equation}
where $\hat c_{aij}$ are {\em point-independent} structure constants.
\end{theor}
\begin{demo}
Deriving  Eq.  (\ref{ffprim1}) with respect to $\xi^a$ in $\xi=0$ we deduce
\begin{equation}\label{fkcf}
f^{i}_{\mu\lambda}k_{a\,;\nu}^{\lambda}-
f^{i}_{\nu\lambda}k_{a\,;\mu}^{\lambda}=\hat c_{aij}f^{j}_{\mu\nu}\,,
\end{equation}
which leads to the explicit form
\begin{equation}
\hat c_{aij}= -\frac{2}{n}\,\varepsilon_{ijl}\, \hat k^l_a\,,\quad
\hat k^l_a=f^{l\,\mu\nu}k_{a\, \mu;\nu}\,.
\end{equation}
Bearing in mind that  $f^i_{\mu\nu;\sigma}=0$  and using
the identity  $f^{i\,\mu\nu}k_{a\,\mu;\nu;\sigma}=
R_{\mu\sigma\nu\,\cdot}^{\,\cdot\,\cdot\,
\cdot\,\lambda}k_{a\,\lambda}f^{i\,\mu\nu}$  \cite{W} and Eq.  (\ref{2Rf0}),
we find that $\nabla_{\sigma}\hat k^i_a=
f^{i\,\mu\nu}k_{a\, \mu;\nu;\sigma}=0$ which means that $\partial_{\sigma}
\hat c_{aij}=0$. Finally, from Eq.  (\ref{fkcf}) we derive the commutation
rules (\ref{Xscs}).
\end{demo}\\
Now we can point out how act the isometries $x\to x'=\phi_{\xi}(x)$ on the
operators $D^i$.
\begin{cor}
The Dirac-type operators $D^i$ produced by any triplet ${\mb f}$
transform under isometries according to the representation $O_{\mb
f}= \{ \hat{\frak R}(\xi)\,|\, (Id,\phi_{\xi})\in I(M_n)\}$ of the
group $I(M_n)$ {\em induced}  by the group $O(3)$. The matrices
$\hat{\frak R}(\xi)$ of this representation have the form
(\ref{rotroot}) and give the transformation rule
\begin{equation}\label{UDUR}
(U_{\xi}D^iU^{-1}_{\xi})[\phi_{\xi}(x)]=\{T(A_{\xi})D^i\overline{T}(A_{\xi})(x)=
\hat{\frak R}_{ij}(\xi)D^j(x)      \,,
\end{equation}
where the action of $U_{\xi}\in spin[S(M_n)]$ is defined by Eq.  (\ref{rep}).
\end{cor}
\begin{demo}
The matrices (\ref{rotroot}) generated as any adjoint representation,
\begin{equation}
\hat {\frak R}(\xi)=e^{i\xi^a{\frak X}_a}\,, \quad
({\frak X}_a)_{ij}=-i\hat c_{aij}\,,
\end{equation}
are point-independent since $\hat c_{aij}$ are structure constants. Moreover,
if we commute Eq.  (\ref{Xscs}) with $X_b$
using Eqs.  (\ref{comX}) and (\ref{sss}) we obtain
$[ {\frak X}_a, {\frak X}_b]=ic_{abc}{\frak X}_c$  concluding
that $O_{\mb f}$ is a well-defined induced representation of $I(M_n)$.
From Eqs.  (\ref{SfS}) and (\ref{ffprim1}) we derive
\begin{equation}\label{STTSR}
(U_{\xi}\Sigma^iU^{-1}_{\xi})(x')=
[T(A_{\xi})\Sigma^i\overline{T}(A_{\xi})](x)= \hat{\frak
R}_{ij}(\xi)\Sigma^j(x)\,,
\end{equation}
which leads to Eq. (\ref{UDUR}) after a commutation with $D$ that is invariant
under $U_{\xi}$.
\end{demo}\\
Hence we have the complete image of the symmetries that preserve
the anticommutation rules of the real D-superalgebra ${\mb d}_{\mb
f}$ in a hyper-K\" ahler manifold. These are encapsulated in the
group $Aut({\mb d}_{\mb f})$ whose transformations are defined by
Eqs.  (\ref{TDT10}), (\ref{TDT20}) and (\ref{UDUR}).
\begin{cor}
The group $Aut({\mb d}_{\mb f})=spin[G_{\mb f}\,\circledS\,
S(M_n)]$ is a representation of the semi-direct product  $G_{\mb
f} \,\circledS\, S(M_n)$ where $G_{\mb f}$ is the invariant
subgroup.
\end{cor}
\begin{demo}
The basis generators of the Lie algebra $aut({\mb d}_{\mb f})$ of the group
$Aut({\mb d}_{\mb f})$ are the operators $\hat s_i$ and $X_a$ that obey the
commutation relations (\ref{comX}), (\ref{sss}) and (\ref{Xscs}).
These operators form a Lie algebra since $\hat c_{aij}$ are
point-independent. In this algebra $g_{\mb f}\sim su(2)$ is an ideal and,
therefore, the corresponding $SU(2)$ subgroup is invariant. However, this result
can be obtained directly taking $(A_{\xi},\, \phi_{\xi})\in S(M_n)$ and
 $(A_{\rho}, id)\in G_{\mb f}$ and evaluating
$(A_{\xi},\, \phi_{\xi})* (A_{\rho},\, id)* (A_{\xi},\, \phi_{\xi})^{-1}=
([A_{\xi}\times (A_{\rho}\times A_{\xi}^{-1})]\circ \phi_{\xi}^{-1}, \,id)=
(A_{\rho'},\, id)$ where, according to (\ref{Axx}), (\ref{LX}) and
(\ref{ffprim1}), we have $\rho'=\rho_i\hat{\frak R}_{ij}(\xi)f^j$.
Consequently, $(A_{\rho'},\, id)\in G_{\mb f}$
which means that $G_{\mb f}\sim SU(2)$ is an invariant subgroup.
\end{demo}\\
Another important consequence of the previous theorem is
\begin{cor}
The basis generators of $spin[S(M_n)]$ and the Dirac-type
operators of the ${\cal N}=4$ D-superalgebra ${\mb d}_{\mb f}$
obey
\begin{equation}
[X_a, D^i]=i\hat c_{aij} D^j\,.
\end{equation}
\end{cor}
\begin{demo}
This formula results commuting Eq. (\ref{Xscs}) with $D$.
\end{demo}\\
Finally, we find an interesting restriction that can be formulated as
\begin{cor}
The minimal condition that $M_n$ allows a hypercomplex structure is to
have an isometry group that includes at least one $O(3)$ subgroup.
\end{cor}
\begin{demo}
The subgroup $O_{\mb f}\sim O(3)$ of $Aut({\mb d}_{\mb f})$ needs at least
three generators $X_a$ satisfying the $su(2)$ algebra. Thus we conclude that
$S(M_n)$ must include one $SU(2)$ group for each different hypercomplex
structure of $M_n$.
\end{demo}\\
This restriction is known in four dimensions  where there exists only three
hyper-K\" ahler manifolds with only one hypercomplex structure and one
subgroup  $O(3)\subset I(M_4)$  \cite{GR1}. These are
given by the Atiyah-Hitchin  \cite{AH}, Taub-NUT and Eguchi-Hanson  \cite{EH}
metrics, the first one being only that does not admit more $U(1)$ isometries
\cite{GR1,BS}. In addition, we have the example of the four-dimensional
Euclidean flat space that has two different triplets and the isometry group
$E(4)$ including the group $O(4)\sim O(3)\times O(3)$.

The theory above can be easily extended to non-K\" ahlerian manifolds having
pairs of adjoint triplets of unit roots. In this case Eq. (\ref{rotroot}) gives
complex-valued orthogonal matrices which oblige us to start with the complex
D-superalgebra $({\mb d}_{\mb f})_c$ (defined over ${\Bbb C}$ instead of ${\Bbb
R}$) and with the complexified groups and Lie algebras  \cite{CVnew}. Then the
group of automorphisms of $({\mb d}_{\mb f})_c$, will be the group $Aut({\mb
d}_{\mb f})_c$, that is a representation of the group $(G_{\mb f})_c
{~\circledS~} S(M_n)$ involving the representation $O_{\mb f}$ of the group
$S(M_n)$ induced by the group of the complex-valued orthogonal matrices
$O(3)_c$. Obviously, the invariant subgroup here is $(G_{\mb f})_c \sim SL(2,
{\Bbb C})$. The minimal condition that $M_n$ allows a pair of adjoint triplets
is the group $S(M_n)$ to include at least one $SL(2, {\Bbb C})$ subgroup since
we need six generators for building the representation $O_{\mb f}\sim O(3)_c$.
The Minkowski spacetime which has a pair of adjoint triplets and $O(3,1)$
isometries is a typical example (see the Appendix B).

\subsection{Discrete symmetries}

In many physical problems the  study of the discrete symmetries could
be also productive. Of course, the results concerning the continuous symmetries
obtained above will be crucial for understanding the structure of the discrete
transformations which relate among themselves the standard Dirac operator and
the Dirac-type ones \cite{K2,CV6,CV7}.

Let us start with the simplest case of an isolated unit root $f$.
\begin{theor}\label{D1}
For any unit root $f$ there exists the discrete group
${\Bbb  Z}_4(f)\subset G_f$ the orbit of which is $\{D,-D,D_f,-D_f\}$.
\end{theor}
\begin{demo}
Using Eqs.  (\ref{T1}) and (\ref{T2}) one observes that the transformations
${\mb 1}$, $U_f=T(\frac{\pi}{2}f)$, $\digamma  =(U_f)^2=T(\pi f)$, and
$(U_f)^3 = T(-\frac{\pi}{2}f)=\digamma  U_f=U_f \digamma  $ form the spinor
representation of
the cyclic group ${\Bbb  Z}_4(f)$. Since  $\digamma  ^2={\mb 1}$, the pair
$({\mb 1}, \digamma  )$ represents
the subgroup ${\Bbb  Z}_2\subset {\Bbb  Z}_4(f)$. According to Eq.  (\ref{T1})
we find that
\begin{equation}\label{DUDU}
D_f=U_f D(U_f)^{-1} \,,
\end{equation}
while the action of the matrix $\digamma $,
\begin{equation}\label{pari}
\digamma  \gamma^{\mu}\digamma =-\gamma^{\mu}\,,
\end{equation}
is independent on the form of $f$ so that this changes the sign of all the
Dirac or Dirac-type  operators.
\end{demo}\\
For a given manifold, $M_n$, the matrix $\digamma $ is uniquely defined up to a
factor $\pm 1$. Thus $\digamma $ is in some sense independent on the discrete
symmetry group where is involved, playing the role of a chiral matrix. For
this reason it is convenient to identify  $\gamma^{ch}=\digamma $ bearing in
mind that then we must have $\overline{\digamma }=\epsilon_{ch} \digamma
=\pm \digamma  $.
When the metric is pseudo-Euclidean then the operators of the spinor
representation of the discrete group ${\Bbb Z}_4(f^*)$ produced by
$f^*\not =f$ have to be written directly using the Dirac adjoint. Indeed, from
Eq.  (\ref{DUDU}) we obtain
\begin{equation}
U_{{f}^*}=(\overline{U_f})^{-1}=\overline{U_f}\,\,\overline{\digamma }\,,
\end{equation}
and similarly for the other operators.
Starting with these elements  the remaining operators of
the cyclic group will be obtained using multiplication  \cite{BR}.

A most interesting case is that of the discrete symmetries of the Dirac-type
operators $D^i$ ($i=1,2,3$) given by the triplet ${\mb f}$ which satisfy Eqs.
(\ref{algf}) \cite{CV7}.
\begin{theor}
The Dirac operators $\pm D$ and the Dirac-type ones
$\pm D^1$, $\pm D^2$, and $\pm D^3$ are related among themselves through the
transformations of the spinor representation of the quaternion group,
${\Bbb Q}({\mb f})\subset G_{\mb f}$.
\end{theor}
\begin{demo}
Let us denote by $U_i=T(\frac{\pi}{2}f^i)$ the operators that, according to
Theorem \ref{D1}, have the properties
\begin{equation}
(U_1)^2=(U_2)^2=(U_3)^2=\digamma \,,\quad \digamma ^2={\mb 1}\,,
\end{equation}
and $\digamma U_i=U_i \digamma $. Furthermore, from Eqs.  (\ref{TgT1}),
(\ref{Lamaf}) and
(\ref{algf}) we deduce
\begin{eqnarray}
&U_1 U_2=U_3\,, \quad U_2 U_3=U_1\,, \quad &U_3 U_1=U_2\,,\\
&U_2 U_1=U_3 \digamma \,, \quad U_3 U_2=U_1 \digamma \,, \quad &U_1 U_3=U_2
\digamma \,,
\end{eqnarray}
and, after a few manipulation, we see that  ${\mb 1}$, $\digamma $, $U_i$ and
$\digamma U_i$ ($i=1,2,3$) form a representation of the dicyclic group
$\left<2,2,2 \right>$ \cite{BR} which is isomorphic with the quaternion subgroup of
$G_{\mb f}$ we denote by ${\Bbb Q}({\mb f})$.
Using Eqs.  (\ref{DUDU}) and (\ref{pari}) we find that
its orbit in the space of the Dirac operators is the desired one.
\end{demo}\\
As expected, the cyclic groups  ${\Bbb Z}_4(f^i)$ are subgroups of
${\Bbb Q}({\mb f})$. For this reason the spinor representation of the group
${\Bbb Q}({\mb f}^*)$ has to be derived from that of
${\Bbb Q}({\mb f})$ using the same method as in the case of cyclic groups.

Hence we conclude that for each isolated unit root $f$ one can define a finite
group ${\Bbb Z}_4(f)$ which is a subgroup of $G_f$ while the triplets
${\mb f}$ produce  more complicated finite discrete groups,
${\Bbb Q}({\mb f})\subset G_{\mb f}$. Since the groups $G_f$ and
$G_{\mb f}$ cannot be embedded in a larger group, the product of two
operators of the spinor representation
of two different discrete groups is, in general, an arbitrary operator which
do not correspond to a transformation of another discrete group ${\Bbb Z}_4$
or ${\Bbb Q}$. In addition, this new operator could transform the standard
Dirac operator in a new operator having  different properties to those of the
Dirac-type ones. Therefore when we restrict ourselves to orbits containing
only the Dirac  and Dirac-type operators we have to consider only the discrete
groups discussed above \cite{CV7}.

Finally we note that in any Dirac theory  the discrete symmetries due
to the existence of the unit roots appear in association with the
transformations of parity (P) and charge conjugation (C). The form of these
transformations depends on the physical meaning of the theory as well as
on the metric signature. In physical spacetimes $M_{d+1}$, described by
Definition
\ref{phys}, these transformations  can be introduced in
a similar way as in QED \cite{CAD}. Thus the parity changes
${\mb x}\to -{\mb x}$ and $\psi({\mb x},t)\to \gamma^0 \psi(-{\mb x},t)$
leaving the Dirac equation invariant. In addition, if there are involved
only  Dirac free
fields on $M_{d+1}$, we can adopt a type of charge conjugation close to that
of QED. The gamma-matrices of $M_{d+1}$ satisfy Eqs. (\ref{ACOM}) and,
consequently, they must be either symmetric or skew-symmetric. Assuming
that there are $s+1$
symmetric gamma-matrices, namely $\gamma^0$ and $s$ matrices with
space-like indices,
$\gamma^{\hat\alpha_1},\,\gamma^{\hat\alpha_2},...,\gamma^{\hat\alpha_s}$,
we observe that the matrix
$C=(-1)^{\frac{s}{2}}\gamma^0\gamma^{\hat\alpha_1}\gamma^{\hat\alpha_2},...,
\gamma^{\hat\alpha_s}$
has the convenient properties $C^{-1}=C^{T}=(-1)^{s}\overline{C}$ and
$C\gamma^{\hat\mu}C^{-1}=(-1)^s (\gamma^{\hat\mu})^T$. With its help one can
define the charge conjugated spinor $\psi^{c}=C{\overline{\psi}}^{\,T}$ of the
spinor $\psi$ and verify that the equation of the free Dirac field remains
invariant under the charge conjugation  $\psi\to \psi^c$. This transformation
is point-independent which means that the vacuum state could be {\em stable}
(or invariant \cite{BD}) in quantum field theories based on free field
equations invariant under this type of charge conjugation. However, the above
definition of the charge conjugation does not hold in Kaluza-Klein theories
where several space dimensions are used for introducing the interaction.

\section{The Euclidean Taub-NUT space}

Involved in many modern studies in physics \cite{TNUT}, the metric of the
Euclidean Taub-NUT space is a self-dual instanton solution
with self-dual Riemann tensor \cite{Ha, AH} of the Euclidean Einstein equations
without cosmological constant. The Taub-NUT space is of interest since beside
isometries there are hidden symmetries giving rise to conserved quantities
associated to {S-K} tensors \cite{GH}. There is a conserved vector, analogous
to the Runge-Lenz vector of the Kepler type problem, whose existence is rather
surprising in view of the complexity of the equations of motion
\cite{GM, GR, FH, CFH}. These hidden symmetries are related with the existence
of four K-Y tensors  generating the {S-K} ones \cite{GR,GR1,JWH,VV1,VV2}. Three
such tensors form in fact a hypercomplex structure giving the
hyper-K\" ahlerian character of this geometry.

The quantum theory in the Euclidean Taub-NUT background has also
interesting specific features in the case of the scalar fields
\cite{FH,CFH,CV1} as well as for fields with spin half where our
results \cite{CV2,CV3,CV4} complete the previous ones
\cite{DIRAC}. In both cases there exist large algebras of
conserved observables \cite{CV5} including the components of the
angular momentum and three components of the Runge-Lenz operator
that lead to six-dimensional dynamical algebras
\cite{GH,FH,CV4,CV5}. Remarkably, the orbital angular momentum has
a special unusual form that generates new harmonics, called
$SO(3)\otimes U(1)$-harmonics \cite{CV1,CV8}, and corresponding
new spherical spinors \cite{CV2}. These will enter in the
structure of the particular solutions of the Klein-Gordon or Dirac
equations giving the discrete quantum modes of the scalar or spin
half particles. Moreover, in the Dirac theory in this geometry,
beside the new Dirac-type operators there are similar operators as
those of the scalar theory but completed with spin terms that help
us to understand the spin effects in K\" ahlerian manifolds.

\subsection{$SO(3)\otimes U(1)$ isometry transformations}

The Euclidean Taub-NUT manifold denoted from now by $M$ is a 4-dimensional
Kaluza-Klein space which has static charts with Cartesian coordinates
$x^{\mu}$ ($\mu, \nu,...=1,2,3,4$) where  $x^{i}$ ($i,j,...=1,2,3$) are
the {\em physical} Cartesian space coordinates while $x^4$ is the Cartesian
extra-coordinate. Taking $\eta=1_4$ we have to use the three-dimensional vector
notations,  $\vec{x}=(x^1,x^2,x^3)$, $r=|\vec{x}|$  and
$dl^{2}=d\vec{x}\cdot d\vec{x}$, for writing the line element
\begin{equation}\label{3(met)}
ds^2=\frac{1}{V(r)}dl^2 + V(r)[dx^4+A^{em}_i(\vec{x})dx^i]^2 \,,
\end{equation}
defined by the specific functions
\begin{equation}\label{3(tn)}
\frac{1}{V}=1+\frac{\mu}{r}\,,\quad A^{em}_{1}=
-\frac{\mu}{r}\frac{x^{2}}{r+x^{3}}\,,
\quad A^{em}_{2}=\frac{\mu}{r}\frac{x^{1}}{r+x^{3}}\,,\quad A^{em}_{3}=0\,.
\end{equation}
The real number $\mu$ is the main parameter of the theory. If one interprets
$\vec{A}^{em}$ as the vector potential (or gauge field) it
results  the magnetic field with central symmetry
\begin{equation}
\vec{B}^{em}\,=\mu\frac{\vec{x}}{r^3}\,.
\end{equation}

Other important charts are those with spherical coordinates
$(r, \theta, \varphi, \chi)$ where  $r, \theta, \varphi$, are commonly related
to the physical Cartesian ones, $x^i$, while the fourth coordinate $\chi$ is
defined by
\begin{equation}\label{34cp}
x^{4}=-\mu(\chi +\varphi)\,.
\end{equation}
In this chart the radial coordinate belongs the radial domain $D_r$ where
$r>0$ if $\mu>0$ or $r>|\mu|$ if $\mu<0$,
while the angular coordinates $\theta,\,\varphi$ cover the sphere $S^{2}$ and
$\chi\in D_{\chi}=[0,4\pi)$.  The line element in spherical coordinates is
\begin{equation}\label{3metsf}
ds^{2}=\frac{1}{V}(dr^{2}+r^{2}d\theta^{2}+
r^{2}\sin^{2}\theta\, d\varphi^{2})+\mu^{2}V(d\chi+\cos\theta\, d\varphi)^{2}\,,
\end{equation}
since
\begin{equation}
A^{em}_{r}=A^{em}_{\theta}=0\,,\quad A^{em}_{\varphi}=\mu(1-\cos\theta)\,.
\end{equation}

The Euclidean Taub-NUT space possesses a special type of isometries
which combines the space transformations with the gauge transformations of the
gauge field $\vec{A}^{em}(\vec{x})$. There are $U(1)_4$ translations
$x^4\to x^{\prime 4}=x^4 +a^4$  which leave the metric invariant if
$a^4$ is a point-independent real constant. Moreover, if one takes
$a^4=a^4(\vec{x})$ an arbitrary function of $\vec{x}$ then these
become gauge transformations preserving the form of the line element only if
one requires $\vec{A}^{em}$ to transform as
\begin{equation}
A^{em}_i(\vec{x})\to A^{\prime\, em}_i(\vec{x})=A^{em}_i(\vec{x})-\partial_i
a^4(\vec{x})\,.
\end{equation}
Thus it is obvious that $U(1)_4$ is an isometry group playing, in addition,
the role of the gauge group associated to the gauge field $\vec{A}^{em}$.
In other respects, this geometry allows an $SO(3)$ symmetry given by usual
{\em linear} rotations of the physical space coordinates,
$\vec{x}\to {\vec{x}\,}'={\frak R}\, \vec{x}$ with ${\frak R}\in SO(3)$, and
the special non-linear transformations of the fourth coordinate,
\begin{equation}\label{3ind}
{\frak R}\,:\quad x^4\to x^{\prime\, 4}=x^4+h({\frak R},\vec{x})  \,,
\end{equation}
produced by a function $h$ depending on ${\frak R}$ and $\vec{x}$ which must
satisfy
\begin{equation}\label{3cond}
h({\frak R}' {\frak R}, \vec{x})=h({\frak R}',{\frak R}\vec{x})+
h({\frak R},\vec{x})\,, \quad h(1_3,\vec{x})=0 \,,
\end{equation}
where $1_3$ is the identity of $SO(3)$. Obviously, this condition
guarantees that Eq. (\ref{3ind}) defines a representation of the
$SO(3)$ group. These transformations preserve the general form of
the line element (\ref{3(met)}) if $\vec{A}^{em}$ transforms {\em
manifestly} covariant under rotations as a vector field, up to a
gauge transformation, $V$ being a scalar. In this way one obtains
a representation of the group $SO(3)\otimes U(1)_4$ whose
transformations,
\begin{eqnarray}
\vec{x}&\to& {\vec{x}\,}'={\frak R}\, \vec{x}\label{3ec1}\\
\left[ {\frak R},a^4(\vec{x}) \right]\,:\,\qquad x^4&\to& x^{\prime\, 4}
=x^4+h({\frak R},\vec{x})+a^4(\vec{x})\label{3ec2}\\
\vec{A}^{em}(\vec{x})&\to& {\vec{A}}^{\prime\,em}({\vec{x}\,}')=
{\frak R}\left\{ \vec{A}^{em}(\vec{x})
-\vec{\partial}\,
[h({\frak R},\vec{x})+ a^4(\vec{x})]\right\}\,,\label{3ec3}
\end{eqnarray}
produced by any ${\frak R}\in SO(3)$ and real function
$a^4(\vec{x})$, {\em combine} isometries and gauge
transformations. Hereby we can separate the isometries requiring
the gauge field to remain unchanged, i.e.
$A^{\prime\,em}_i=A^{em}_i$, for point-independent parameters
$a^4$. According to Eq. (\ref{3ec3}), this condition can be
written as
\begin{equation}\label{3hA}
\vec{\partial}\, h({\frak R},\vec{x})=\vec{A}^{em}(\vec{x})
-{\frak R}^{-1}\vec{A}^{em}({\frak R}\vec{x}) \,,
\end{equation}
defining the {\em specific} function $h$ corresponding to the gauge field
$\vec{A}^{em}$.
\begin{rem}\label{isom}
The isometry transformations of the Euclidean Taub-NUT space,
$x\to x'=  \phi_{{\frak R},a^4}(x)$, are three-dimensional
rotations and  $x^4$ translations  that transform
$x=(\vec{x},x^4)$  into  $x'=({\vec{x}\,}',x^{\prime\,4})$
according to Eqs. (\ref{3ec1}) and (\ref{3ec2}) restricted to
point-independent values of $a^4$, while the function $h$ is
defined by Eq. (\ref{3hA}).
\end{rem}
These transformations form the isometry group
$I(M)=SO(3)\otimes U(1)_4$ of the Euclidean Taub-NUT space $M$, the
universal covering group of which is the external symmetry group
$S(M)=SU(2)\otimes U(1)_4$. What is remarkable here is that the
representation of $I(M)$ carried by $M$ mixes up linear transformations
with non-linear ones involving the function $h$.

The study of this type of representation is important since it
governs the transformation laws of the vectors and tensors under isometries
that are the starting points in deriving conserved quantities through the
Noether theorem. However, the properties of the isometries will
be better understood if we know the analytical expression of the function $h$.
This may be found combining the integration of the equations
(\ref{3hA}) with some algebraic properties resulted from the condition
(\ref{3cond}). The main point is to show that the transformation rule
(\ref{3ind}) of the fourth coordinate of $M$ is given by a representation
of the isometry group induced by  one of its subgroups. We recall that the
$SO(3)$ subgroup of $I(M)$ has three independent one-parameter subgroups,
$SO_i(2)$, $i=1,2,3$, each one including rotations ${\frak R}_i(\alpha)$,
of angles
$\alpha\in [0,2\pi)$ around the axis $i$. With this notation any rotation
${\frak R}\in SO(3)$ in the usual Euler parametrization reads
${\frak R}(\alpha,\beta,\gamma)={\frak R}_3(\alpha){\frak R}_2(\beta)
{\frak R}_3(\gamma)$.

We start with the observation that the special form of the gauge field
(\ref{3(tn)}) does not depend on $x^4$ and has a special form such that
all the rotations of the subgroup $SO_3(2)$ satisfy
\begin{equation}\label{3hatR}
{\frak R}_3 \vec{A}^{em}(\vec{x})= \vec{A}^{em}({\frak R}_3 \vec{x})\,,
\quad \forall \,  {\frak R}_3\in SO_3(2)\,.
\end{equation}
In these conditions we adopt
\begin{defin}
The subgroup $H(M)=SO_3(2)\otimes U(1)_4\subset I(M)$ is
the {\em little group} associated to $\vec{A}^{em}$.
\end{defin}
In what follows we are interested to exploit the existence of the little group
focusing on the rotations ${\frak R}_3\in SO_3(2)$. According to Eqs.
(\ref{3hA})
and (\ref{3hatR}) it results that
\begin{equation}\label{3hath}
h({\frak R}_3,\vec{x})\equiv \hat h({\frak R}_3)
\end{equation}
is point-independent being defined only on $SO_3(2)$. Then the condition
(\ref{3cond}) becomes
\begin{equation}\label{3abe}
\hat  h({\frak R}_3{\frak R}_3')=\hat h({\frak R}_3) +
\hat h({\frak R}_3')\,, \quad \forall
~{\frak R}_3,\,{\frak R}_3' \in SO_3(2) \,,
\end{equation}
which means that the set  $\{\hat h({\frak R}_3)\,|\, {\frak R}_3\in SO_3(2)\}$
forms a one-dimensional representation of the $SO_3(2)$ group provided
$\hat h(1_3)=0$. This representation is non-trivial
(with $\hat h({\frak R}_3)\not=0$
when ${\frak R}_3\not=1_3$) only if we assume that
\begin{equation}\label{3hRa}
\hat h[{\frak R}_3(\alpha)]= {\rm const.}\,  \alpha  \,.
\end{equation}

These properties suggests us to write the function $h$ using rotations in the
Euler parametrization and the chart with spherical
coordinates where the differential equations could be simpler since
$h({\frak R},\vec{x})=h({\frak R},\theta,\varphi)$ does not depend on the
radial coordinate $r$.
\begin{theor}\label{Teh}
In the chart with spherical coordinates the solution of the system
(\ref{3hA}) with the condition (\ref{3hRa}) reads
\begin{eqnarray}
h[{\frak R}(\alpha,\beta,\gamma), \theta,\varphi]&=&-\mu (\alpha+\gamma)
\nonumber\\
&&-2\mu\, {\rm arctan}\left[\frac{\sin(\varphi+\gamma)}
{\cot\frac{\theta}{2}\cot\frac{\beta}{2}-\cos(\varphi+\gamma)}\right]\,,
\label{3FIN}
\end{eqnarray}
for any ${\frak R}\in SO(3)$.
\end{theor}
\begin{demo}
According to Eqs. (\ref{3hath}) and (\ref{3abe}), we can write
\begin{eqnarray}
h[{\frak R}(\alpha,\beta,\gamma), \vec{x}]&=&\hat h[{\frak R}_3(\alpha)]
+h[{\frak R}_2(\beta)
{\frak R}_3(\gamma), \vec{x}] \nonumber\\
&=&h[{\frak R}_2(\beta),{\frak R}_3(\gamma)\vec{x}]+
\hat h[{\frak R}_3(\alpha+\gamma)] \,,
\end{eqnarray}
pointing out that the central problem is to integrate the system (\ref{3hA}) in
spherical coordinates for the particular case of ${\frak R}={\frak R}_2(\beta)$.
Denoting  $h[{\frak R}_2(\beta),\vec{x}]\equiv h(\beta, \theta,\varphi)$,
after a few manipulation we find that Eqs. (\ref{3hA}) are equivalent with
\begin{eqnarray}
\partial_{\theta}h(\beta,\theta,\varphi)&=&-\mu\, \frac{\sin\varphi\,\sin\beta}
{1+\cos\theta\,\cos\beta-\sin\theta\,\cos\varphi\,\sin\beta}\,,\label{3cucu1}\\
\partial_{\varphi}h(\beta,\theta,\varphi)&=&\mu\, \frac{(1-\cos\theta)(1-
\cos\beta)-\sin\theta\,\cos\varphi\,\sin\beta}
{1+\cos\theta\,\cos\beta-\sin\theta\,\cos\varphi\,\sin\beta}\,.\label{3cucu2}
\end{eqnarray}
The integration of this system  gives $h(\beta,\theta,\psi)$ up to
some arbitrary integration constants resulting from Eq.  (\ref{3hRa})
with $const.=-\mu$, as it is shown in Ref. \cite{CV8}.
\end{demo}

The last step here is to show that the function $h({\frak R},\theta,\varphi)$
can be easily found by using the technique of induced representations
in the chart with spherical coordinates $(r,\theta,\varphi,\chi)$
where $\theta$ and $\varphi$ are the Euler angles of
the rotation giving $\vec{x}={\frak R}(\theta,\varphi,0)\vec{x}_o$ from
$\vec{x}_o=(0,0,r)$. After an arbitrary rotation
${\frak R}(\alpha,\beta,\gamma)\in I(M)$ we arrive to the chart with the new
coordinates $(r, \theta', \varphi', \chi')$ among them the first three
are the spherical coordinates of the transformed vector
\begin{eqnarray}
{\vec{x}\,}'&=&{\frak R}(\varphi',\theta',0)\vec{x}_o={\frak R}(\alpha,\beta,
\gamma)\vec{x}
\nonumber\\
&=&{\frak R}(\alpha,\beta,\gamma){\frak R}(\varphi,\theta,0)
\vec{x}_o\,.\label{3transx}
\end{eqnarray}
In this context the previous theorem allows us to understand the meaning of
the transformation rule of the fourth spherical coordinate \cite{CV8}.
\begin{cor}
The spherical coordinate $\chi$ transforms under rotations according
to a representation {\em induced} by the natural representation of the
group $SO_3(2)\subset H(M)$ such that the transformed spherical
coordinates  satisfy
\begin{equation}\label{RsfRsf}
{\frak R}(\varphi',\theta',\chi')=
{\frak R}(\alpha,\beta,\gamma){\frak R}(\varphi,\theta,\chi) \,,
\end{equation}
for any ${\frak R}(\alpha,\beta,\gamma)\in SO(3)$.
\end{cor}
\begin{demo}
If we assume that this is true we find the transformation rule
of the induced representation
\begin{eqnarray}\label{indR3}
{\frak R}_3(\chi'-\chi)&=&{\frak R}^{-1}(\varphi',\theta',0)
{\frak R}(\alpha,\beta,\gamma)
{\frak R}(\varphi, \theta, 0)\nonumber\\
&=&{\frak R}^{-1}(\varphi'-\alpha,\theta',0)
{\frak R}_2(\beta){\frak R}(\varphi+\gamma, \theta, 0)\,.
\label{3final}
\end{eqnarray}
Furthermore, we express this equation in terms of $SU(2)$ transformations
corresponding to all the particular rotations involved therein and calculate
the transformed coordinates and
\begin{equation}
h({\frak R},\theta,\varphi)=-\mu (\chi'+\varphi'  -\chi-\varphi)\,,
\end{equation}
that is just the function $h$ given by Theorem \ref{Teh}.
\end{demo}\\
Thus we have shown that the transformation law of the fourth spherical
coordinate is given by the induced representation (\ref{indR3}).

\subsection{The angular momentum and related operators}

In the quantum theory on the Euclidean Taub-NUT background the basic operators
are introduced using the geometric quantization.  Now when we know the closed
form of the function $h$ we can calculate directly the components of the
Killing vectors and the generators of the natural representation of the
isometry group  using only group theoretical methods.

The generators of the natural representation of $I(M)$ are the (orbital)
differential operators (\ref{genL}). They can be calculated starting with a set
of parameters $\xi^a$  and the  functions  $\phi_{{\frak R}, a^4}\equiv
\phi_{\xi}(x)=x'(x,\xi)$
that give the Killing vectors $k_a$ according to Eq. (\ref{ka}). In the case of
our isometry group we take the first three parameters, $\xi^i$, the Cayley-Klein
parameters of the rotations ${\frak R}(\vec{\xi})\in SO(3)$  and we denote
$\xi^4=a^4$.
Then we find that the generator of the $U(1)_4$ translations is the fourth
component of the momentum operator, $P_4=-i\partial_4$ since $k_{(4)}^i=0$
and $k_{(4)}^4=1$.
\begin{theor}\label{tLLL}
The $SO(3)$ generators of the natural representation are the components of
the orbital angular momentum operator:
\begin{eqnarray}
L_1&=&-i(x^2\partial_3-x^3\partial_2)+i\mu\,\frac{x^1}{r+x^3}\partial_4\,,
\nonumber\\
L_2&=&-i(x^3\partial_1-x^1\partial_3)+i\mu\,\frac{x^2}{r+x^3}\partial_4\,,
\label{LLL}\\
L_3&=&-i(x^1\partial_2-x^2\partial_1)+i\mu\,\partial_4\,.\nonumber
\end{eqnarray}
\end{theor}
\begin{demo}
The first terms of the angular momentum correspond to the usual linear
transformation  ${\vec{x}\,}'={\frak R}\,\vec{x}$ giving the components
$k_{(j)}^i=\varepsilon_{ijk}x^k$ of the first three Killing vectors in the
basis of the Cartesian natural frame. The contributions due to $h$ have to be
calculated according to Eqs. (\ref{ka}) and (\ref{3ind}) starting with
\begin{equation}
\left.{\frac{\partial}{\partial\beta} h(\beta,\theta,\phi)}\right|_{\beta=0}=
-\mu \frac{\sin\theta\sin\phi}{1+\cos\theta}\,.
\end{equation}
Then, denoting $\xi^2=\beta$ for $\alpha=\gamma=0$ and $\xi^3=\alpha$ for
$\beta=\gamma=0$ and using a simple rotation of angle $\pi/2$ around the third
axis we find
\begin{equation}\label{hxi}
\left.k_{(1,2)}^4=\frac{\partial h({\frak R},\vec{x})}{\partial\xi^{1,2}}
\right|_{\xi=0}=-
\mu\frac{x^{1,2}}{r+x^3}\,,\quad
\left.k_{(3)}^4=\frac{\partial h({\frak R},\vec{x})}{\partial\xi^3}
\right|_{\xi=0}=- \mu\,.
\end{equation}
Finally from Eq.  (\ref{genL}) we obtain the operators (\ref{LLL}) corresponding
to the Caley-Klein parameters $\xi^i$.
\end{demo}

In the Cartesian charts one can choose a diagonal gauge suitable for physical
interpretation. This is given by the gauge fields $\hat e^{\hat\alpha}$ and
$e_{\hat\alpha}$ having the non-vanishing components \cite{BuCh}
\begin{eqnarray}
&&\hat e^{i}_{j}=\frac{1}{\sqrt{V}}\,\delta_{ij}\,, \quad \hat
e^{4}_{i}=\sqrt{V}A^{em}_{i}\,, \quad
\hat e^{4}_{4}=\sqrt{V}\,, \nonumber\\
&&e^{i}_{j}=\sqrt{V}\delta_{ij}\,,\quad
e^{4}_{i}=-\sqrt{V}A^{em}_{i}\,,\quad
e^{4}_{4}=\frac{1}{\sqrt{V}} \,,\label{ehe}
\end{eqnarray}
in the natural Cartesian frame. This gauge fixing defines local orthogonal
frames where the  Killing vectors $k_{(i)}$ yield by the previous theorem have
the components
\begin{equation}
k_{(j)}^i=\frac{1}{\sqrt{V}}\,\varepsilon_{ijk}x^k \,,\quad k_{(i)}^4= -\mu
\frac{x^i}{r}\sqrt{V}\,,
\end{equation}
which covariantly transform under linear $SO(3)$ rotations. This
behavior is rather surprising in view of the fact that the fourth coordinate
transforms under rotations according to the induced representation (\ref{3ec2}).
The explanation of this remarkable phenomenon is given by
\begin{theor}\label{t23}
In the local frames of the Euclidean Taub-NUT space defined by the gauge
(\ref{ehe}) the representation $vect[S(M)]$ of the group
$S(M)=SU(2)\otimes U(1)_4$ is just the {\em linear fundamental}
representation of the group $I(M)=SO(3)\otimes U(1)_4$.
\end{theor}
\begin{demo}
The representation of the $U(1)_4$ translations is anyway the usual one such
that the problem here is to calculate the behavior under rotations.
Starting with the isometries $\phi_{\vec{\xi}}=\phi_{{\frak R}(\vec{\xi}),
a^4=0}$
and $(A_{\vec{\xi}},\, \phi_{\vec{\xi}})\in
SU(2)\subset S(M)$ and using Eqs. (\ref{3hA}) and (\ref{ehe}), we find that
the matrices defined by Eq. (\ref{Axx}) are point-independent having the
non-vanishing matrix elements
$\Lambda^{i\,\cdot}_{\cdot\,j}[A_{\vec{\xi}}(x)]={\frak R}_{ij}(\vec{\xi})$
and
$\Lambda^{4\,\cdot}_{\cdot\,4}[A_{\vec{\xi}}(x)]=1$. Thus we see that in
the local frames given by this
gauge the fourth component of any vector behaves as a scalar
under rotations.
\end{demo}\\
We specify that here it is crucial to consider the group $S(M)$ instead of
$I(M)$ since only the transformations of the external symmetry group preserve
this gauge showing off the $SO(3)$ symmetry as a {\em global} one.

In this context one can correctly define the three-dimensional
physical momentum $\vec{P}$ whose  components in the above defined local
frames are
\begin{equation}
{P}_i= -i\frac{1}{\sqrt{V}}\,e_i^{\mu}\partial_{\mu}=
-i(\partial_i-{A}^{em}_i\partial_{4}) \,,
\end{equation}
obeying the following commutation relations
\begin{equation}
[P_{i},P_{j}]=i\varepsilon_{ijk}B_{k}^{em}P_{4}\,,\quad [P_{i},P_{4}]=0\,,\quad
[L_i, \,P_j]=i\varepsilon_{ijk}P_{k}\,,
\end{equation}
which indicate that $\vec{P}$  behaves as a vector under rotations. In
addition, the angular momentum can be written in covariant form as
\begin{equation}\label{3(angmom)}
\vec{L}\,=\,\vec{x}\times\vec{P}-\mu\frac{\vec{x}}{r}P_4\,.
\end{equation}
The scalar quantum mechanics in the Taub-NUT geometry \cite{CFH} is based on
the Schr\" odinger or Klein-Gordon equations  involving the  static operator
\begin{equation}
\Delta=-\nabla_{\mu}g^{\mu\nu}\nabla_{\nu}
=V{\vec{P}\,}^{2}+\frac{1}{V}{P_{4}}^{2} \,,
\end{equation}
which is either proportional with the Hamiltonian operator of the Schr\" odinger
theory  or represents the static part of the Klein-Gordon operator \cite{CV5}.
In both cases we are interested to find operators commuting with $\Delta$ since
these give rise to the conserved quantities with physical significance.

The Euclidean Taub-NUT space is a hyper-K\" ahler manifold possessing a triplet
of real unit roots (i.e., a hypercomplex structure),
${\mb f}=\{ f^{(1)},\,f^{(2)},\, f^{(3)}\}$, defined as
\begin{equation}\label{trip}
f^{(i)}= f^{(i)}_{\hat\alpha \hat\beta} \hat e^{\hat\alpha}\land
\hat e^{\hat\beta}= 2\hat e^i\land \hat
e^4-\varepsilon_{ijk} \hat e^j\land \hat e^k \,,
\end{equation}
where the 1-forms $ \hat e^{\hat \alpha}=\hat e^{\hat
\alpha}_{\mu}dx^{\mu}$ are defined by the tetrads (\ref{ehe}). In
addition, there exists a fourth K-Y tensor,
\begin{equation}\label{3fy}
f^Y = f^Y_{\hat\alpha \hat\beta}\hat e^{\hat\alpha}\land \hat
e^{\hat\beta}=
 \frac{x^i}{r}f^{(i)} +\frac{2 x^i}{\mu V}\varepsilon_{ijk}
\,{\hat e}^j\land \,{\hat e}^k\,,
\end{equation}
which is not covariantly constant. The presence of $f^Y$ is due to the
existence of the hidden symmetries of the Euclidean Taub-NUT geometry which are
encapsulated in three non-trivial {S-K} tensors  and interpreted as the
components of the so-called Runge-Lenz vector of the Euclidean Taub-NUT problem.
These {S-K} tensors  can be expressed as symmetrized products of K-Y tensors
\cite{VV2,VV1},
\begin{equation}\label{3kff}
k_{(i)\mu\nu} = {\mu\over 4}(f^Y_{~\mu\lambda}f^{(i)\lambda}_{~~\nu} +
f^Y_{~\nu\lambda}f^{(i)\lambda}_{~~\mu})+{1\over 2\mu}(k_{(4)\mu}k_{(i)\nu} +
k_{(4)\nu}k_{(i)\mu}) \,,
\end{equation}
and with their help one defines the vector operator
\begin{equation}
\vec{K}\,=-\frac{1}{2}\nabla_{\mu}\vec{k}^{\mu\nu}\nabla_{\nu}=
\frac{1}{2}\left(\vec{P}\times \vec{L}-
\vec{L}\times \vec{P}\right)-
\mu \frac{\vec{x}}{r}
\left(\frac{1}{2}\Delta-P_4^{2}\right)\,,
\end{equation}
which play the same role as the Runge-Lenz vector operator in the usual quantum
mechanical Kepler problem \cite{CFH}. This transforms as a vector under
the rotations of $vect[S(M)]$ such that one can write the following complete
system of commutation relations
\begin{eqnarray}
\left[ L_{i},\, L_{j} \right]& =& i \varepsilon_{ijk}\,L_{k}\,,\nonumber\\
\left[ L_{i},\, K_{j} \right]& =& i \varepsilon_{ijk}\,K_{k}\,,\label{LLKK} \\
\left[ K_{i},\, K_{j} \right]& =& i \varepsilon_{ijk}\,L_{k}B^2\,,\nonumber
\end{eqnarray}
where $B^2 ={P_4}^2-\Delta$. The operators $L_i$ and $K_i$ commute
with $B$ since they commute with $\Delta$ and $P_4$. Moreover, it
is known \cite{FH} that the operators
\begin{eqnarray}
&&C_1=\vec{L}^2B^2+\vec{K}^2=\mu^2P_4^2B^2+\frac{\mu^2}{4}
\left(B^2+{P_4}^2\right)^2-B^2 \label{C1C2}\nonumber\\
&&C_2=\vec{L}\cdot
\vec{K}=-\frac{\mu^2}{2}P_4(B^2+P_4^2)\,,\label{C1C21}
\end{eqnarray}
play the role of Casimir operators for the open algebra (\ref{LLKK}).  With
their help we can define the new Casimir operators,
\begin{equation}\label{CBCB}
C^{\pm}=C_1\pm 2B C_2 +B^2=\frac{\mu^2}{4}(P_4\mp B)^4=B^2(N\pm \mu P_4)^2\,.
\end{equation}
where $N$ is the operator whose eigenvalues are just the values of the
principal quantum number of the discrete energy spectra \cite{ILA}. We recall
that $K^i$ commute with $\Delta$ grace to the factorization  (\ref{3kff}) which
eliminates the quantum anomaly.

When one goes to the chart with spherical coordinate a special
attention must be paid to the meaning of Eq. (\ref{34cp}) which
shows that the fourth spherical coordinate $\chi$ is {\em
translated} with the angular coordinate $\varphi$ \cite{CV8}. This
translation is rather unusual being performed by the unitary
operator
\begin{equation}
U(\varphi)=e^{i\varphi P_{\chi}} \,,
\end{equation}
where $P_{\chi}=-\mu P_4=-i\partial_{\chi}$ replaces the Cartesian operator $P_4$.
Since the differential and local operators are defined often in the coordinate
representation of a given chart, we must take care when we change the chart.
\begin{rem}
The coordinate representation of the  Cartesian chart must be transformed into
the {\em equivalent} coordinate representation of the spherical chart
transforming each operator $X$ defined in the Cartesian chart
into the equivalent operator
\begin{equation}
X^{sph}=U(\varphi)XU^{\dagger}(\varphi)
\end{equation}
of the spherical chart.
\end{rem}
Thus, for example, the components of the orbital angular momentum
(\ref{LLL}) in the spherical chart and canonical basis
(with $L_{\pm}=L_{1}\pm i L_{2}$) become
\begin{eqnarray}
L_{3}^{sph}&=&-i\partial_{\varphi}\,,\label{3tl1}\\
L_{\pm}^{sph}&=&e^{\pm i\varphi}\left[\pm\,\partial_{\theta}+
i\left(\cot\theta\,\partial_{\varphi}-
\frac{1}{\sin\theta}\,\partial_{\chi}\right)\right]\,.\label{3tl2}
\end{eqnarray}
Many other operators including the Runge-Lenz vector will take new forms in
the representation of the spherical coordinates but preserving their
commutation relations. However, in current calculations when we do not work
simultaneously with both these representations of the operator algebra we drop
out the superscript above, denoting the equivalent operators with the same
symbol.

\subsection{Scalar quantum modes and dynamical algebras}

The special form of the $SO(3)$ generators (\ref{3tl1}) and
(\ref{3tl2}) leads to new spherical harmonics that permit to
separate the spherical variables in the static Klein-Gordon equation,
$\Delta {\frak U}_E=E^2 {\frak U}_E$, giving the eigenfunctions of the
operator $\Delta$ which represents the squared Hamiltonian operator of the
relativistic theory of the scalar field without the explicit mass term.
\begin{defin}
The {\em central regular} modes of the scalar field on $M$ are given by
the eigenfunctions of the complete set of commuting operators
$\{ \Delta,\, P_{\chi},\,{\vec{{L}}\,}^{2},\,{L}_{3}\}$.
\end{defin}
The corresponding eigenvalues  $E^2,\,q,\, l(l+1)$ and $m$ determine the
eigenfunctions
\begin{equation}
{\frak U}_{E,l,m}^q(r,\varphi, \theta,\chi)=\frac{1}{r}{\frak f}_{E,q,l}(r)
Y^{q}_{l,m}(\theta, \varphi, \chi)  \,,
\end{equation}
which have separated variables.
\begin{defin}
We call $SO(3)\otimes U(1)$ spherical harmonics the  functions $Y^{q}_{l,m}$
defined on the compact domain  $S^{2}\times D_{\chi}$, which satisfy the
eigenvalue problems
\begin{eqnarray}
{\vec{{L}}\,}^{2}Y_{l,m}^{q}&=&l(l+1)\,Y_{l,m}^{q}\,,\label{3(lp)}\\
{L}_{3}Y_{l,m}^{q}&=&m\,Y_{l,m}^{q}\,,\label{3(l3)}\\
P_{\chi}Y_{l,m}^{q}&=&q\,Y_{l,m}^{q}\,,\label{3(l0)}
\end{eqnarray}
and the orthonormalization condition
\begin{eqnarray}
&&\left<Y_{l,m}^{q},Y_{l',m'}^{q'}\right>=
\int_{S^2}d(\cos\theta)d\varphi\,\int_{0}^{4\pi}d\chi\,
{Y_{l,m}^{q}(\theta, \varphi, \chi)}^{*}\,
Y_{l',m'}^{q'}(\theta, \varphi, \chi)\nonumber\\
&&~~~~~~~~~~~~~~~~~~~~~~~~~~~~~=\delta_{l,l'}\delta_{m,m'}
\delta_{q,q'}\,.\label{3(spy)}
\end{eqnarray}
\end{defin}
These functions form a basis of the Hilbert space of square integrable
functions on $S^2\times D_{\chi}$ since the set of commuting operators
$\{{\vec{{L}}\,}^{2},{L}_{3},P_{\chi}\}$ is complete in this space.

In Ref. \cite{CV1} we pointed out that the $SO(3)\otimes U(1)$-harmonics,
are new spherical harmonics. The usual boundary conditions on
$S^{2}\times D_{\chi}$ require $l$ and $m$ to be integer numbers and
$q=0,\pm 1/2,\pm 1,...$ \cite{CFH} but, in general, $q$ can be any real number.
Solving  Eqs. (\ref{3(l3)}) and (\ref{3(l0)}) we get
\begin{equation}\label{3(Y)}
Y_{l,m}^{q}(\theta,\varphi,\chi)=\frac{1}{4\pi}\, \Theta_{l,m}^{q}(\cos\theta)
e^{im\varphi}e^{iq\chi} \,,
\end{equation}
where the function $\Theta_{l,m}^{q}$ must satisfy Eq. (\ref{3(lp)})
and the normalization condition
\begin{equation}\label{3(np)}
\int_{-1}^{1}d(\cos\theta)\left|\Theta_{l,m}^{q}(\cos\theta)\right|^{2}=2\,,
\end{equation}
resulted from Eq. (\ref{3(spy)}). This problem has solutions for all the values
of the quantum numbers obeying $|q|-1<|m|\le l$ when one founds \cite{CV1}
\begin{eqnarray}
\Theta_{l,m}^{q}(\cos\theta)&=&\frac{\sqrt{2l+1}}{2^{|m|}}\left[\frac{
(l-|m|)!\,(l+|m|)!}{
\Gamma(l-q+1)\Gamma(l+q+1)}\right]^{\frac{1}{2}}\label{3(fin)}\label{3(Tet)}\\
&&\times \left(1-\cos\theta\right)^{\frac{|m|-q}{2}}
\left(1+\cos\theta\right)^{\frac{|m|+q}{2}}\,
P_{l-|m|}^{(|m|-q,\,|m|+q)}(\cos\theta)\,.\nonumber
\end{eqnarray}
For $m=|m|$ the $SO(3)\otimes U(1)$ harmonics are given by (\ref{3(Y)})
and (\ref{3(fin)}) while for $m<0$ we have to use the obvious formula
\begin{equation}
Y_{l,-m}^{q}=(-1)^{m}\left(Y_{l,m}^{-q}\right)^{*}\,.
\end{equation}
When the boundary conditions allow half-integer quantum numbers $l$ and $m$
then we say that the functions defined by Eqs. (\ref{3(Y)}) and (\ref{3(fin)})
(up to a suitable factor) represent $SU(2)\otimes U(1)$ harmonics.
Thus we have obtained a non-trivial generalization of the  spherical
harmonics of the same kind as the spin-weighted spherical harmonics \cite{NP}
or those studied in \cite{HARM}. Indeed, if $l$, $m$ and $q=m'$ are either
integer or half-integer numbers then we have
\begin{equation}
Y_{l,m}^{m'}(\theta,\varphi,\chi)=\frac{\sqrt{2l+1}}{4\pi}D^{l}_{m,m'}
(\varphi,\theta, \chi) \,,
\end{equation}
where $D_{m,m'}^{l}$ are the matrix elements of the irreducible representation
of weight $l$ of the $SU(2)$ group corresponding to the rotation of Euler
angles $(\varphi, \theta, \chi)$. What is new here is that our harmonics
are defined for any real number $q$. For this reason these are useful
in solving some actual physical problems \cite{Gosh}. Notice that similar
spherical harmonics were used recently in \cite{Mard} under the name of
ring-shaped harmonics.

Turning back to the scalar modes on $M$ and using the identity
\begin{equation}
\vec{P}^2=-\partial_r^2-\frac{2}{r}\partial_r +\frac{1}{r^2}\vec{L}^2-
\frac{1}{r^2}P_{\chi}^2 \,,
\end{equation}
we find that the radial wave functions ${\frak f}$ satisfy the radial equation
\begin{equation}\label{(kgrad)}
\left[-\frac{d^2}{dr^2}+\frac{l(l+1)}{r^2}-\frac{a}{r}\,\right]
{\frak f}_{E,q,l}(r)=-b^2\, {\frak f}_{E,q,l}(r)\,,
\end{equation}
whose parameters $a=\mu\left[E^{2}-2{\hat q^2}\right]$ and
$b^2={\hat q}^2-E^2$ depend on the eigenvalues of $P_4$ denoted by
$\hat q=-q/{\mu}$. Notice that here $b^2$ is the eigenvalue of the operator
$B^2$ for given $E$ and $q$. Looking for the particular solutions of the radial
equation on the non-compact domain $D_r$ we have to select either square
integrable functions with respect to the radial scalar product \cite{CV1}
\begin{equation}\label{(scprod1)}
\left<{\frak f}_{E,q,l},{\frak f}_{E',q,l}\right>=\int_{D_{r}}dr \,\left|1+
\frac{\mu}{r}
\right| {\frak f}_{E,q,l}(r)^{*} {\frak f}_{E',q,l}(r)\,,
\end{equation}
or solutions that behave as tempered distributions on $D_r$. One obtains thus a
Kepler-like problem similar to the well-known one of the non-relativistic
quantum mechanics, the only differences being the parametrization and the form
of the scalar product. The particular solution of Eq.  (\ref{(kgrad)}) can be
written in terms of the confluent hypergeometric function as
\begin{equation}
{\frak f}_{E,q,l}(r)=N_{E,q,l}\, r^{l+1}e^{-2br}F(s,2l+2,2br)\,,
\quad  s=l+1-\frac{a}{2b}\,,
\end{equation}
where $N_{E,q,l}$ is the normalization constant. It is easy to show that for
$\mu>0$ the
radial wave functions are not square integrable and, therefore, the energy
spectrum is continuous in the domain $E\ge |\hat q|$. The case of $\mu<0$ is
most interesting since beside the mentioned continuous energy spectrum  there
is a discrete spectrum for $a>0$ and $b^2>0$ when $0<E<|\hat q|$. Indeed, then
the quantization condition $s=-n_{r}, \,n_{r}=0,1,2,...$ gives the usual
formula  $a=2n |b|$ where $n=n_{r}+l+1$ is the {\em principal} quantum number.
Hereby one finds the energy levels
\begin{equation}\label{(een)}
{E_{n}}^2=\frac{2}{\mu^2}\left[n\sqrt{n^{2}-q^{2}}-(n^{2}-q^{2})\right] \,,
\end{equation}
for all $n> |q|>0$ \cite{CFH}. This spectrum is countable and
finite since $\lim_{n\to \infty} E_{n}=\hat q$. We observe that
these levels are degenerate depending on the quantum numbers $l$
and $m$ which satisfy $|q|-1<l\le n-1$ and $|q|-1<m\le l$.
\begin{rem}
The condition $E>0$ guarantees that there are no zero modes and,
therefore, the operator $\Delta$ is invertible.
\end{rem}

Another problem is to define the generators of the dynamical algebras
corresponding to different spectral domains. This can be done as
in the case of the standard quantum Kepler problem {\em rescaling}
the operators $K_i$ in order to close up the commutation relations (\ref{LLKK}).
For given values of $E$ and $\hat q$ the rescaled operators
\begin{equation}\label{Ri}
{K}^{re}_{i}=\left\{
\begin{array}{lllll}
{ B}^{-1}{ K}_{i}&{\rm for}& \mu<0&{\rm and}&E<|\hat q|\\
{ K}_{i}&{\rm for}&{\rm any}~ \mu&{\rm and}&E=|\hat q|\\
\pm i{ B}^{-1}{ K}_{i}&{\rm for}&{\rm any}~ \mu &{\rm and}& E>|\hat q|
\end{array}\right.
\end{equation}
and $L_i$ ($i=1,2,3$) generate either a  representation of the
$o(4)$ algebra for $\mu<0$ and discrete energy spectrum  or a
representation of the $o(3,1)$ algebra for the continuous spectrum
in the domain $E>|\hat q|$. A special case is that of the
dynamical algebra  $e(3)$ which corresponds only to the ground
energy of the continuous spectrum, $E=|\hat q|$. In general, the
dynamical algebras may help us to analyze the structure of the
spaces of the eigenvectors of $\Delta$ seen as carrier spaces of
several irreducible representations of the dynamical algebra
\cite{FH}.

Here it is worth pointing out that the concrete representations of
these algebras are rather unusual. For example, in the case of the
discrete energy spectrum the dynamical algebra $so(4)=su(2)\oplus
su(2)$ has the canonical generators $L_i$ and $K_i^{re}=K_i
B^{-1}$. With their help one can construct the operators
$A^i_{\pm}=\frac{1}{2}(L_i\pm K_i^{re})$ representing the
generators of the pair of $su(2)$ subalgebras of $so(4)$
\cite{FH}. The Casimir operators (\ref{C1C2}) and (\ref{C1C21})
allow one to calculate the eigenvalues $a_+(a_+ +1)$ and $a_-(a_-
+1)$ of the operators ${\vec{A}_{+}}^2$ and respectively
${\vec{A}_{-}}^2$. However, the surprise is that one obtains the
$su(2)$ weights $a_{\pm}=\frac{1}{2}(n\pm q-1)$ which are no
longer integer or half-integer numbers as in the usual theory of
linear representations. Thus, for a given $q$, the eigenspace of
the energy level $E_n$ appears as the carrier space of a
representation of the $so(4)$ algebra having the non-standard
$su(2)$ weights $(\frac{n+q-1}{2},\frac{n-q-1}{2})$. For the
dynamical algebras $so(3,1)$ or $e(3)$ the problem is more
complicated since their unitary representations are no longer
finite-dimensional. In our opinion, these arguments indicate that
the problem of defining the irreducible representations of the
dynamical algebras remains open.

Finally we note that there are other interesting scalar modes that do not
present  central symmetry. Thus we have show that the
complete set of commuting operators $\{ \Delta, P_{\chi}, K_3, L_3 \}$ gives
{\em axial} modes that can be completely solved in parabolic coordinates
\cite{CV1}. The discrete axial modes are determined by the set of
eigenvalues ${E_n}^2, q,\kappa$ and $m$ that depend on the integer numbers
$n_1,\, n_2=0,1,2,...$ and $m$ (obeying $|m|>|q|-1$) which give the principal
quantum number $n=n_1+n_2+|m|+1$ and the eigenvalue $\kappa=|b|(n_2-n_1-q)$
of the operator $K_3$.

\subsection{Conserved  Dirac and Pauli operators}

For building the Dirac theory we consider the Cartesian chart, the usual
four-dimensional space of the Dirac spinors, $\Psi$, and the Dirac matrices
$\gamma^{\hat\alpha}$, that  satisfy $\{ \gamma^{\hat\alpha},\,
\gamma^{\hat\beta} \} =2\delta_{\hat\alpha \hat\beta}$,
in the following  representation
\begin{equation}\label{3(gammai)}
\gamma^i = -i
\left(
\begin{array}{cc}
0&\sigma_i\\
-\sigma_i&0
\end{array}\right)\,,\quad
\gamma^4 =
\left(
\begin{array}{cc}
0&{\mb 1}_2\\
{\mb 1}_2&0
\end{array}\right)\,,
\end{equation}
where $\sigma_i$ are the Pauli matrices. In addition we take $\gamma={\mb 1}$
and denote by $\gamma^5=\gamma^1\gamma^2\gamma^3\gamma^4={\rm diag}({\mb 1}_2,
-{\mb 1}_2)$ the chiral matrix ($\gamma^5=\gamma^{ch}=\digamma$) which is
just the matrix $\gamma^0$ of the Kaluza-Klein theory explicitly depending on
time  \cite{CV2}. In this representation adopted here all the gamma-matrices
are self-adjoint with respect to the Dirac adjoint ($\overline{X}=X^{+}$) and
the Euclidean metric is of positive signature for a pure space-like manifold.
Therefore, it is convenient to change some phase factors of the operators we
define here as indicated in Remark \ref{rem1}.

Let us start with the {\it standard} Dirac operator
without explicit mass term  defined now as
$D =\gamma^{\alpha}\nabla_{\alpha}$  \cite{CV2,CV3}.
This is related to the Hamiltonian operator  \cite{CV2,CV4}
\begin{equation}\label{HH}
H =\gamma^5 D
=\left(
\begin{array}{cc}
0&{\mb \alpha}^{*}\\
{\mb \alpha}&0
\end{array}\right)
=\left(
\begin{array}{cc}
0&V\pi^{*}\frac{1}{\sqrt{V}}\\
\sqrt{V}\pi&0
\end{array}\right)\,,
\end{equation}
that, after a little calculation, can be  expressed in terms of
$\pi = {\sigma}_{P}-iV^{-1}P_{4}$
and $\pi^* = {\sigma}_{P}+iV^{-1}P_{4}$ depending on
$\sigma_P=\vec{\sigma}\cdot\vec{P}$. These  operators obey
\begin{equation}\label{daapipi}
\Delta= {\mb \alpha}^{*}{\mb \alpha}=V\pi^{*}\pi\,.
\end{equation}
We specify that here the star superscript is a mere notation that does not
coincide with the Hermitian conjugation at the level of the Pauli operators
which enter in the structure of  the basic Dirac operators. The Hamiltonian
operator that is the central piece of the Dirac theory has remarkable
properties.
\begin{theor}\label{invers}
The spectrum of the Hamiltonian operator coincides with the energy spectrum of
the operator $\Delta$.
\end{theor}
\begin{demo}
From Eqs.  (\ref{HH}) and (\ref{daapipi}) we see that the eigenvalue problem
$H{\psi}_E=E{\psi}_E$ is solved by the Dirac spinors
${\psi}_E=({u}_E,\,E^{-1}{\mb \alpha}{u}_E)^T$ if the first Pauli spinor
${u}_E$ satisfies $\Delta {u}_E=E^2 {u}_E$. Consequently, $E$
is an eigenvalue of $H$ only if $E^2$ is an eigenvalue of the
operator $\Delta$.
\end{demo}\\
The main consequence is that the operator $H$ is {\em invertible} since there
are no zero modes.

The  operators we intend to study here are operators of the
Dirac theory which {\em commute} with the Hamiltonian operator (\ref{HH}).
\begin{defin}
We say that the Dirac operators which commute with $H$ are conserved.
\end{defin}
We denote by
${\mb D}=\{ X\,|\, [X,H]=0\}$ the algebra of the conserved Dirac  operators
observing that they can be related to Pauli operators commuting with
$\Delta$ which form the algebra ${\mb P}=\{\hat X\,|\, [\hat X, \Delta]=0 \}$
where we include the orbital operators  having this property. All these
operators are considered as conserved operators in the sense of the
Klein-Gordon theory. Notice that the Pauli operators are interesting here since
they are involved in different versions of the dyon theory  \cite{DYON}
which may be compared to our approach.
\begin{theor}
The  Pauli blocks, $\hat X^{(ab)}$  ($a,b =1,2$), of any
operator
\begin{equation}\label{Xab}
X=\left(
\begin{array}{cc}
\hat X^{(11)}&\hat X^{(12)}\\
\hat X^{(21)}&\hat X^{(22)}
\end{array}\right)\,\in {\mb D} \,,
\end{equation}
must satisfy the conditions
$\hat X^{(21)}={\mb \alpha}\hat X^{(12)}{\mb \alpha}\Delta^{-1}$
and $\hat X^{(11)},\,\hat X^{(12)}{\mb \alpha},\,
{\mb \alpha}^{*}\hat X^{(21)}\in {\mb P}$.
\end{theor}
\begin{demo}
From $[X,H]=0$ it results the equivalent system
\begin{eqnarray}
&&\hat X^{(22)}{\mb \alpha}= {\mb \alpha}\hat X^{(11)}\,,\quad
{\mb \alpha}^{*}\hat X^{(22)}=
\hat X^{(11)}{\mb \alpha}^{*}\label{X1}\,, \\
&&\hat X^{(12)}{\mb \alpha}= {\mb \alpha}^{*} \hat X^{(21)}
\,,\quad
{\mb \alpha}\hat X^{(12)}=
\hat X^{(21)}{\mb \alpha}^{*} \,, \label{X2}
\end{eqnarray}
giving $\hat X^{(21)}$ and $[\hat X^{(11)},\Delta]=
[\hat X^{(12)}{\mb \alpha},\Delta]=
[{\mb \alpha}^{*}\hat X^{(21)},\Delta]=0$\,.
\end{demo}\\
We observe that  possible solutions of Eqs. (\ref{X1}) and (\ref{X2}) are
the diagonal operators
\begin{equation}
{\cal D}(\hat X)=\left(
\begin{array}{cc}
\hat X&0\\
0&{\mb \alpha}\hat X\Delta^{-1}{\mb \alpha}^{*}
\end{array}\right) \,,
\end{equation}
where $\hat X\in {\mb P}$. Particularly, for $\hat X={\mb 1}_2$ we obtain the
projection operator
\begin{equation}\label{id}
I ={\cal D}({\mb 1}_2)=\left(
\begin{array}{cc}
{\mb 1}_2&0\\
0&{\mb \alpha}\Delta^{-1}{\mb \alpha}^{*}
\end{array}\right) \,,
\end{equation}
on the space $\Psi_D = I \Psi $ in which the eigenspinors $\psi_E$ of $H$ form
a (generalized) basis. This projection operator splits the algebra
${\mb D}={\mb D}_{0}\oplus {\mb D}_{1}$ in two subspaces of the projections
$XI\in {\mb D}_{0}$ and $X({\mb 1}-I)\in {\mb D}_{1}$ of all $X\in {\mb D}$.
\begin{theor}
The subalgebra ${\mb D}_1$ is an ideal in ${\mb D}$.
\end{theor}
\begin{demo}
According to Eqs. (\ref{X1}) and (\ref{X2}) we find that the projections of
two arbitrary operators $X,\,Y\in {\mb D}$ satisfy
$(XI)(YI)=(XY)I$ and $[X({\mb 1}-I)](YI)=0$ which lead to the conclusion that
${\mb D}_{0}$ is a subalgebra while ${\mb D}_{1}$ is even an ideal in
${\mb D}$. Obviously, $I$ is the identity operator of ${\mb D}_{0}$.
\end{demo}

In  \cite{CV2} we introduced  the ${\cal Q}$-operators
defined as
\begin{equation}
{\cal Q}(\hat X)=\left\{ H\,,\,\left(
\begin{array}{cc}
\hat X&0\\
0&0
\end{array}
\right) \right\}
=\left(
\begin{array}{cc}
0&\hat X{\mb \alpha}^*\\
{\mb \alpha}\hat X&0
\end{array}
\right)\,,
\end{equation}
where $\hat X$ may be any Pauli operator. However, if $\hat
X\in {\mb P}$ then ${\cal Q}(\hat X)\in {\mb D}_{0}$ since
$[{\cal Q}(\hat X),H]=0$ and ${\cal Q}(\hat X)I= {\cal Q}(\hat X)$.
If $\hat X={\mb 1}_2$ we obtain just the Hamiltonian operator
$H={\cal Q}({\mb 1}_2)\in {\mb D}_0$. Consequently, the inverse of $H$ with
respect
to $I$ can be represented as $H^{-1}={\cal Q}(\Delta^{-1})$.
The mappings ${\cal D} : {\mb P}\to {\mb D}_{0}$ and
${\cal Q} : {\mb P}\to {\mb D}_{0}$ are linear and have the following
algebraic properties
\begin{eqnarray}
{\cal D}(\hat X){\cal D}(\hat Y)&=&{\cal D}(\hat X \hat Y)\,,\\
{\cal Q}(\hat X){\cal Q}(\hat Y)&=&{\cal D}(\hat X \hat Y\Delta)\,,\\
{\cal D}(\hat X){\cal Q}(\hat Y)&=&
{\cal Q}(\hat X){\cal D}(\hat Y)={\cal Q}(\hat X \hat Y)\,,
\end{eqnarray}
for any  $\hat X,\, \hat Y \in {\mb P}$. Moreover, the relations
$[\gamma^5,\, {\cal D}(\hat X)]=0$ and
$\{\gamma^5,\, {\cal Q}(\hat X)\}=0$
show us that, according to the usual terminology  \cite{TH},
${\cal D}$ and $\gamma^5{\cal D}$ are {\em even} Dirac operators while
${\cal Q}$ and $\gamma^5{\cal Q}$ are {\em odd} ones. We note that there
are many other odd or even operators which do not have such forms.

Since $I$ is the projection operator on the space of the Dirac
spinors $\Psi_D$ we adopt
\begin{defin}
We say that the projection $IXI$ of any Dirac operator $X$ represents the
{\em physical part} of $X$.
\end{defin}
One can convince ourselves that if $X\in {\mb D}$ then
\begin{equation}\label{WW}
IXI\equiv XI={\cal D}(\hat X^{(11)})+{\cal Q}(\hat X^{(12)}{\mb \alpha}
\Delta^{-1})\,,
\end{equation}
which means that all the operators of ${\mb D}_{0}$ can be written in terms of
${\cal D}$ or ${\cal Q}$-operators. Thus the action of $X$ reduces to
that of the Pauli operators involved in (\ref{WW}) allowing us to rewrite
the problems of the Dirac theory in terms  of Pauli operators
 \cite{CV3,CV4}. Indeed, it is easy to show that the action of any  operator
$X\in {\mb D}$ on $\psi_{E}\in \Psi_D$  is
\begin{equation}\label{cue}
 X\psi_{E}=XI\psi_{E}=\left(
\begin{array}{r}
\hat{\cal P}_{E}(X)\,u_{E}\\
E^{-1}{\mb \alpha}\hat{\cal P}_{E}(X)\,u_{E}
\end{array}\right)\,,
\end{equation}
where, by definition,
\begin{equation}\label{PE}
\hat{\cal P}_{E}(X)=\hat X^{(11)}+ E^{-1}\hat X^{(12)}{\mb \alpha} \in {\mb P}
\end{equation}
is the Pauli operator {\em associated} to $X$.
Since the mapping $\hat{\cal P}_{E}: {\mb D}\to {\mb P}$ is linear and
satisfies  $\hat{\cal P}_{E}(X)=  \hat{\cal P}_{E}(XI)$  it results that
${\rm Ker}\,\hat{\cal P}_{E}={\mb D}_{1}$.
In other respects,  Eqs. (\ref{X1}) and (\ref{X2}) lead to
the  important property
\begin{equation}
\hat{\cal P}_{E}(XY)=  \hat{\cal P}_{E}(X)\hat{\cal P}_{E}(Y)\,, \quad \forall
X,Y\in {\mb D}\,.
\end{equation}
which guarantees  that  $\hat{\cal P}_{E}$ preserves  the algebraic
relations, mapping any algebra or superalgebra of ${\mb D}_0$ into an
{\em isomorphic} algebra or superalgebra of ${\mb P}$, with the same
commutation and anticommutation rules.

The conclusion here is that only the diagonal conserved Dirac operators
can be correctly associated to  conserved Pauli operators independent
on $E$.  However, the off-diagonal operators can be transformed at any
time in diagonal ones using the multiplication with $H$ or $H^{-1}$.
For example,  $H$ itself which is off-diagonal is related to the diagonal
operators $H^2={\cal D}(\Delta)$ or  $I$. Thus each  Dirac operator from
${\mb D}$ can be brought in a diagonal form  associated with an operator
from ${\mb P}$.

\subsection{The operators of the Dirac theory}

The simplest operators of ${\mb D}$ which commute with $H$, $D$, and $\gamma^5$
are the generators (\ref{X}) of the representation $spin[S(M)]$ carried by
the space $\Psi$. As observed before, the expressions of these operators
are strongly dependent on the gauge fixing.
\begin{theor}
In the gauge (\ref{ehe}) the spinor fields transform {\em manifestly
covariant} under the transformations of the group $S(M)=SU(2)\otimes U(1)_4$.
\end{theor}
\begin{demo}
Using the components of the Killing vectors given by Theorem \ref{tLLL}
and calculating the functions $(\ref{Om})$ we find that in this gauge
the rotation generators of $spin[S(M)]$ are the standard components of the
{\em total} angular momentum
\begin{equation}
{\cal J}_i=L_i+S_i\,, \quad S_i=\textstyle{\frac{1}{2}\varepsilon_{ijk}S^{jk}
=\frac{1}{2}}{\rm diag}(\sigma_i,\sigma_i)\,,
\end{equation}
with point-independent spin operators,  \cite{CV2}. In the same way
one can show that the $U(1)_4$ generator, $P_4$, does not get a spin term.
\end{demo}\\
Hence it results that the representation $spin[S({M})]$ is {\em reducible}
being a sum of two irreducible representations carried by spaces of
two-dimensional Pauli spinors  where the components
of the total angular momentum are  $J_i=L_i +\frac{1}{2}\sigma_i$. This
means that we can put
\begin{equation}\label{JI}
{\cal J}_{i}I= {\cal D}(J_{i})={\cal D}(L_{i}) + \textstyle\frac{1}{2}
{\cal D}(\sigma_{i}) \,,
\end{equation}
where both the orbital and the spin terms  {\em separately} commute with $H$
since $L_{i}$ and $\sigma_{i}$ commute with $\Delta$.

The triplet ${\mb f}$ defined by Eq.  (\ref{trip}) gives rise to the spin-like
operators
\begin{equation}\label{3sl}
\Sigma^{(i)}=\frac{i}{4}\hat f^{(i)}_{\hat{\alpha}\hat\beta}
\gamma^{\hat{\alpha}}\gamma^{\hat\beta}=\left(
\begin{array}{cc}
\sigma_i&0\\
0&0
\end{array}\right)\,,
\end{equation}
and, according to Eq.  (\ref{DDS}), produce the  Dirac-type operators
 \cite{CV2}
\begin{equation}\label{3Dto}
D^{(i)}= -f^{(i)}_{\mu\nu}
\gamma^{\nu}\nabla^{\mu}=i[D,\,\Sigma^{(i)}]= -i\left(
\begin{array}{cc}
0&\sigma_i {\mb \alpha}^*\\
{\mb \alpha}\sigma_i&0
\end{array}\right)=-i {\cal Q}(\sigma_i) \,,
\end{equation}
which anticommute with $D$ and $\gamma^5$.
The operators $D$ and $D^{(i)}$, $i=1,2,3$, form the basis of the ${\cal N}=4$
superalgebra ${\mb d}_{\mb f}\subset {\mb D}_0$ with the same anticommutation
relations as Eq. (\ref{4sup}). According to the general theory presented in the
previous section, the spinor representation the of group
$(G_{\mb f})\sim SU(2)$ is generated by the operators
$\hat s_i=\frac{1}{2}\Sigma^{(i)}$ satisfying  the commutation rules
$(\ref{sss})$ and  $[{\cal J}_i ,\, \hat s_j]=i\varepsilon_{ijk} \hat s_k$.
They generates  the $SU(2)$  matrices (\ref{TxiS}),
\begin{equation}\label{3Uxi}
T({\rho})=e^{-i\vec{\rho}\cdot \vec{\Sigma}}=\left(
\begin{array}{cc}
\hat U(\vec{\rho})&0\\
0&{\mb 1}_2
\end{array}\right)\,,
\end{equation}
where $\vec{\rho}={\alpha} \vec{\nu}$ ( $\vec{\nu}^2=1$) and $\hat U$ is the
$SU(2)$ transformation
\begin{equation}\label{3SU2}
\hat U(\vec{\rho})=e^{-i\vec{\rho}\cdot\vec{\sigma}}=
{\mb 1}_2 \cos{\alpha} -i\vec{\nu}\cdot\vec{\sigma}\sin\alpha\,.
\end{equation}
These matrices give the transformations  (\ref{TDT10}) and
(\ref{TDT20}). The group $Aut({\mb d}_{\mb f})$ is completed by
the transformations produced by isometries, governed by
 the real-valued orthogonal matrices  of the group $O_{\mb f}$ defined by Eq.
(\ref{rotroot}). Using the results of Theorem \ref{t23}, it is
straightforward  to calculate these matrices in the gauge
(\ref{ehe}) obtaining $\hat{\frak R}(\xi)={\frak R}^T$ for any
isometry $\phi_{\xi}=\phi_{{\frak R}, a^4}$. This explains why
$D^i$ behave under rotations as vector components  while $D$
remains invariant. In other respects, from Eq. (\ref{3Uxi})  we
find that the discrete transformations of the group ${\Bbb Q}({\mb
f})$ are ${\mb 1}$,  $\gamma^5$ and the  sets of matrices
$U_{(k)}={\rm diag} (i\sigma_{k},{\mb 1}_2)$ and
$\gamma^5U_{(k)}$.

In current calculations, when we are not interested to exploit
the  ${\cal N}=4$ superalgebra,  it is indicated to use
the simpler operators
\begin{equation}
Q_i= i H^{-1}D^{(i)}=H^{-1}{\cal Q}(\sigma_{i})={\cal D}(\sigma_{i}) \,,
\end{equation}
instead of $D^{(i)}$. However, in this case the fourth partner of the
operators $Q^i$ is rather trivial since this is just $I$. Therefore, these
operators form a representation of the
quaternion units (or of the algebra of Pauli matrices) with values in
${\mb D}_{0}$,
\begin{equation}\label{Pau}
Q_{i}\, Q_{j}=\delta_{ij}I+i\varepsilon_{ijk} Q_{k}\,,
\end{equation}
producing an evident ${\cal N}=3$ superalgebra.

The corresponding Dirac-type operator of the last K-Y tensor, $f^Y$, calculated
according to the general rule (\ref{df}) with a suitable phase factor ($i$),
was obtained in  \cite{CV3}.  This has the form
\begin{equation}\label{dy1}
D^Y=-{\cal Q}(\sigma_r)
+\frac{2i}{\mu\sqrt{V}}\left(
\begin{array}{cc}
0&\lambda\\
-\lambda&0
\end{array}
\right)\,,
\end{equation}
where the Pauli operators $\sigma_r=\vec{\sigma}\cdot \vec{x}/r$ and
$\lambda=\vec{\sigma}\cdot(\vec{x}\times \vec{P})+{\mb 1_2}
=\sigma_{L}+{\mb 1_2}+\mu\sigma_{r}P_{4}$
have the properties
\begin{equation}
\{\sigma_{r},\lambda\}=0 \,,\quad [\sigma_{r},\sigma_{P}]=
\frac{2i}{r}\,\lambda \,,
\end{equation}
\begin{equation}\label{silam}
\sigma_{P}\lambda=-\lambda\sigma_{P}=
\frac{i}{2}\,\vec{\sigma}\cdot(\vec{P}\times\vec{L}-\vec{L}\times\vec{P})
-\frac{i\mu}{r}\lambda P_{4} \,,
\end{equation}
that help one to find the equivalent forms
reported in  \cite{CV3}  and verify that $D^Y$ commutes with $H$ and
$P_{4}$ and anticommutes with ${D}$ and $\gamma^5$. Moreover,
after a little calculation,  we obtain the remarkable identity
\begin{equation}
\mu P_{4}\left[D^{Y}+
{\cal Q}(\sigma_r)\right]=\left\{H,\,\hat\Lambda\right\} \,,
\end{equation}
involving the operator $\hat\Lambda={\rm diag}(\lambda,\lambda)$ that is a
particular version of an  operator proposed by Biedenharn  \cite{BID}.
This is not conserved but
$\hat \Lambda^{2}=\vec{{\cal J}}^2-\mu^{2}{P_{4}}^{2}+\frac{1}{4}{\mb 1}$ has
this property. Furthermore, we observe that, according to Eq.
(\ref{silam}), the physical part of $D^Y$  can be put in the form
\begin{equation}
D^{Y}I={\cal Q}\left(-\sigma_{r}+\frac{2i}{\mu}\lambda\pi\Delta^{-1}\right)
={\cal Q}(\sigma^{Y}\Delta^{-1}) \,,
\end{equation}
where
\begin{equation}
\sigma^{Y}=\frac{2}{\mu}\left[\sigma_{K}+(\sigma_{L}+{\mb 1}_2)P_{4}\right]
\end{equation}
is a new  conserved Pauli operator  associated to
\begin{equation}
Q^Y=H D^{Y}=H D^{Y}I= {\cal D}(\sigma^{Y})\in {\mb D}_0.
\end{equation}
We note that the Pauli operators
$\sigma_L=\vec{\sigma}\cdot \vec{L}$ and
$\sigma_K=\vec{\sigma}\cdot \vec{K}$ are conserved and satisfy
\begin{equation}\label{alabala}
\{\sigma_K,\,\sigma_L+{\mb 1}_2\}= 2\vec{L}\cdot\vec{K}\,,\quad
\{\sigma_r,\,\sigma_L+{\mb 1}_2\}= -2\mu P_4\,.
\end{equation}

As in the case of the Klein-Gordon theory, we can define  the components of
the conserved Runge-Lenz operator of the Dirac theory  \cite{CV3,CV4} giving
directly their physical parts,
\begin{equation}\label{KI}
{\cal K}_{i}I=\frac{i\mu}{4}\left\{ HD^Y,\,H^{-1}D^{(i)}\right\}
+ \frac{i}{2}({\cal B}-P_4)H^{-1} D^{(i)}-{\cal J}_i I P_4 \,,
\end{equation}
where ${\cal B}^{2}={P_{4}}^{2}-H^2$. Since  ${\cal B}^2 I={\cal D}(B^2)$,
we can express
\begin{equation}\label{Kas}
{\cal K}_{i}I= {\cal D}(\hat K_{i})\,,\quad
\hat K_{i}=K_{i} +\frac{\sigma_{i}}{2}\,B\in {\mb P}\,.
\end{equation}
Furthermore, we obtain the following commutation relations
\begin{eqnarray}
\left[ {\cal J}_{i},\, {\cal J}_{j}\right]&=&i\varepsilon _{ijk}{\cal J}_{k}
\,,\nonumber\\
\left[ {\cal J}_{i},\, {\cal K}_{j}\right]&=&i\varepsilon _{ijk}{\cal K}_{k}
\,,\label{algJK}\\
\left[ {\cal K}_{i},\, {\cal K}_{j}\right]&=&i\varepsilon _{ijk}{\cal J}_{k}
{\cal B}^{2}\,,\nonumber
\end{eqnarray}
and  the commutators with the operators $Q_i$ \cite{CV5},
\begin{eqnarray}\label{QJK}
\left[ {\cal J}_{i},\, {Q}_{j}\right]&=&i\varepsilon _{ijk}{Q}_{k}
\,,\nonumber\\
\left[ {\cal K}_{i},\, {Q}_{j}\right]&=&i\varepsilon _{ijk}{Q}_{k}
{\cal B}\,.
\end{eqnarray}
The algebra (\ref{algJK}) does not close as a Lie algebra because
of the factor ${\cal B}^2$. The dynamical algebras of the Dirac
theory have to be obtained as in the scalar case by replacing this
operator with its eigenvalue $\hat q^2-E^2$ and rescaling the
operators ${\cal K}_{i}$. One obtains thus the same dynamical
algebras as those governing the scalar modes but in different
representations \cite{CV4}.

Other operators related to $\vec{\cal K}$ are the Casimir
operators of the open algebra (\ref{algJK}) defined as
\begin{equation}\label{Cas}
{\cal C}_1= \vec{\cal J}^2{\cal B }^2 + \vec{\cal K}^2\,, \quad {\cal C}_2=
\vec{\cal J}\cdot \vec{\cal K}\,.
\end{equation}
 In addition, we can introduce a new useful
Casimir-type operator
\begin{equation}\label{4QI}
Q=Q I=\frac{\mu}{2}\,Q^Y + ({\cal B}-P_4) {\cal D}(\sigma_L+{\bf
1}_2)= {\cal D}[\sigma_K+(\sigma_L+{\bf 1}_2)B]\,,
\end{equation}
related to $Q^Y$. This satisfies the simple algebraic relations,
\begin{equation}\label{QQKJ}
[Q, {\cal J}_i]=0\,, \quad [Q, {\cal K}_i]=0\,,\quad \{ Q, Q_i
\}=2({\cal K}_i +{\cal J}_i{\cal B})I\,,
\end{equation}
and  the identity
\begin{equation}\label{Q2}
Q^2=\frac{\mu^2}{4}({P_4}I-{\cal B})^4\,,
\end{equation}
resulting from Eqs. (\ref{C1C2}), (\ref{C1C21}) and (\ref{4QI}). Moreover,
using Eqs. (\ref{CBCB}) we find two new operators that can be put in a closed
form,
\begin{eqnarray}\label{CCC}
&&{\cal C}^+={\cal C}_1+2{\cal B}{\cal C}_2+{\cal B}^2=(Q+{\cal B})^2\,,\\
&&{\cal C}^-={\cal C}_1-2{\cal B}{\cal C}_2+{\cal B}^2=\frac{\mu^2}{4}\,(P_4
I+{\cal B})^4 \,.
\end{eqnarray}
The operators $Q$ and ${\cal C}^+$ are  Casimir operators only for the algebra
(\ref{algJK}) but $Q^2$ and ${\cal C}^-$ are general Casimir operators since
they commute with any other conserved Dirac operator.

Finally we observe that we can take over the operator $N$ of the scalar theory
since the Dirac and the Klein-Gordon particles have the same energy spectrum.
This offers us the opportunity  to introduce the new Casimir operator
\begin{equation}\label{MNP}
M=(N+\mu P_4)^2 I \,\in {\bf D}_0
\end{equation}
 that allows us to write
 \begin{equation}\label{QQM}
\{ Q,\,Q\}=2{\cal B}^2 M \,.
\end{equation}
as it results from Eqs. (\ref{CBCB}) and (\ref{Q2}).

We conclude that the conserved operators of the Dirac
theory can be associated with conserved Pauli or Klein-Gordon operators
produced by the same geometrical objects. A brief image of these
relationships is given in the table below.
\begin{center}
\begin{tabular}{cccccc}
\hline
geometric       &nature   &symmetry&Dirac   &Pauli   &Klein-Gordon\\
object          &         &        &operator&operator&operator\\
&&&&&\\
\hline
&&&&&\\
$f_{\mu\nu}^{(i)}$&K-Y tensor&*     &$Q_{i}$&$\sigma_{i}$&-\\
$f_{\mu\nu}^{Y}$&K-Y tensor&*      &$Q^Y$&$\sigma^{Y}$&-\\
$k_{(4)}^{\mu}$&K vector&$U(1)_4$&$P_{4}$&$P_{4}$&$P_{4}$\\
$k_{(i)}^{\mu}$&K vector&$SO(3)$&${\cal J}_{i}$&$J_{i}$&$L_{i}$\\
$k_{(i)}^{\mu\nu}$&S-K tensor& hidden&${\cal K}_{i}$
&$\hat{K}_{i}$&$K_{i}$\\
\hline
\end{tabular}
\end{center}

\subsection{The quantum modes of the Dirac field}

The large collection of conserved observables we  have presented
above will help us to select  different complete sets of commuting
observable which should define static quantum modes. Here we
restrict ourselves to discuss only the sets including the
operators $H$ and $P_{4}$ (or $P_{\chi}$) for which we need to
introduce three new operators in order to completely determine the
quantum modes with given energy $E>0$ and $\hat q$. These
operators can be selected at the level of the associated Pauli
operators since, according to Theorem \ref{invers}, the eigenvalue
problem $H\psi_E=E\psi_E$ is solved by the spinors
$\psi_E=(u_{E},\, E^{-1}{\mb \alpha}u_{E})^T$ depending on the
Pauli spinors $u_E$ which satisfy $\Delta u_E=E^2 u_E$. Therefore,
we may construct well-defined spinors $u_E$  starting with the
operators $\Delta$ and $P_{\chi}$ and adding three more operators
from ${\mb P}$ which have to fill in the set of the commuting
operators able to determine the spinors $u_E$ as common
eigenspinors. In this way the problem is solved since the second
Pauli spinor of $\psi_E$ is $E^{-1}{\mb \alpha} u_E$. The last
step is to identify the operators from ${\mb D}$ whose physical
parts form the set of commuting Dirac operators associated to the
Pauli ones.

In what follows we shall define two types of quantum modes with
given energy starting with the simplest ones for which $u_E$ has
separated variables.
\begin{defin}
We say that the common eigenspinors of the set of commuting
operators $\{H,\,P_{\chi},\,\vec{\cal J}^{2},\,{\cal J}_{3},\,
{\cal Q}(\sigma_{L}+{\mb 1}_2)\}$ corresponding to the eigenvalues
$E,\, q,\, j(j+1),\, m_{j}$ and $\pm E(j+\frac{1}{2})$. define the
{\em central} modes.
\end{defin}
Therefore, $u_E$ must be the common eigenspinor of
the set $\{\Delta,\, P_{\chi} ,\,\vec{J}^2 ,\,J_3 ,\, \sigma_{L}+{\mb 1}_2 \}$
corresponding to the eigenvalues
$E^{2},\, q,\, j(j+1),\, m_{j}$ and $\pm (j+\frac{1}{2})$

This problem can be solved only defining new spherical spinors involving
our previously presented $SO(3)\otimes U(1)$  spherical harmonics  \cite{CV2}.
Following the traditional method  \cite{TH}  we defined the
spherical spinors  $\Phi^{\pm}_{q,j,m_{j}}(\theta,\varphi,\chi)$
as the common  eigenspinors of the eigenvalue problems
\begin{eqnarray}
P_{\chi}\,\Phi^{\pm}_{q,j,m_{j}}&=& q\,\Phi^{\pm}_{q,j,m_{j}}\,, \\
\vec{ J}^{2}\,\Phi^{\pm}_{q,j,m_{j}}&=& j(j+1)\,\Phi^{\pm}_{q,j,m_{j}}
\,, \\
 J_{3}\,\Phi^{\pm}_{q,j,m_{j}}&=& m_{j}\,\Phi^{\pm}_{q,j,m_{j}}\,, \\
(\sigma_{L}+{\mb 1}_{2})\,\Phi^{\pm}_{q,j,m_{j}}&=& \pm(j+1/2)\,
\Phi^{\pm}_{q,j,m_{j}}\,.\label{3kkk}
\end{eqnarray}
These spinors are, in addition, eigenfunctions of
${\vec{{L}}}^{2}$ corresponding
to the eigenvalues $l(l+1)$ with $l=j\pm\frac{1}{2}$. For $j=l+\frac{1}{2}
>|q|-\frac{1}{2}$ we have
 \cite{TH,CV1}
\begin{equation}
\Phi^{+}_{q,j,m_{j}}=\frac{1}{\sqrt{2j}}\left(
\begin{array}{l}
\sqrt{j+m_{j}}\, Y^{q}_{j-\frac{1}{2},m_{j}-\frac{1}{2}}\\
\sqrt{j-m_{j}}\, Y^{q}_{j-\frac{1}{2},m_{j}+\frac{1}{2}}
\end{array}\right) \,,
\end{equation}
while for $j=l-\frac{1}{2}>|q|-\frac{3}{2}$ we get
\begin{equation}
\Phi^{-}_{q,j,m_{j}}
=\frac{1}{\sqrt{2j+2}}\left(
\begin{array}{l}
\sqrt{j-m_{j}+1}\, Y^{q}_{j+\frac{1}{2},m_{j}-\frac{1}{2}}\\
-\sqrt{j+m_{j}+1}\, Y^{q}_{j+\frac{1}{2},m_{j}+\frac{1}{2}}
\end{array}\right)\,.
\end{equation}
These spherical spinors  are
orthonormal since the $SO(3)\otimes U(1)$ harmonics are orthonormal with
respect to the angular scalar product (\ref{3(spy)}).
From  the second of Eqs. (\ref{alabala}) it results the useful property
\begin{equation}\label{pmmp}
\sigma_{r}\Phi^{\pm}_{q,j,m_{j}}=
\pm\lambda^{q}_{j}\,\Phi^{\pm}_{q,j,m_{j}}
+\sqrt{1-(\lambda^{q}_{j})^{2}}\,\Phi^{\mp}_{q,j,m_{j}} \,,
\end{equation}
where  $\lambda^{q}_{j}=q/(j+\frac{1}{2})$. Note that a
similar formula is reported in Ref. \cite{HARM1}.

The energy eigenspinors of the central modes can be expressed in
terms of these new spherical spinors as  \cite{CV2}
\begin{eqnarray}
\psi^{\pm}_{E, q,j,m_{j}}&=&\left(
\begin{array}{r}
u_{E,q,j,m_{j}}^{\pm}\\
E^{-1}{\mb \alpha}u_{E,q,j,m_{j}}^{\pm}
\end{array}\right) \nonumber \\
&=& N_{E,q,j}\,\frac{1}{r}\left[
\begin{array}{r}
{\frak f}^{\pm}_{E,q,j}\,\Phi^{\pm}_{q,j,m_{j}}~~\\
iE^{-1}\sqrt{V}\left({\frak h}^{\pm} _{E,q,j}\,\Phi^{\pm}_{q,j,m_{j}}+
{\frak g}^{\pm} _{E,q,j}\,\Phi^{\mp}_{q,j,m_{j}}\right)
\end{array}\right]\,,\label{sscc}
\end{eqnarray}
where the radial functions ${\frak f}^{\pm}_{E,q,j}\equiv
{\frak f}_{E,q, l_{\pm}}$ are the solutions of the radial equation
(\ref{(kgrad)}) for $l=l_{\pm}=j\mp \frac{1}{2}$.
The other radial functions have to be calculated using Eq. (\ref{pmmp})
and taking  into account that
\begin{equation}
i\sigma_{P}=\sigma_{r}\left(\partial_{r}+\frac{1}{r}-\frac{\sigma_{L}+1}{r}
\right)+\frac{1}{r}P_{\chi} \,.
\end{equation}
In this way we obtain the equations
\begin{eqnarray}
{\frak g}^{\pm}_{E,q,j}&=&\sqrt{1-(\lambda^{q}_{j})^{2}}\left(-\frac{d}{dr}
\pm\frac{j+\frac{1}{2}}{r}\right){\frak f}^{\pm}_{E,q,j}\,,\\
{\frak h}^{\pm}_{E,q,j}&=&\lambda^{q}_{j}\left(\mp\frac{d}{dr}
+\frac{j+\frac{1}{2}}{\mu V}\right){\frak f}^{\pm}_{E,q,j}\,,
\end{eqnarray}
which lead to the identity
\begin{equation}
{\frak h}^{\pm}_{E,q,j}=\frac{q}{\mu}\,{\frak f}^{\pm}_{E,q,j}\pm
\frac{\lambda^{q}_{j}}{\sqrt{1-(\lambda^{q}_{j})^2}}\,{\frak g}^{\pm}_{E,q,j}\,.
\end{equation}

The axial modes can be constructed in a similar manner.
\begin{defin}
The common eigenspinors of the set $\{H,\,P_4,\,{\cal
K}_{3},\,{\cal J}_{3},\, {\cal Q}(\sigma_3)\}$ corresponding to
the eigenvalues $E$, $\hat q$, $\hat\kappa$, $m_{j}$ and $E\sigma$
define the {\em axial} modes.
\end{defin}
Now $u_E$ is the common eigenspinor of the set
$\{ \Delta$, $P_{4}$, $\hat K_3$,  $J_3, \sigma_3\}$ corresponding to the
eigenvalues
$E^2$, $\hat q$, $\hat\kappa$ and $m_j$ and $\sigma$. These eigenspinors can be
calculated in parabolic coordinates as  axial solutions of the
Klein-Gordon (or Schr\" odinger equation  \cite{CV1}) multiplied
with the constant eigenspinors of $\frac{1}{2}\sigma_3$ having the eigenvalues
$\sigma=\pm \frac{1}{2}$. For example, solving the problem of the discrete
axial modes  \cite{CV1} one finds that the scalar parts are eigenfunctions of
the set $\{ \Delta, P_{4}, K_3, L_3 \}$ corresponding to the eigenvalues
$E^2,\, \hat q,\, \kappa$ and $m$ that give those of the Dirac modes as
$ \hat\kappa=\kappa+|b|\sigma$ and $m_{j}=m+\sigma$.

In other respects, the algebraic method based on the dynamical
algebras may help one to obtain  new results even if the rescaling
of the Runge-Lenz operator has to be done as in the scalar case
for the same spectral domains of the Kepler-type problem. If we
define
\begin{equation}
{\cal R}_{i}=\left\{
\begin{array}{lllll}
{\cal B}^{-1}{\cal K}_{i}&{\rm for}& \mu<0&{\rm and}&E<|\hat q|\\
{\cal K}_{i}&{\rm for}&{\rm any}~ \mu&{\rm and}&E=|\hat q|\\
i{\cal B}^{-1}{\cal K}_{i}&{\rm for}&{\rm any}~ \mu &{\rm and}& E>|\hat q|
\end{array}\right. \,,
\end{equation}
then the operators ${\cal J}_i$ and ${\cal R}_i$ ($i=1,2,3$) will generate
either a  representation of the $o(4)$ algebra for the discrete energy
spectrum in the domain $E<|\hat q|$ or a representation of the $o(3,1)$
algebra for  continuous spectrum in the domain $E>|\hat q|$. The  dynamical
algebra  $e(3)$  corresponds only to the ground energy of the continuous
spectrum, $E=|\hat q|$.

For each set of eigenvalues $E$ and $q$ one obtains a specific
particular representations of one of the above defined dynamical
algebras \cite{CV7}. However,  in the scalar case we have seen
that the representations of these  algebras are no longer usual
linear ones since there appear non-standard $su(2)$ weights. In
the Dirac theory we shall meet a similar situation. In order to
illustrate this phenomenon let us focus on the representations of
the $o(4)$ dynamical algebra of the discrete quantum modes. In the
Dirac theory the dynamical algebra is the same as in the scalar
case but its representations are generated by the operators ${\cal
J}_{i}$ and ${\cal R}_{i}= {\cal B}^{-1}{\cal K}_{i}$ which have
spin terms. According to Eq. (\ref{cue}), these representations
are equivalent with those generated by the associated Pauli
operators $J_{i}$ and $R_{i}=B^{-1}\hat K_{i}$ acting upon the
first Pauli spinor. However, since
$J_{i}=L_{i}+\frac{1}{2}\sigma_{i}$ and
$R_{i}=K_{i}^{re}+\frac{1}{2}\sigma_{i}$, we can assume that the
Dirac discrete modes are governed by the {\em reducible}
representation
\begin{equation}\label{desc}
\textstyle{\left(\frac{n+q-1}{2},\frac{n-q-1}{2}\right)\otimes
\left(\frac{1}{2},0\right)=
\left(\frac{n+q}{2},\frac{n-q-1}{2}\right)\oplus
\left(\frac{n+q}{2}-1, \frac{n-q-1}{2}\right)\,.}
\end{equation}
The Casimir operators, $\hat C_{1}=\vec{J}^2+{\vec{R}}^2$ and
$\hat C_2=\vec{J}\cdot \vec{R}$ take now the eigenvalues, $\hat
c_{1}$ and $\hat c_{2}$, such that $\hat c_{1}-2\hat c_{2}=(n-q)^2
-1$ for both irreducible representations while $\hat c_{2}$ takes
the value $(n+\frac{1}{2})(q+\frac{1}{2})$ for the representation
$(\frac{n+q}{2},\frac{n-q-1}{2})$ and $
(n-\frac{1}{2})(q-\frac{1}{2})$ for the representation
$(\frac{n+q}{2}-1,\frac{n-q-1}{2})$. This suggests us to introduce
the new Pauli operator
\begin{equation}
\hat C=2\hat C_{2}-\frac{1}{2}{\mb 1}_2=\{\sigma_R,\sigma_L+{\bf
1}_2\}+
 \sigma_{R}+\sigma_{L}+{\mb 1}_2\,,
\end{equation}
where we denoted
$\sigma_{R}=B^{-1}\sigma_K=\vec{\sigma}\cdot\vec{K}^{re}$. This
operator allows us to distinguish between the irreducible
representations resulted from the decomposition (\ref{desc}). The
advantage is that it takes the simpler eigenvalues $c=2nq\pm(n+q)$
for these representations. The new Dirac operator associated to
$\hat C$ is ${\cal C}={\cal D}(\hat C)=2\vec{\cal J}\cdot
\vec{\cal R}-\frac{1}{2}{\mb 1}$.

The operator ${\cal C}$ can be used for defining new quantum
modes. For example, we may choose the set $\{H,\,P_4,\,{\cal
C},\,\vec{\cal J}^{2},\,{\cal J}_{3}\}$ that determines new
central modes whose eigenspinors correspond to the eigenvalues
$E_{n}$, $\hat q$, $c$, $j(j+1)$ and $m_{j}$. Another possibility
is to take the set $\{H,\,P_4,\,{\cal C},\,{\cal R}_{3},\,{\cal
J}_{3}\}$ of new axial modes corresponding to the eigenvalues
$E_{n}$, $\hat q$, $c$, $m_{r}$ and $m_{j}$. Since neither ${\cal
Q}(\sigma_{L}+{\mb 1}_2)$ nor $Q_3$ do not commute with ${\cal
C}$, it results  that these modes do not have eigenspinors with
separated variables. Therefore these must be linear combinations
of the spinors of the original central or axial modes.

In Ref. \cite{CV4} we tried to derive the spinors of these modes
starting wrongly with the particular case of the discrete energy
spectrum with $q=0$ that does not have physical meaning.
Therefore, we must specify that the results obtained in this
manner are not correct. Nevertheless, this mistake does not affect
other results we have obtained since this arises in an isolated
sector being rather of academic interest. The major problem here
is to investigate the nature of the unusual representations
involving non-standard $su(2)$ weight. New quantum modes could be
seriously studied only after we shall understand the subtle
mechanisms of these representations. We would like to consider
this problem elsewhere when we shall have all the technical
elements we need for defining correctly new quantum modes of the
Dirac field in the Euclidean Taub-NUT background.

\subsection{Infinite-dimensional  superalgebra}

Now we may ask how could be organized the very rich set of conserved Dirac
operators we constructed above. There are many commutation and anticommutation
relations that can not be ignored such that it seems that the suitable
structure may be a superalgebra.  Thus we start with the simplest algebraic
structure.
\begin{defin}
 ${\cal S}_0$ is the open superalgebra generated by the operators
\begin{equation}
  \{I, M, {\cal J}_i, {\cal K}_i, Q, Q_i\}\subset {\bf D}_0
\end{equation}
which satisfy Eqs. (\ref{algJK}), (\ref{QJK}), (\ref{QQKJ}) and (\ref{QQM}).
\end{defin}
We observe that, as in the non-relativistic quantum Kepler problem, there are
algebraic relations which remain open because of the factors ${\cal B}$.
Therefore, we are forced  to embed all the above ingredients in an {\em
infinite-di\-men\-sional} superalgebra constructed  in the same manner as the
infinite algebra of Ref. \cite{DD}. The difference is that here we  have a
superalgebra with generators of bosonic or fermionic type.

\begin{defin} ${\cal S}$ is the infinite-dimensional superalgebra  generated
by the countable set of operators
\begin{equation}
\{I_n,M_n,J^i_n,K^i_n,Q_n,Q^i_n\}\,,\quad  n=0,1,2,...
\end{equation}
among them the set of operators
\begin{equation}\label{defEJK}
I_n=I{\cal B}^n\,,\quad M_n=M{\cal B}^n  \,, \quad J_n^i={\cal J}_i{\cal
B}^n\,,\quad K_n^i={\cal K}_i{\cal B}^n\,,
\end{equation}
form the {\em bosonic} sector while the supercharges of the {\em fermionic}
sector are
\begin{equation}\label{defQQ}
Q_n=Q {\cal B}^n\,,\qquad Q_n^i={Q}_i{\cal B}^n \,.
\end{equation}
\end{defin}
The operators $I_n$ and $M_n$ are Casimir-type operators commuting between
themselves and with all the operators of the bosonic or fermionic sectors.
Then, according to Eqs. (\ref{algJK}) and (\ref{defEJK}), we obtain the
following non-trivial commutators of the bosonic sector
\begin{eqnarray}
\left[ J^i_n, J^j_m \right] &=& i\varepsilon_{ijk}J^{k}_{n+m}\,,\\
\left[ J^{i}_n, K^{j}_m \right] &=& i\varepsilon_{ijk}K^k_{n+m}\,,\\
\left[ K^i_n, K^j_m \right] &=& i\varepsilon_{ijk}J^k_{n+m+2}\,,\label{sic}
\end{eqnarray}
while from Eqs. (\ref{Pau}),(\ref{QQKJ}) and (\ref{QQM})  we deduce the
anticommutators of the fermionic sector,
\begin{eqnarray}
\{ Q^i_n,Q^j_m \}&=&2\delta_{ij}I_{n+m}\,,\\
\{ Q_n, Q^i_m \}&=&2( K^i_{n+m} + J^i_{n+m+1})\,,\label{cuc}\\
\{Q_n, Q_m \}&=& 2\, M_{m+n+2}\,.\label{muc}
\end{eqnarray}
The commutations relations between the bosonic and fermionic operators are
\begin{eqnarray}
\left[ Q_n,J^j_m \right]=0 \,,&\quad&
[ Q^i_n,J^j_m ]=i\varepsilon_{ijk}Q^k_{n+m}\,,\\
\left[ Q_n,K^j_m \right]=0\,,&\quad& [ Q^i_n,K^j_m
]=i\varepsilon_{ijk}Q^k_{n+m+1}\,.\label{cucu}
\end{eqnarray}

Thus we constructed the infinite-dimensional superalgebra  ${\cal S}$ which
represents the {\em minimal} closed dynamical algebraic structure of the Dirac
theory on Taub-NUT manifolds. We observe that the typical algebraic structure
related to the Kepler problem is the infinite-dimensional algebra ${\cal A}$
generated by $\{J^i_n,K^i_n\}$ which is a subalgebra in ${\cal S}$.

\subsection{Twisted loop superalgebras}

Now we intend to show  that the superalgebra ${\cal S}$ can be seen as a {\em
twisted} Kac-Moody superalgebra such that its subalgebra ${\cal A}$ should be a
twisted loop algebra of the usual $so(4)$ algebra, in the sense of Ref.
\cite{DD}.

First we introduce the auxiliary finite-dimensional superalgebra ${\cal W}_0$
generated by the operators $\{ E, F, A^i, B^i,G, G^i\}$. We assume that $E$ and
$F$ commute with any other generator and that the generators $\{A^i,B^i\}$
satisfy the $so(4)$ algebra,
\begin{equation}\label{algAB}
[ {A}^{i},\, {A}^{j}]=i\varepsilon _{ijk}{A}^{k} \,,\quad [ {A}^{i},\,
{B}^{j}]=i\varepsilon _{ijk}{B}^{k} \,,\quad [ {B}^{i},\, {B}^{j}]=i\varepsilon
_{ijk}{A}^{k}\,.
\end{equation}
The operators $G$ and $G^i$ are fermionic supercharges obeying
\begin{equation}\label{algAB1}
\{ G^i, G^j\}=2\delta_{ij}E\,, \quad \{ G, G^i\}=2\,(A^i+B^i)\,,\quad\left\{ G,
G\right\}=2\,F
\end{equation}
and the commutation relations
\begin{eqnarray}
 [ A^{i},\, G ]=0\,,&\quad &[ {A}^{i},\, {G}^{j} ]=i\varepsilon _{ijk}{G}^{k} \,,\nonumber\\
 {[ B^{i},\, G ]}=0\,,&\quad &[ {B}^{i},\, {G}^{j} ]=i\varepsilon
 _{ijk}{G}^{k}\,.\label{algAB2}
\end{eqnarray}
This superalgebra has simple finite-dimensional representations as we briefly
present in the Appendix C.

Furthermore, we consider the corresponding Kac-Moody infinite loop superalgebra
${\cal W}$ generated by the operators $\{ E_n, F_n, A_n^i, B_n^i, G_n,
G_n^i\}$, $n\in {\Bbb Z}$, with the following properties
\begin{eqnarray}
\left[ {A}^{i}_n,\, {A}^{j}_m\right]&=&i\varepsilon_{ijk}{A}^{k}_{n+m}
\,,\qquad \left\{ G^i_n, G^j_m\right\}=2\delta_{ij}E_{n+m}\,,\nonumber\\
\left[ {A}^{i}_n,\, {B}^{j}_m\right]&=&i\varepsilon_{ijk}{B}^{k}_{n+m}\,,
\qquad \left\{ {G}_n,\, {G}^{j}_m\right\}=2\,(A^i_{n+m}+B^i_{n+m})\,,\label{algABn}\\
\left[ {B}^{i}_n,\, {B}^{j}_m\right]&=&i\varepsilon_{ijk}{A}^{k}_{n+m}\,,
\qquad \left\{ {G}_n,\, {G}_m\right\}=2\, {F}_{n+m}\,,\nonumber
\end{eqnarray}
and
\begin{eqnarray}
 [ A^{i}_n,\, G_m ]=0\,,&\quad &[ {A}^{i}_n,\, {G}^{j}_m ]
 =i\varepsilon _{ijk}{G}^{k}_{n+m} \,,\nonumber\\
 {[ B^{i}_n,\, G_m ]}=0\,,&\quad &[ {B}^{i}_n,\, {G}^{j}_m ]=i\varepsilon
_{ijk}{G}^{k}_{n+m}\,,
\end{eqnarray}
understanding that the generators $E_n$ and $F_n$, $n\in {\Bbb Z}$, commute
with any other generator of ${\cal W}$.

The next step is to define
\begin{defin}
The involution automorphism $\tau : {\cal W}\to {\cal W}^{\tau}$ is the mapping
selecting the countable subset of operators
\begin{equation}\label{setop}
\{ E_{2n}, F_{2n},A_{2n}^i, B_{2n+2}^i, G_{2n+2}, G_{2n}^i\}\,,\quad n\in {\Bbb
Z} \,,
\end{equation}
which generates the superalgebra ${\cal W}^{\tau}\subset {\cal W}$.
\end{defin}
The algebraic properties of this superalgebra are given by the commutation
relations of the bosonic sector,
\begin{eqnarray}
\left[ {A}^{i}_{2n},\, {A}^{j}_{2m}\right]&=&i\varepsilon_{ijk}{A}^{k}_{2(n+m)}\,,\nonumber\\
\left[ {A}^{i}_{2n},\, {B}^{j}_{2m+2}\right]&=&i\varepsilon_{ijk}{B}^{k}_{2(n+m)+2}\,,\label{algABns}\\
\left[ {B}^{i}_{2n+2},\,
{B}^{j}_{2m+2}\right]&=&i\varepsilon_{ijk}{A}^{k}_{2(n+m+2)}\,,\nonumber
\end{eqnarray}
the anticommutation relations of the fermionic sector
\begin{eqnarray}
\left\{G^i_{2n}, G^j_{2m}\right\}&=&2\delta_{ij}E_{2(n+m)}\,,\nonumber\\
\left\{G_{2n+2}, G^j_{2m}\right\}&=&2\,(A_{2(n+m+1)}+B_{2(n+m)+2})\,,\label{algABns1}\\
\left\{G_{2n+2}, G_{2m+2}\right\}&=&2\,F_{2(n+m+2)} \,,\nonumber
\end{eqnarray}
and the commutation relations among both sectors,
\begin{eqnarray}
 [ A^{i}_{2n},\, G_{2m+2} ]=0\,,&\quad &[ {A}^{i}_{2n},\, {G}^{j}_{2m} ]
 =i\varepsilon _{ijk}{C}^{k}_{2(n+m)}
\,,\nonumber\\
 {[ B^{i}_{2n+2},\, G_{2m+2} ]}=0\,,&\quad &[ {B}^{i}_{2n+2},\, {G}^{j}_{2m} ]
 =i\varepsilon _{ijk}{C}^{k}_{2(n+m+1)}
\,.\label{algABns2}
\end{eqnarray}
In this way we have constructed the {\em twisted} loop superalgebra ${\cal
W}^{\tau}$ the positive part of which (with $n\ge 0$)  will be denoted by
${\cal W}^{\tau}_+$.

Now we can demonstrate the superalgebra ${\cal S}$ represents a twisted loop
superalgebra related to ${\cal W}^{\tau}_+$.
\begin{theor}
The mapping $\phi\, :\, {\cal W}^{\tau}_+ \to {\cal S}$ defined by
\begin{equation}\label{IJKQ}
I_n=\phi(E_{2n})\,,\quad M_n=\phi(F_{2n})\,,\quad J^i_n=\phi(A^i_{2n})\,,\quad
K^i_n=\phi(B^i_{2n+2})\,,
\end{equation}
and
\begin{equation}\label{IJKQ1}
{Q}_n=\phi(G_{2n+2})\,,\qquad Q^i_n=\phi(G^i_{2n})\,,\quad n=0,1,2,...
\end{equation}
is an {\em homomorphism}.
\end{theor}
\begin{demo}
Indeed, if we consider, for example, the last of Eqs. (\ref{sic}) we can write
\begin{eqnarray}
\left[\phi(B^i_{2n+2}),\phi(B^j_{2m+2}) \right]&=&\left[ K^i_n,
K^j_m \right]=i\varepsilon_{ijk}J^k_{n+m+2}\\
&=&i\varepsilon _{ijk}\phi({A}^{k}_{2(n+m+2)})
=\phi(\left[B^i_{2n+2},B^j_{2m+2} \right])\,.\nonumber
\end{eqnarray}
In this manner one can demonstrate step by step that for any pair of
generators, $X$ and $Y$, of ${\cal W}^{\tau}_+$ we have either $[\phi(X),
\phi(Y)]=\phi([X,Y])$ or $\{\phi(X), \phi(Y)\}=\phi(\{X,Y\})$. Reversely, if we
start with Eqs. (\ref{IJKQ}) and (\ref{IJKQ1}) supposing that $\phi$ is an
homomorphism, then we recover the superalgebra ${\cal S}$. For example, the
last of Eqs. (\ref{cucu}) results from
\begin{eqnarray}
\left[ Q^i_n, K^j_m \right]&=& \left[\phi(G^i_{2n}),\phi(B^j_{2m+2}) \right]
=\phi(\left[G^i_{2n},B^j_{2m+2} \right])\nonumber\\
 &=&i\varepsilon
_{ijk}\phi({G}^{k}_{2(n+m+1)}) =i\varepsilon_{ijk}Q^k_{n+m+1}\,.
\end{eqnarray}
In this way, we bring arguments that our superalgebra ${\cal S}$ can be seen as
a twisted loop superalgebra.
\end{demo}

It finally should be mentioned that the above construction of the twisted loop
superalgebras could be regarded differently. The connection between the set of
operators (\ref{defEJK}), (\ref{defQQ}) and (\ref{setop}) can be realized
directly assigning grades to each operator \cite{daboul} as follows:
\begin{eqnarray}
&&E_{2n}:=I{\cal B}^n\,,\quad F_{2n}:=M{\cal B}^n  \,,
\quad A_{2n}^i:={\cal J}_i{\cal B}^n\,,\nonumber\\
&&B_{2n+2}^i:={\cal K}_i{\cal B}^n\,,\quad G_{2n+2}:=Q {\cal B}^n\,,\quad
G_{2n}^i:={Q}_i{\cal B}^n \,.
\end{eqnarray}
Thus we achieve a graded loop superalgebra of the Kac-Moody type and the sum of
the grades is conserved under (anti)commutations.

\subsection{Remarks}

Here we constructed the infinite-dimensional superalgebra ${\cal S}$ starting
with the finite-dimensional open superalgebra ${\cal S}_0$ formed only by
conserved operators commuting with $I$, $H$, $P_4$ and the whole set of Casimir
operators freely generated by these three operators.

In ${\cal S}_0$ we explicitly used two Casimir operators, namely $I$ and $M$.
As mentioned, $I$ is the projector on the physical spinor subspace playing the
role of identity operator. More interesting is the operator $M$ since this
depends on $N$ which is in some sense similar with the operator
$(-2H_K)^{-1/2}$ of the $so(4,2)$ dynamical algebra of the quantum Kepler
problem governed by the non-relativistic Hamiltonian operator
$H_K=-\frac{1}{2}\Delta - r^{-1}$ \cite{BARUT}. We remind the reader that the
$so(4,2)$ dynamical algebra of the Kepler problem contains not only conserved
operators but even operators that do not commute with $H_K$. In this case, the
conserved operators, i. e. the angular momentum and the Runge-Lenz vector
operator, are {\em orthogonal}  and generate an {\em open} algebra that can be
rescaled obtaining thus the dynamical algebra $o(4)\subset so(4,2)$ of the
discrete energy spectrum. The first Casimir operator of $o(4)$, that reads
$C_K^1=(-2H_K)^{-1}-I$, has a similar form with our operator $M$. However, the
second Casimir operator of $o(4)$ vanishes while our operator ${\cal C}_2$,
given by Eq. (\ref{Cas}), is different from zero since the vector operators
$\vec{\cal J}$ and $\vec{\cal K}$ are not orthogonal.

In these circumstances we can say that the subalgebra (\ref{algJK}) of ${\cal
S}_0$ corresponds to the open algebra that gives the $o(4)$ dynamical algebra
of the Kepler problem. This explains why our twisted loop superalgebra was
constructed in a similar way as that of the Kepler case \cite{DD}. The main
difference between these two theories is that ${\cal S}_0$ is an open
superalgebra containing the supercharges $Q$ and $Q_i$ that naturally arise
from the very special geometry of the Euclidean Taub-NUT space. For this reason
we were forced to include, in addition,  the bosonic Casimir operator $M$ for
writing down Eq. (\ref{QQM}). We specify that this is more than a simple
artifice since the resulting infinite superalgebra ${\cal S}$ is a twisted loop
superalgebra  arising from  a coherent algebraic structure, namely the
superalgebra ${\cal W}_0$ the representations of which are presented in
Appendix.

In other respects, it is clear that the operators $Q$ and $Q_i$ appear only in
the Dirac theory on ${\frak M}$ since they are in fact Dirac-type operators.
Therefore, it is interesting to compare our results with relativistic systems
with spin half whose non-relativistic limit is the quantum Kepler problem.
Thus, in the case of the Dirac electron in external Coulomb field there exists
a hidden symmetry even if one does not have a conserved Runge-Lenz operator.
This symmetry is related to another operator, called the Johnson-Lippmann
operator \cite{JL}, that is a scalar conserved operator. In the
non-relativistic limit this becomes the projection of the usual Runge-Lenz
vector operator of the Kepler problem on the electron spin direction \cite{KK}.
In our approach, we can say that the Johnson-Lippmann operator is just the
supercharge $Q$ whose first term  given by Eq. (\ref{4QI}) is ${\cal
D}(\vec{\sigma}\cdot \vec{K})$.

Finally we note that our open superalgebra ${\cal S}_0$ could be enlarged
adding non-conserved operators that can be either leader operators or operators
related to the manifest supersymmetry \cite{CV2} of our Hamiltonian $H$.
However, this problem will be considered elsewhere.

\section{Concluding comments}

To conclude, the Dirac theory in curved space gives rise to a large collection
of conserved operators associated with standard Killing vectors or hidden
symmetries due to S-K or K-Y tensors. For example, in the case of the Euclidean
Taub-NUT space the conserved operators are involved in interesting and
non-trivial algebraic structures as dynamical algebras. The natural way to
organize the large collection of conserved operators is to arrange them in a
graded loop superalgebra of the Kac-Moody type. Further work must be done to
describe the involution automorphism which is needed to define the twisting in
connection with the graded loop algebra of the Kac-Moody type.

We believe that K-Y tensors deserve further attention. They are
involved in a multitude of different topics such as conformal S-K or
K-Y tensors,
non-standard
supersymmetries, quantum anomalies, index theorems, etc.
So far gravitational anomalies have proved to be absent for scalar
fields for spaces
admitting K-Y tensors and it would be valuable to know this persist in
the case of the full quantum field theories on curved spaces.
Concerning the axial anomaly and its connection with the index of the
Dirac operators the role of K-Y tensors
is not obvious. The topological properties of the spaces are more
important in comparison with non-standard symmetries.

\subsection*{Acknowledgments} The authors are grateful to Laszlo
Feher for interesting and useful discussions on closely related
subjects.
\setcounter{section}{0}\renewcommand{\thesection}{\Alph{section}}
\setcounter{equation}{0} \renewcommand{\theequation}
{A.\arabic{equation}}
\section{K\" ahlerian geometries}

Let us consider the manifold $M_n$ ($n=2k$) and its tangent fiber bundle,
${\cal T}(M_{n})$,  assuming that $M_n$ is equipped with a
{\em  complex structure} that is a particular bundle automorphism
$h: {\cal T}(M_{n}) \to {\cal T}(M_{n})$ which satisfies
$\left<h\right>^2=-{\mb I}$ and is covariantly constant.
Notice that the matrix of $h$ in local frames is an orthogonal
point-dependent transformation of the gauge group $G(\eta)$.
With its help one gives the following definition \cite{LM,GM}:
\begin{defin}
A Riemannian metric $g$ on $M_n$ is said K\" ahlerian if $h$ is
pointwise orthogonal, i.e., $g(hX,hY)=g(X,Y)$ for all $X,Y\in
{\cal T}_x(M_{n})$ at all points $x$.
\end{defin}
In local coordinates, $h$ is a skew-symmetric second rank tensor with
real-valued components, $h_{\mu\nu}=-h_{\nu\mu}$,
which obey $g_{\mu\nu}h^{\mu\,\cdot}_{\cdot\,\alpha}
h^{\nu\,\cdot}_{\cdot\,\beta}=g_{\alpha\beta}$. This gives rise to the
symplectic form $\tilde\omega=\frac{1}{2}h_{\nu\mu}dx^{\nu}\land dx^{\mu}$
(i.e., closed and non-degenerate).
Alternative definitions can be formulated starting with both, $g$ and
$\tilde\omega$, which have to satisfy the K\" ahler relation
$\tilde\omega(X,Y)=g(X,hY)$ \cite{GM}.

A {\em hypercomplex structure} on $M_n$ is an ordered triplet $H = (h^1, h^2,
h^3)$ of complex structures on $M_n$ satisfying Eq. (\ref{algf}).
In Lie algebraic terms, the matrices ${1 \over 2} \left<h^j\right>$ realize
the $su(2)$ algebra.
\begin{defin}
A hyper-K\" ahler manifold is a manifold whose Riemannian metric is
K\"ahlerian with respect to each different complex structures
$h^1, h^2$ and $h^3$.
\end{defin}

Our unit roots, $f$, are defined in a similar way as the complex structures
with the difference that the unit roots are automorphisms of the
{\em complexified} tangent bundle,
$f: {\cal T}(M_{n})\otimes{\Bbb C} \to {\cal T}(M_{n})\otimes {\Bbb C}$.
Therefore, $f$ have complex-valued components and the transformation
matrix $\left<f\right>$ is of the complexified group $G_{c}(\eta)$. Thus
it is clear that the real-valued unit roots are complex structures as defined
above. The families of unit roots may differ from the hypercomplex structures
but have the same algebraic properties given by Eq. (\ref{algf}).

The passing from the complex structures to unit roots is productive
from the point of view of the Dirac theory since in this way
one can introduce families of unit roots generating superalgebras of
Dirac-type operators even in manifolds which do not admit complex structures.
The Minkowski spacetime is a typical example.

\setcounter{equation}{0} \renewcommand{\theequation}
{B.\arabic{equation}}

\section{The Minkowski spacetime}

In the Minkowski spacetime, $M_{3+1}$, with the metric $\eta=(1,-1,-1,-1)$ we
consider the gauge of the {\em inertial} frames,
$e^{\mu}_{\nu}=\hat e^{\mu}_{\nu}=\delta^{\mu}_{\nu}$, ($\mu,\nu,...=0,1,2,3$).
In this gauge we use the chiral representation of the Dirac matrices
(with off-diagonal $\gamma=\gamma^0$ \cite{TH})
where the standard Dirac operator reads
\begin{equation}
D=i\gamma^{\mu}\partial_{\mu}=\left(
\begin{array}{cc}
0&i(\partial_t+\vec{\sigma}\cdot\vec{\partial})\\
i(\partial_t-\vec{\sigma}\cdot\vec{\partial})&0
\end{array}\right)=\left(
\begin{array}{cc}
0&D^{(+)}\\
D^{(-)}&0
\end{array}\right)\,,
\end{equation}
and the generators of the spinor representation of the group
${\mb G}(\eta) = SL(2,{\Bbb C})$ take the form
$S^{ij}=\varepsilon_{ijk} S_k=\frac{1}{2}\varepsilon_{ijk}{\rm diag}
(\sigma_k, \sigma_k)$ and $S^{i0}=\frac{i}{2}{\rm diag}(\sigma_i,-\sigma_i)$.

The isometries of $M_{3+1}$ are just the transformations
$x'=\Lambda(\omega)x -a$ of the Poincar\' e group,
${\cal P}_{+}^{\uparrow} = T(4)~\circledS~ L_{+}^{\uparrow}$  \cite{W}.
If we denote by $\xi^{(\mu\nu)}=\omega^{\mu\nu}$ the $SL(2,\Bbb C)$ parameters
and by $\xi^{(\mu)}=a^{\mu}$ those of the translation group $T(4)$, then
we obtain the standard basis generators
\begin{eqnarray}
X_{(\mu)}&=&i\partial_{\mu} \,,\\
X_{(\mu\nu)}&=&i(\eta_{\mu\alpha}x^{\alpha}
\partial_{\nu}- \eta_{\nu\alpha}x^{\alpha}\partial_{\mu})+
S_{\mu\nu} \,,
\end{eqnarray}
which show us that in this gauge $\psi$ transforms manifestly covariant.
On the other hand, it is clear that the group
$S(M_{3+1})= \tilde{\cal P}^{\uparrow}_{+}
\sim T(4)~\circledS~ SL(2,\Bbb C)$
is just the universal covering group of $I(M_{3+1})={\cal P}_{+}^{\uparrow}$.
In applications it is convenient to denote
${\cal J}_i=\frac{1}{2}\varepsilon_{ijk}X_{jk}$ and  ${\cal K}_i=X_{0i}$.

The Minkowski spacetime possesses a pair of adjoint triplets \cite{K2}. The
unit roots of the first triplet, ${\mb f}$, have the non-vanishing
complex-valued  components \cite{K2}
\begin{eqnarray}
f^{(1)}_{23}=1\,,\quad&f^{(2)}_{31}=1\,,&\quad f^{(3)}_{12}=1\,,\\
f^{(1)}_{01}=i\,,\quad&f^{(2)}_{02}=i\,,&\quad f^{(3)}_{03}=i\,,
\end{eqnarray}
giving rise to the spin-like operators
\begin{equation}
\Sigma^{(i)}=\frac{1}{2}f^{(i)}_{\mu\nu}S^{\mu\nu}=\left(
\begin{array}{cc}
\sigma_i&0\\
0&0
\end{array}\right) \,,
\end{equation}
and to the Dirac-type operators
\begin{equation}\label{Dto}
D^{(i)}=i[D,\,\Sigma^{(i)}]=\left(
\begin{array}{cc}
0&-i\sigma_i D^{(+)}\\
iD^{(-)}\sigma_i&0
\end{array}\right) \,,
\end{equation}
which {\em anticommute} with each other as well as with $D$ and $\gamma^0$.

The operators $D$ and $D^i$ form a basis for the superalgebra
$({\mb d}_{\mb f})_c$ defined now over ${\Bbb C}$ since the isometries will
give rise to complex-valued orthogonal transformations among $D^{(i)}$. The
spinor representation of group $SL(2, {\Bbb C})\subset S(M_{3+1})$ is generated by
${\cal J}_i$ and ${\cal K}_i$ while that of the group
$(G_{\mb f})_c\sim SL(2, {\Bbb C})$ is
generated by the operators $\hat s_i=\frac{1}{2}\Sigma^{(i)}$ and
$\hat r_i =-\frac{i}{2}\Sigma^{(i)}$. All these generators satisfy usual
$sl(2, {\Bbb C})$ commutation rules and
\begin{eqnarray}
&[{\cal J}_i ,\, \hat s_j]=i\varepsilon_{ijk} \hat s_k\,,\quad
&[{\cal K}_i,\, \hat s_j]=i\varepsilon_{ijk} \hat r_k\,,\nonumber\\
&[{\cal J}_i,\, \hat r_j]=i\varepsilon_{ijk} \hat r_k\,,\quad
&[{\cal K}_i,\, \hat r_j]=-i\varepsilon_{ijk} \hat s_k\,.\label{ubauba}
\end{eqnarray}
The next step is to construct the group
 $O_{\mb f}$ of the complex-valued orthogonal
matrices defined by Eq. (\ref{rotroot}) that read
\begin{equation}
\hat{\frak R}_{ij}(\omega)=\textstyle\frac{1}{4}
f^{(i)\,\alpha\beta}
\Lambda^{\mu\,\cdot}_{\cdot\,\alpha}(\omega)
\Lambda^{\nu\,\cdot}_{\cdot\,\beta}(\omega)
f^{(j)}_{\mu\nu}\,.
\end{equation}
These form a representation of the group
$I(M_{3+1})= {\cal P}^{\uparrow}_{+}$ {\em induced} by the group
$O(3)_c$. Of course, the translations have no effects in this
representation remaining only with the transformations  $\Lambda(\omega)
\in O(3,1)$. These give rise to non-trivial matrices $\hat{\frak R}(\omega)$
as, for example,
\begin{equation}
\hat{\frak R}(\varphi)=\left(
\begin{array}{ccc}
1&0&0\\
0&\cos\varphi&\sin\varphi\\
0&-\sin\varphi&\cos\varphi
\end{array}\right)\,,\quad
\hat{\frak R}(\alpha)=\left(
\begin{array}{ccc}
1&0&0\\
0&\cosh\alpha&i\sinh\alpha\\
0&-i\sinh\alpha&\cosh\alpha
\end{array}\right)\,,
\end{equation}
calculated for non-vanishing parameters $\omega_{23}=\varphi$ (a rotation
around $x^1$) and respectively $\omega_{01}=\alpha$ (a boost along $x^1$).
Hereby we see that $O_{\mb f}\sim O(3)_c$ which requires
linear structures defined over ${\Bbb C}$ instead of ${\Bbb R}$ as we used in
the case of the hyper-K\" ahler manifolds. Thus we have all the ingredients we
need to write down the action of the transformations of the group
$Aut({\mb d}_{\mb f})_c$. We note that Eqs. (\ref{ubauba}) show that
$(G_{\mb f})_c$ is an invariant subgroup.

The second triplet is ${\mb f}^*$ for which all the spinor quantities are
just the Dirac conjugated of those of ${\mb f}$. The corresponding spin-like
operators are
\begin{equation}
\overline{\Sigma^{(i)}}=\frac{1}{2}\left(f^{(i)}_{\mu\nu}\right)^{*}S^{\mu\nu}
=\left(
\begin{array}{cc}
0&0\\
0&\sigma_i
\end{array}\right) \,,
\end{equation} which means that the representations  $spin({G}_{\mb f})$ and
$spin({G}_{{\mb f}^*})$ of $SU(2)$ act {\em separately} on the left and
right-handed parts of the Dirac spinor. Moreover, it is interesting to observe
that $\Sigma^{(i)}+\overline{\Sigma^{(i)}}=2 S^i$. This perfect balance between
the chiral sectors is due to the fact that the operator
$D^{(+)}D^{(-)}=D^{(-)}D^{(+)}$ commutes with $\sigma_i$.

The discrete symmetry is given by two representations of the quaternion group
acting on each of both chiral sectors. On the left-handed sector acts the
group  ${\Bbb Q}({\mb f})$ represented by the operators ${\mb 1}$,
$\digamma=\gamma^5={\rm diag}(-{\mb 1}_2,{\mb 1}_2)$,
$U_{(i)}={\rm diag}(i\sigma_i, {\mb 1}_2)$ and  $\gamma^5 U_{(i)}$.
The operators of the group of the right-handed sector, ${\Bbb Q}({\mb f}^*)$,
can be obtained using the Dirac adjoint and taking into account that here
$\overline{\gamma^5}=-\gamma^5$. The resulted operators,
${\rm diag}({\mb 1}_2, \pm i\sigma_i)$, act only on the right-handed sector.

Hence it is clear that each chiral sector has its own sets of unit roots
defining Dirac-type operators. These are  the spheres ${S}^2_{\mb f}$ of the
left-handed sector and ${S}^2_{{\mb f}^*}$ of the right-handed one. Since there are
no other independent unit roots, the whole set of unit roots  of the Minkowski
spacetime is ${\mb R}_1(M_{3+1})={S}^2_{\mb f}\bigcup {S}^2_{{\mb f}^*}$. The
continuous and discrete symmetry groups of the Dirac-type operators are
defined separately on each of these two spheres.

\setcounter{equation}{0} \renewcommand{\theequation} {C.\arabic{equation}}

\section{The superalgebra ${\cal W}_0$}

Here we would like to show that a fundamental representation of the
superalgebra ${\cal W}_0$ arises from a particular representation of the
$so(4)$ Lie algebra.

We start with a finite-dimensional representation, $\rho$, of this algebra
generated by the linear operators $\{A^i_{\rho},B^i_{\rho}\}$ defined on the
space ${\frak M}_{\rho}$ and obeying Eqs. (\ref{algAB}). The identity operator
of on ${\frak M}_{\rho}$ is denoted by $1_{\rho}$.  The $su(2)\times su(2)$
content of the $so(4)$ algebra can be pointed out in the new basis
$\{J^{i}_{\rho\,+},J^{i}_{\rho\,-}\}$ given by the operators
$J^{i}_{\rho\,\pm}= \frac{1}{2}(A_{\rho}^i\pm B_{\rho}^i)$ that satisfy the
$su(2)$ commutation relations,
\begin{equation}\label{algJJ}
[ {J}^{i}_{\rho\,+},\, {J}^{j}_{\rho\,+}]=i\varepsilon _{ijk}{J}^{k}_{\rho\,+}
\,,\quad [ {J}^{i}_{\rho\,-},\, {J}^{j}_{\rho\,-}]=i\varepsilon
_{ijk}{J}^{k}_{\rho\,-} \,,\quad [ {J}^{i}_{\rho\,+},\, {J}^{j}_{\rho\,-}]=0\,.
\end{equation}
The representation $\rho=(j^+,j^-)$ is completely determined by the $su(2)$
weights defined by the Casimir operators
$\vec{J}^2_{\rho\,\pm}=j^{\pm}(j^{\pm}+1)\,1_{\rho}$. However, here we have to
consider, in addition, the usual Casimir operators
$C_{\rho\,1}=\vec{A}^2_{\rho}+\vec{B}^2_{\rho}$ and
$C_{\rho\,2}=\vec{A}_{\rho}\cdot\vec{B}_{\rho}$ or the new ones
\begin{equation}
C_{\rho \pm}=C_{\rho\, 1}\pm 2 C_{\rho\, 2} +1_{\rho}=4 \vec{J}_{\rho\, \pm}^ 2
+ 1_{\rho} = (2j^{\pm}+1)^2\, 1_{\rho} \,.
\end{equation}
We note that when $j^+=j^-$ then $\vec{A}_{\rho}$ and $\vec{B}_{\rho}$ are
orthogonal, as in the case of the dynamical algebra of the quantum Kepler
problem.

Our purpose is to construct the superalgebra ${\cal W}_0$ in the carrier space
${\frak M}={\frak M}_{\rho}\otimes {\frak M}_{(\frac{1}{2},0)}$ of the
reducible representation $(j^+,j^-)\otimes (\frac{1}{2},0)$ given by arbitrary
weights $j^{\pm}$ taking positive real values. The representation,
$(\frac{1}{2},0)$, is generated by the operators $\hat A^i=\frac{1}{2}\sigma_i$
and  $\hat B^i=\frac{1}{2}\sigma_i$ acting in the two-dimensional space ${\frak
M}_{(\frac{1}{2},0)}$ where the identity operator is $1_2$. In these
circumstances we define first the identity operator on ${\frak M}$,
$E=1_{\rho}\otimes 1_2$, and the Casimir-type operator $F=C_{\rho\,+}\otimes
1_2$. The $so(4)$ generators of this representation are
\begin{equation}
{A}^i=A^i_{\rho}\otimes 1_2 +{\textstyle \frac{1}{2}}1_{\rho} \otimes
\sigma_i\,,\quad {B}^i=B^i_{\rho}\otimes 1_2 +{\textstyle
\frac{1}{2}}1_{\rho}\otimes \sigma_i\,.
\end{equation}
Moreover, we introduce the supercharges
\begin{equation}
G_i=1_{\rho}\otimes \sigma_i\,, \quad G=\vec{A}_{\rho}\otimes
\vec{\sigma}+\vec{B}_{\rho}\otimes \vec{\sigma}+E\,,
\end{equation}
so that $G^2=F$. Now it is a simple exercise to show that the operators
\begin{equation}
\{E,F, A^i,B^i,G,G^i\}
\end{equation}
 satisfy Eqs. (\ref{algAB}), (\ref{algAB1}) and
(\ref{algAB2}).

The conclusion is that the superalgebra ${\cal W}_0$ can be realized in the
carrier space of any reducible representation $\rho\otimes (\frac{1}{2},0)$ of
the $so(4)$ algebra.

\end{document}